\documentclass[aps,twocolumn]{revtex4-1}
\usepackage{times}
\usepackage{graphicx}
\usepackage{amsfonts}
\usepackage{amsmath, amsthm, amssymb}
\usepackage{dsfont}
\usepackage{color}
\usepackage{empheq}
\usepackage{standalone}
\usepackage[most]{tcolorbox}
\usepackage{physics}
\usepackage{bm}
\usepackage[utf8]{inputenc}
\usepackage{amssymb,amsmath,amsfonts,amscd}
\usepackage{mathrsfs}
\usepackage{siunitx}
\usepackage{commath}
\usepackage{graphicx}      
\graphicspath{ {Images/} } 
\usepackage{braket}
\usepackage{bm}
\usepackage{empheq}
\usepackage[most]{tcolorbox}
\usepackage{physics}
\usepackage{placeins}

\usepackage{hyperref} 
\hypersetup{
    colorlinks=true,
    linkcolor=blue,    
    citecolor=blue,    
    urlcolor=blue,     
}

\begin{document}

\title{Eliashberg study of superconductivity induced by interfacial coupling to antiferromagnets}
\author{Even Thingstad}
\thanks{These authors contributed equally to this work}
\affiliation{\mbox{Center for Quantum Spintronics, Department of Physics, Norwegian University of Science and Technology,}\\NO-7491 Trondheim, Norway}

\author{Eirik Erlandsen}
\thanks{These authors contributed equally to this work}
\affiliation{\mbox{Center for Quantum Spintronics, Department of Physics, Norwegian University of Science and Technology,}\\NO-7491 Trondheim, Norway}
 
\author{Asle Sudb\o}
\email[Corresponding author: ]{asle.sudbo@ntnu.no}
\affiliation{\mbox{Center for Quantum Spintronics, Department of Physics, Norwegian University of Science and Technology,}\\NO-7491 Trondheim, Norway}

\begin{abstract}
We perform Eliashberg calculations for magnon-mediated superconductivity in a normal metal, where the electron-magnon interaction arises from interfacial coupling to antiferromagnetic insulators. In agreement with previous studies, we find $p$-wave pairing for large doping when the antiferromagnetic interfaces are uncompensated, and $d$-wave pairing close to half-filling when the antiferromagnetic interfaces are compensated. However, for the $p$-wave phase, we find a considerable reduction in the critical temperature compared to previous weak-coupling results, as the effective frequency cutoff on the magnon propagator in this case is found to be much smaller than the cutoff on the magnon spectrum. The $d$-wave phase, on the other hand, relies less on long-wavelength magnons, leading to a larger effective cutoff on the magnon propagator. Combined with a large density of states close to half-filling, this might allow the $d$-wave phase to survive up to higher critical temperatures. Based on our findings, we provide new insight into how to realize interfacially induced magnon-mediated superconductivity in experiments.
\end{abstract}

\pacs{Valid PACS appear here}

\maketitle

\section{Introduction}\label{Section:Introduction}

For conventional superconductors, the fluctuations responsible for Cooper-pairing of electrons are provided by phonons \cite{Bardeen1957}. As the role of the phonons is simply to introduce attractive interaction between electrons, superconductivity can in principle arise from exchange of any bosonic quasiparticle that is able to provide a similar attractive interaction \cite{Schlawin2019, Kavokin2016, Laussy2010, Takada1978}. One alternative that has received much attention is exchange of paramagnetic spin-fluctuations \cite{Scalapino1999, Moriya2003}. The idea is that the spins in a paramagnet, close to magnetic ordering, can act like a medium that can be polarized by the spin of an electron. Another electron can then interact with the polarized medium, giving rise to an effective interaction between the electrons. The quasiparticle mediating the interaction, the paramagnon, represents a damped spin-wave propagating in an ordered patch of the paramagnet \cite{Berk1966, Doniach1966}.\\
\indent The paramagnon exchange mechanism has been proposed to be closely related to the superconductivity of heavy fermion materials~\cite{cyrot1986, Scalapino1986, Miyake1986} and high-$T_c$ cuprates~\cite{Monthoux1991, Monthoux1992}. In the context of the Hubbard model, paramagnon exchange has been found to give rise to $p$-wave superconductivity for small isotropic Fermi surfaces, and $d$-wave superconductivity closer to half-filling \cite{Scalapino1986}. This $d$-wave superconductivity arises from antiferromagnetic fluctuations, so that the interaction is peaked at finite momentum. Although the spin-singlet $s$-wave channel is repulsive, the \(d\)-wave channel is then able to become attractive by taking advantage of sign changes in the gap function~\cite{Scalapino1999}. \\
\indent In these systems, superconductivity arises from interactions between fermions due to their own collective spin excitations \cite{Monthoux1991, Monthoux1992,Abanov2003}. Spin-fluctuation mediated superconductivity may also occur in heterostructures with itinerant fermions proximity coupled to the spins of insulating materials~ \cite{Kargarian2016, Gong2017, Fjaerbu2018, Hugdal2018, Fjaerbu2019, Erlandsen2019_Enhancement, Erlandsen2020_Schwinger, Erlandsen2020_TI, Hugdal2020}. Since the spins and the itinerant fermions are then separate degrees of freedom, this  provides a simpler context to study superconductivity mediated by spin-fluctuations.\\
\indent Magnon-mediated superconductivity induced in a normal metal (NM) due to proximity-coupling to a magnetic insulator has so far been investigated within a weak-coupling BCS framework \cite{Fjaerbu2018, Fjaerbu2019, Erlandsen2019_Enhancement, Erlandsen2020_Schwinger}. The first case to be considered was a NM coupled to ferromagnetic insulators, which was found to give rise to $p$-wave pairing \cite{Fjaerbu2018}. Similarly, for a NM coupled to an antiferromagnetic insulator (AFMI), $p$-wave solutions were obtained for large dopings by exploiting the inherent squeezing of antiferromagnetic magnons \cite{Kamra2019_Antiferromagnetic} by coupling the conduction electrons in the NM asymmetrically to the two sublattices of the AFMI \cite{Erlandsen2019_Enhancement}. This sublattice coupling asymmetry suppresses sublattice interferences in the pairing potential, which are very unfavorable for the $p$-wave phase. A general asymmetry of this type can be realized by employing an antiferromagnetic interface where both sublattices are exposed (compensated interface), but further breaking the sublattice symmetry by using an antiferromagnetic material with two different atoms on the two sublattices. The particularly relevant case of coupling to only one of the two sublattices is, however, achieved through an uncompensated antiferromagnetic interface where only one of the two sublattices is exposed \cite{Nogues1999, Nogues2005, Stamps2000}. \\
\indent For the case of a compensated antiferromagnetic interface, the magnons live in a Brillouin zone which is reduced compared to the electron Brillouin zone. This introduces electron-magnon scattering processes of two types: regular and Umklapp~\cite{Takei2014, Fjaerbu2017}. In the regular processes, the electrons are scattered with a momentum within the first magnon Brillouin zone. In the Umklapp processes, on the other hand, the outgoing electron receives an additional momentum corresponding to a magnon reciprocal space lattice vector.
The Umklapp processes are of little relevance for the small Fermi surfaces considered in Ref.\! \cite{Erlandsen2019_Enhancement}, but closer to half-filling they have been predicted to give rise to $d$-wave superconductivity in a normal metal sandwiched between two compensated antiferromagnetic interfaces \cite{Fjaerbu2019}. Analogously to the case of paramagnon exchange in the Hubbard model, the $d$-wave pairing arises from a repulsive $s$-wave channel and an interaction that is peaked at finite momentum.\\
\indent We also note that a normal metal coupled to a compensated antiferromagnetic interface is similar to a single material with antiferromagnetically ordered localized spins and itinerant electrons treated as separate degrees of freedom, considered e.g.\! in Refs.\! \cite{Vonsovskii1963, Chen1988, Shimahara1994}. While Ref.\! \cite{Vonsovskii1963} simply found the spin singlet $s$-wave channel to be repulsive for magnon-mediated pairing, Ref.\! \cite{Chen1988} also considered the spin triplet channel and found $p$-wave superconductivity due to their treatment not probing the interference effects discussed in Ref.\! \cite{Erlandsen2019_Enhancement}. Ref.\! \cite{Shimahara1994}, on the other hand, found that two-magnon scattering processes were dominant for small Fermi surfaces due to the strong destructive interference for one-magnon processes, while spin singlet $d$-wave pairing driven by one-magnon processes could be possible for larger Fermi surfaces. \\ 
\indent A notable difference between the electron-phonon coupling in common weak-coupling superconductors and the electron-magnon coupling considered in the present study, is the behaviour of the coupling matrix element in the limit of small momentum transfers. Since the electron-phonon coupling represents a coupling between electrons and spatial fluctuations of ion densities, it vanishes at zero momentum. In contrast, the coupling between the spins of itinerant electrons and the localized spins of the magnetic insulator is local, and therefore constant in momentum space. For the magnon-mediated superconductivity discussed in the above references, this allows processes with small scattering momentum and small magnon frequencies to dominate the superconducting pairing. In turn, these small momentum processes can compensate for the relatively small interfacial coupling strength of order $10\, \textrm{meV}$ \cite{Kajiwara2010, Fjaerbu2018}, which is typically smaller than the energy scale for the electron-phonon coupling giving rise to phonon-mediated superconductivity~\cite{Mazzola2017, Coleman2015}. \\
\indent When the dominant contributions to the pairing arise from long-wavelength magnons, one should expect that it may no longer be reasonable to use the cutoff on the boson spectrum as the characteristic boson energy setting the energy scale for the critical temperature. This is not captured in simple BCS theory, which does not consider the frequency dependence of the bosonic fluctuation spectrum responsible for pairing. 
Furthermore, renormalization of both electrons and bosons is neglected in BCS theory, and these effects could turn out to play a more essential role here. Although BCS theory explains phonon-mediated superconductivity in weak-coupling superconductors reasonably well, a more detailed analysis may be required when other pairing mechanisms are involved.\\ 
\indent In this paper, we therefore investigate superconductivity induced in a NM by interfacial coupling to antiferromagnetic insulators using an Eliashberg theory framework. In addition to exploring how the existing results change when the electron renormalization and the proper frequency dependence of the electron-magnon interaction are taken into account, we also study the effect of magnon renormalization and discuss the importance of vertex corrections. Instead of focusing only on regular \cite{Erlandsen2019_Enhancement} or Umklapp processes \cite{Fjaerbu2019}, we simultaneously take both types of processes into account and examine how the superconductivity varies with both chemical potential and asymmetry in the coupling to the two sublattices of the antiferromagnet.\\
\indent In agreement with earlier results, we find a $p$-wave phase for large sublattice coupling asymmetry and large doping, and a $d$-wave phase for small sublattice coupling asymmetry and small doping. For the $p$-wave phase, the critical temperature is considerably reduced compared to previous weak-coupling studies due to the reduction of the effective magnon frequency cutoff. However, the $d$-wave phase is found to be less reliant on exchange of long-wavelength magnons. This leads to a larger effective cutoff. Near half-filling, the reduction in the contributions from long-wavelength magnons for the $d$-wave phase can be compensated by a larger density of states, opening up for the possibility of larger critical temperatures. For a strongly nested Fermi-surface, however, one needs to consider  e.g.\! the possibility of a competing spin-density wave instability. 
Moreover, while a sufficiently large gap in the magnon spectrum may be necessary to protect the ordering of the magnet upon inclusion of magnon renormalization, the net effect on possible critical temperatures is found to be small. \\
\indent In Sec. \ref{Section:model}, we present the model of our system. In \ref{Section:EliashbergTheory}, we outline the Eliashberg theory for magnon-mediated superconductivity. We further derive the Fermi surface averaged Eliashberg equations in Sec. \ref{Section:FS_average}, and present results for these equations in Sec.\! \ref{Section:solve}. In Sec.\! \ref{Section:MagnonRenormalization} we move on to the effect of renormalization of the magnons. Finally, we discuss the validity of the results, as well as additional neglected effects in \ref{Section:Discussion}, and experimental considerations in \ref{Section:ExperimentalConsiderations}, before we summarize in Sec.\! \ref{Section:Summary}. Additional details, as well as a discussion of the role of vertex corrections can be found in the appendices. \\

\section{Model}\label{Section:model}
\begin{figure}[t] 
    \begin{center}
        \includegraphics[width=0.8\columnwidth,trim= 2.0cm 12.8cm 3.0cm 7.0cm,clip=true]{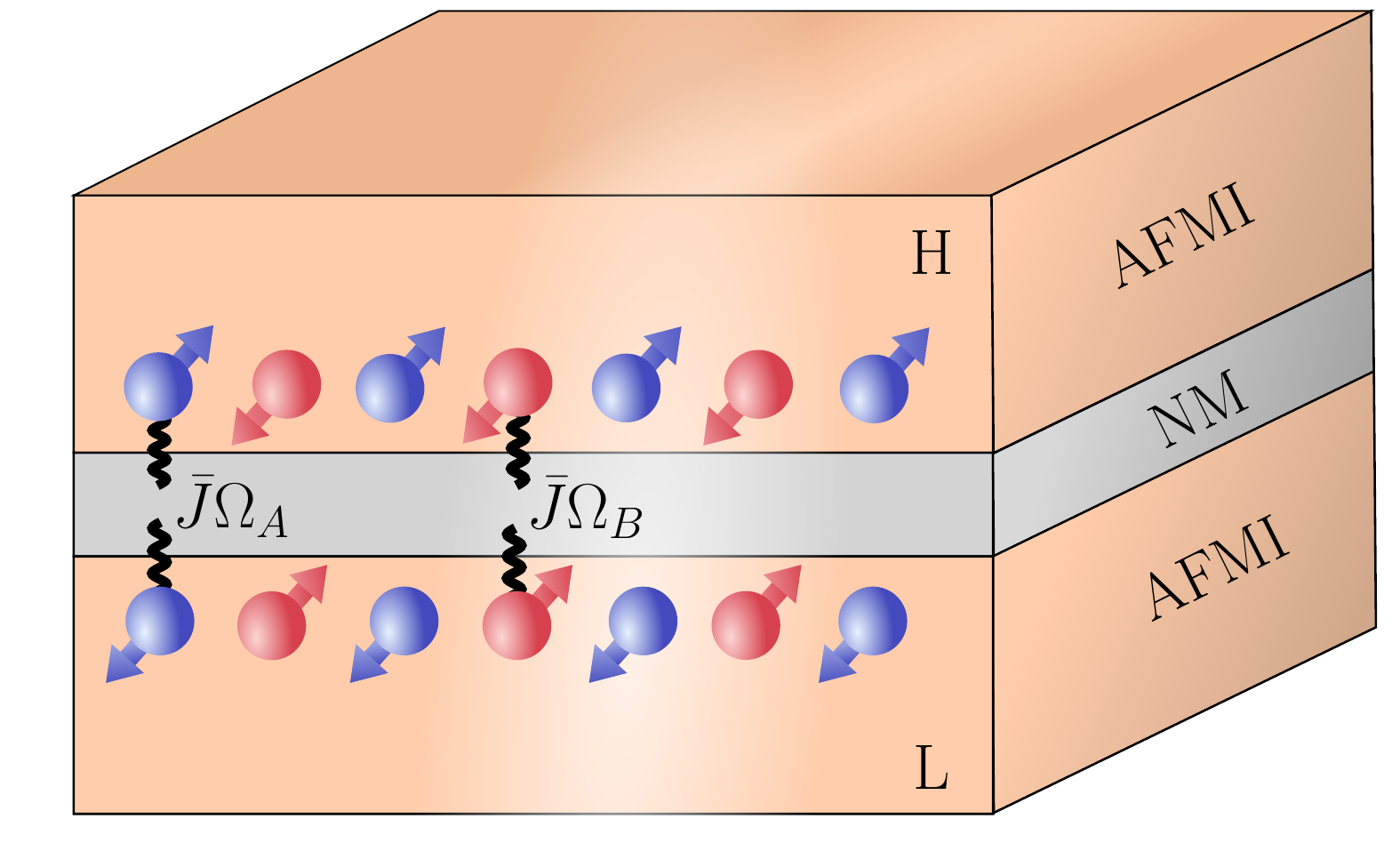}
    \end{center}
    \caption{A trilayer consisting of a normal metal (NM) layer sandwiched between two antiferromagnetic insulator (AFMI) layers. The $A$ and $B$ sublattices of the AFMIs consist of the blue and red lattice sites, respectively. The two AFMIs are oppositely ordered so that the spins associated with a specific sublattice are oppositely oriented for the highest (H) and lowest (L) AFMI. The coupling to the $A$ sublattices of both AFMIs is taken to be of equal strength ($\bar{J}\Omega_A$), and similarly for the $B$ sublattices, so that the itinerant electrons in the NM experience no net magnetic field. The coupling to the $A$ sublattices is however allowed to differ from the coupling to the $B$ sublattices.}
    \label{fig:model}
\end{figure}

\begin{figure}[b] 
    \begin{center}
        \includegraphics[width=0.8\columnwidth]{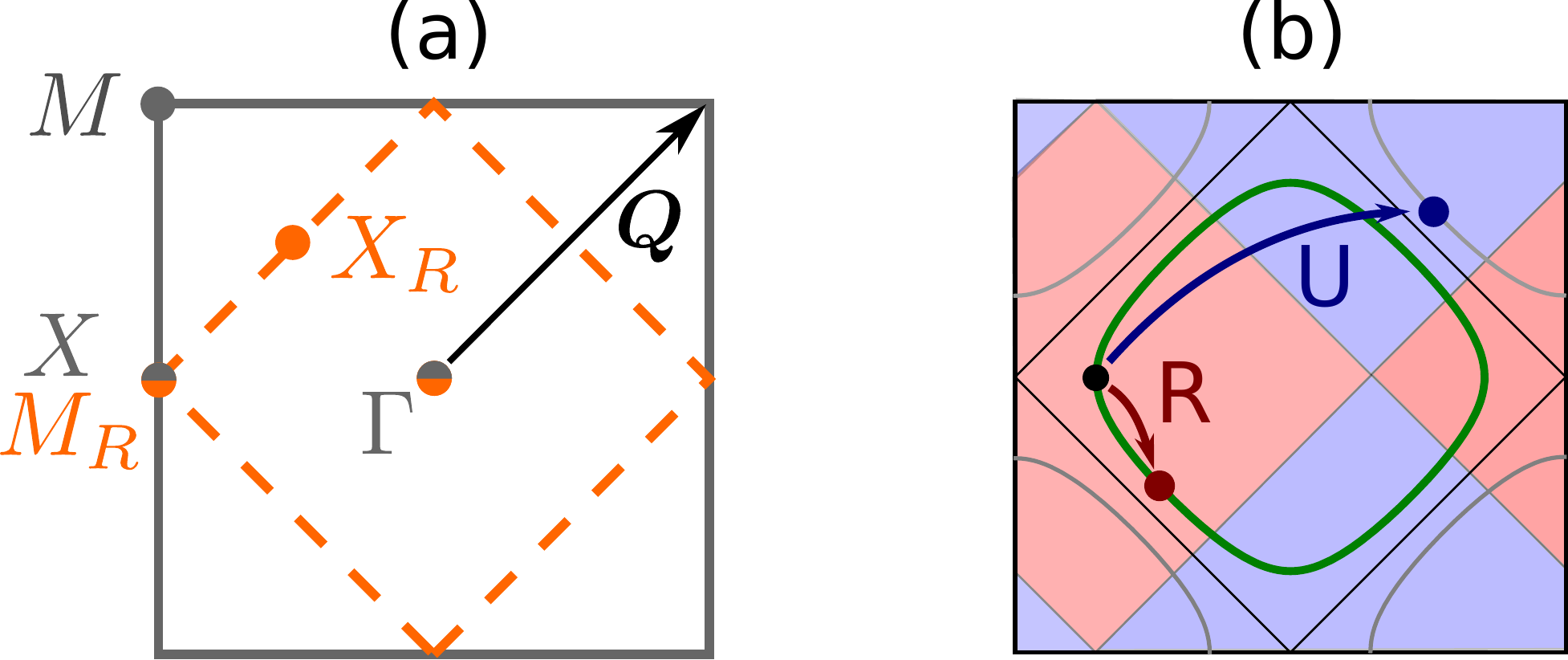}
    \end{center}
    \caption{(a) Electron (grey) and magnon (orange) Brillouin zones with labelling of high symmetry points. We refer to the magnon Brillouin zone as the reduced Brillouin zone (RBZ). The antiferromagnetic ordering vector \(\bm{Q}\) is also indicated.  (b) Fermi surface (green) at moderate doping. Electrons can be scattered from \(\bm{k}\) (black) to points \(\bm{k} + \bm{q}\) inside the shaded red part of the Brillouin zone through regular processes, and to points \(\bm{k} + \bm{q} + \bm{Q}\) in the shaded blue part of the Brillouin zone through Umklapp processes.}
    \label{fig:BrillouinZone}
\end{figure}

We consider a trilayer heterostructure consisting of a normal metal sandwiched between two antiferromagnets, as shown in Fig.~\ref{fig:model}. The experimental realization of the system would consist of a thin NM layer between two thicker AFMI layers. 
For simplicity, we model the system using two-dimensional lattice models for the three distinct layers. We assume that the antiferromagnets have staggered magnetic order along the $z$-direction in spin space, and that this order is opposite in the two antiferromagnets. In general, the spin space $z$-direction can be either in-plane or out-of-plane in real space for our model. \\
\indent We model the system with the Hamiltonian {\(H = H_\mathrm{NM} + H_\mathrm{AFMI} + H_\mathrm{int}\)}, where 

\begin{subequations}
\begin{align}
H_\mathrm{NM} = -  \sum_{ ij,\sigma } t_{ij} c_{i\sigma}^\dagger c_{j \sigma} - \mu \sum_{i\sigma} c_{i\sigma}^\dagger c_{i\sigma},
\label{eq_hamiltonian_nm}
\\
H_\mathrm{AFMI} = \sum_{ij, \eta} J_{ij} \bm{S}_{i\eta} \cdot \bm{S}_{j\eta}
- K \sum_{i,\eta} (S_{i\eta}^z)^2,
\label{eq_hamiltonian_afmi}
\\
H_\mathrm{int} = - 2\bar{J}  \sum_{\eta,\Upsilon}\sum_{i\in \Upsilon} \Omega_{\Upsilon}^{\eta} c_{i}^\dagger \bm{\sigma} c_{i} \cdot \bm{S}_{i\eta},
\label{eq_hamiltonian_interaction}
\end{align}
\end{subequations}

\noindent and the terms describe the normal metal, the antiferromagnetic insulators, and the interfacial coupling between the materials. The sums over \(i,j\) denote sums over lattice sites, the sum over \(\eta\in\{H,L\}\) denotes a sum over the the two antiferromagnetic insulators, and the sum over \(\Upsilon\in\{A,B\}\) denotes a sum over the sublattices.  All three layers are modelled by square lattices with periodic boundary conditions. 
In the normal metal, our model describes spinful electrons with annihilation and creation operators \(c_{i\sigma}\) and \(c_{i\sigma}^\dagger\) for an electron on site \(i\) with spin \(\sigma\). 
The electron chemical potential is expressed as $\mu$, and \(t_{ij}\) is the hopping amplitude, which we set to \(t\) for nearest neighbours and zero otherwise. 
The AFMIs in our model consist of localized lattice site spins, where \(\bm{S}_{i\eta}\) denotes the spin on site \(i\) in antiferromagnet \(\eta\). The exchange coupling between the spins on lattice sites \(i\) and \(j\) 
is \(J_{ij}\), which we assume to take the value \(J_1 > 0\) for nearest neighbour and \(J_2\) for next-nearest neighbour sites. Moreover, \(K > 0\) denotes the easy axis anisotropy,
The interfacial coupling between the materials is included as an effective exchange interaction $\bar{J}$ between the lattice site spins in the antiferromagnets and the spins of the conduction band electrons that are confined to the normal metal \cite{Kajiwara2010, Takahashi2010, Bender2015, Fjaerbu2018, Fjaerbu2019}. We use 
the notation \(c_i = (c_{i\uparrow}, c_{i\downarrow})^T\), and have taken \(\bm{\sigma}\) to denote the Pauli matrix vector in spin space. In order to be able to introduce asymmetry in the coupling between the normal metal and the two sublattices of the antiferromagnets, we have included a dimensionless, sublattice- and layer-dependent, parameter \(\Omega_{\Upsilon}^{\eta}\) in the interaction Hamiltonian \cite{Erlandsen2019_Enhancement, Erlandsen2020_TI, Erlandsen2020_Schwinger}. In order to eliminate any magnetic fields, we will focusing on equal coupling to the two antiferromagnets \cite{Fjaerbu2019}, and therefore let \(\Omega_{\Upsilon}^{\eta} \equiv \Omega_{\Upsilon}\). In the following, we set $\hbar = a = 1$, with $a$ being the lattice constant.\\
\indent The normal metal Hamiltonian can be diagonalized to obtain 

\begin{equation}
H_\mathrm{NM} = \sum_{\bm{k} \in \square, \sigma} \xi_{\bm{k}}\, c^\dagger_{\bm{k} \sigma} c_{\bm{k} \sigma},
\end{equation}

\noindent where the quasimomentum sum runs over the full Brillouin zone, we have defined \(\xi_{\bm{k}} = \epsilon_{\bm{k}} - \mu\), and the single particle electron dispersion relation is given by \(\epsilon_{\bm{k}} = -2t(\cos k_x + \cos k_y)\).\\ 
\indent To determine the eigenexcitations of the antiferromagnetic insulator, we introduce the linearized Holstein-Primakoff transformation to represent the spins in terms of bosons \({a}_{i\eta}\) and \(b_{i\eta}\) on the two sublattices of the system. Further, introducing the Fourier transformed operators \(a_{\bm{q}\eta}\) and \(b_{\bm{q}\eta}\), one may diagonalize the AFMI Hamiltonian using a Bogoliubov transformation 

\begin{subequations}
\begin{align}
a_{\bm{q}\eta} &= u_{\bm{q}} \alpha_{\bm{q}\eta} + v_{\bm{q}} \beta_{-\bm{q}\eta}^\dagger, \\
b_{-\bm{q}\eta}^\dagger &= u_{\bm{q}} \beta_{-\bm{q}\eta}^\dagger 
+ v_{\bm{q}} \alpha_{\bm{q}\eta},
\end{align}
\end{subequations}
as detailed in Appendix \ref{Section:Antiferromagnet}. By suitable choice of coherence factors \(u_{\bm{q}}\) and \(v_{\bm{q}}\), the AFMI Hamiltonian takes the form

\begin{equation}
H_\mathrm{AFMI} = \sum_{\bm{q} \in \diamondsuit, \eta} \omega_{\bm{q}} (\alpha_{\bm{q}\eta}^\dagger \alpha_{\bm{q}\eta} + \beta_{\bm{q}\eta}^\dagger \beta_{\bm{q}\eta}),
\end{equation}

\noindent with eigenmagnon operators \(\alpha_{\bm{q}\eta}\) and \(\beta_{\bm{q}\eta}\), magnon dispersion \(\omega_{\bm{q}}\), and where the quasimomentum \(\bm{q}\) runs over the reduced Brillouin zone, as illustrated in Fig.~\ref{fig:BrillouinZone} (a).\\ 
\indent As shown in Refs.~\cite{Erlandsen2019_Enhancement, Erlandsen2020_TI}, the electron-magnon coupling in this system in general consists of staggered and net magnetic fields, as well as electron scattering processes of both regular and Umklapp type. In our case, all net and staggered magnetic fields from the two opposing antiferromagnetic layers cancel. 

\newpage
The interaction Hamiltonian then takes the form 

\begin{align}
H_\mathrm{int} = 
V \sum_{\substack{\bm{k} \in \Box\\ \bm{q} \in \diamondsuit}} 
\Big[
M_{\bm{q}}^R c_{\bm{k} + \bm{q}, \downarrow}^\dagger c_{\bm{k},\uparrow} 
+ M_{\bm{q}}^U c_{\bm{k} + \bm{q}+\bm{Q}, \downarrow}^\dagger c_{\bm{k},\uparrow} \nonumber \\
+ (M_{-\bm{q}}^R)^\dagger c_{\bm{k} + \bm{q}, \uparrow}^\dagger c_{\bm{k},\downarrow} 
+(M_{-\bm{q}}^U)^\dagger c_{\bm{k} + \bm{q}+\bm{Q}, \uparrow}^\dagger c_{\bm{k},\downarrow} 
\Big],
\end{align}
where we have defined the magnon operators $M^{\kappa}_{\bm{q}} = M^{\kappa}_{\bm{q}H} + M^{\kappa}_{\bm{q}L}$ with 
\begin{subequations}
\begin{align}
M^{\kappa}_{\bm{q}H} &=  \Omega_A a_{\bm{q} H}   + \kappa \,\Omega_B  b_{-\bm{q} H}^\dagger, 
\\
M^{\kappa}_{\bm{q}L} &= 
\Omega_A a_{-\bm{q} L}^\dagger
+ \kappa \,\Omega_B b_{\bm{q}L}.
\end{align}
\end{subequations}

\noindent Here, \(\kappa \in \{R,U\}\) is an index characterizing whether the corresponding electron scattering process is of regular or Umklapp type, which we associate with the values \(R \rightarrow +1\) and \(U \rightarrow -1\) in the definition of \(M_{\bm{q}}^{\kappa}\). Examples of regular and Umklapp scattering processes are shown Fig.~\ref{fig:BrillouinZone}\:(b). We have also defined the momentum shift vector \(\bm{Q} = \pi (\hat{x} +  \hat{y})\) occurring in the Umklapp scattering processes, and the interaction strength parameter \(V \equiv -2\bar{J}\sqrt{S/N}\), where $S$ is the spin quantum number of the AFMI lattice site spins, and \(N\) the number of lattice sites.

In terms of the eigenmagnon operators \(\alpha_{\bm{q}\eta},\beta_{\bm{q}\eta}\), we may also express the magnon operators \(M_{\bm{q}}^{\kappa}\) as

\begin{align}
\begin{aligned}
M_{\bm{q}}^{\kappa} = \big( \Omega_A u_{\bm{q}} &+ \kappa\, \Omega_B v_{\bm{q}} \big)\big( \alpha_{\bm{q}H} + \alpha_{-\bm{q}L}^\dagger\big)\\
+ &\big( \Omega_A v_{\bm{q}} + \kappa\ \Omega_B u_{\bm{q}} \big) \big(\beta_{-\bm{q}H}^\dagger, + \beta_{\bm{q}L}\big),
\end{aligned}
\end{align}

\noindent so that we may think of the magnon operators \(M_{\bm{q}}^{\kappa}\) as linear combinations of antiferromagnetic eigenmagnon operators with a given spin and momentum. 

\section{Eliashberg theory}
\label{Section:EliashbergTheory}

\subsection{Magnon propagators}

Since the magnon operators in the electron-magnon interaction only occur in the particular linear combinations \(M_{\bm{q}}^{\kappa}\), the propagators of \(M_{\bm{q}}^{\kappa}\) will be key building blocks in our Eliashberg theory. In the imaginary time formalism, we therefore define the magnon propagator

\begin{align}
\mathcal{D}^{\kappa\kappa'} (\bm{q}, \tau) = - \langle T_\tau M_{\bm{q}}^{\kappa} (\tau) (M_{\bm{q}}^{\kappa'} )^\dagger (0) \rangle,
\end{align}
where $T_{\tau}$ is the time-ordering operator and the expectation value is computed with the full Hamiltonian. In the non-interacting theory, one may utilize the eigenmagnon propagators to show that

\begin{align}
\mathcal{D}_0^{\kappa \kappa'}(\bm{q}, i\nu_m) = 
& -2 A_e^{\kappa \kappa'}\!(\bm{q}) \frac{2\omega_{\bm{q}} }{\nu_m^2 + \omega_{\bm{q}}^2},
\label{eq_magnonPropagator}
\end{align}
where $\nu_m = 2m\pi/\beta$ is a bosonic Matsubara frequency, and \(\beta\) the inverse temperature. The boosting factors \(A_{e}^{\kappa\kappa'}(\bm{q})\) are given by

\begin{subequations}
\begin{align}
\hspace{-0.29cm}A^{RR}_e(\bm{q}) 
&=
\frac{1}{2} [ (\Omega_A u_{\bm{q}}\! +\! \Omega_B v_{\bm{q}} )^2 + (\Omega_A v_{\bm{q}} + \Omega_B u_{\bm{q}} )^2],
\\
\hspace{-0.29cm}A^{UU}_e(\bm{q}) 
&=
\frac{1}{2} [ (\Omega_A u_{\bm{q}} - \Omega_B v_{\bm{q}} )^2 \!+\! (\Omega_A v_{\bm{q}} - \Omega_B u_{\bm{q}} )^2],
\\
\hspace{-0.29cm}A^{RU}_e(\bm{q}) &= A^{UR}_e(\bm{q}) = 
\frac{1}{2} (\Omega_A^2 - \Omega_B^2)(u_{\bm{q}}^2 + v_{\bm{q}}^2).
\end{align}
\label{eq_evenBoostingFactors}
\end{subequations}

\noindent Here, \(u_{\bm{q}}\) and \(v_{\bm{q}}\) are the magnon coherence factors, arising from the Bogoliubov transformation, discussed in Appendix \ref{Section:Antiferromagnet}. Inspecting the boosting factor corresponding to regular scattering processes, we see that it coincides with the boosting factor occurring from the canonical transformation used to obtain the effective interaction potential in Ref.\! \cite{Erlandsen2019_Enhancement}.\\
\indent From the expressions for the regular and Umklapp boosting factors \(A_e^{RR}(\bm{q})\) and \(A_e^{UU}(\bm{q})\), it is clear that in addition to contributions from only the \(A\) and \(B\) sublattices proportional to factors of \(\Omega_A^2\) and \(\Omega_B^2\), there are in general also interferences between contributions from the two sublattices. Since \(u_{\bm{q}}\) is typically positive and \(v_{\bm{q}}\) is typically negative, as discussed in Appendix \ref{Section:Antiferromagnet}, we typically expect destructive interference in the regular process boosting factor \(A_e^{RR}(\bm{q})\)~\cite{Erlandsen2019_Enhancement} and constructive interference in the Umklapp process boosting factor \(A_e^{UU}(\bm{q})\). The significance of these interference effects is controlled by the asymmetry in the coupling to the two sublattices, where we find the strongest sublattice interferences when we couple equally to both sublattices, and that all interference effects are removed when we couple to only one sublattice. The mixed propagator boosting factors \(A_e^{RU}\) and \(A_e^{UR}\) do not experience similar interferences. 

\subsection{Spinor representations}

To study magnon-mediated superconductivity, we now construct the Eliashberg theory for the system. To do this, we first introduce the Nambu spinor 

\begin{equation}
\psi_{\bm{k}} = 
\begin{pmatrix}
c_{\bm{k} \uparrow} \\
c_{\bm{k} \downarrow} \\
c_{-\bm{k}\uparrow}^\dagger \\
c_{-\bm{k} \downarrow}^\dagger \\
c_{\bm{k} + \bm{Q} \uparrow} \\
c_{\bm{k} + \bm{Q} \downarrow} \\
c_{-\bm{k} - \bm{Q} \uparrow}^\dagger \\
c_{-\bm{k} - \bm{Q} \downarrow}^\dagger \\
\end{pmatrix}.
\end{equation}

\noindent The corresponding Green's function can then in general be written as the \(8\times 8\) matrix 

\begin{align}
G(\bm{k}, \bm{k}', \tau) = - \langle T_\tau \psi_{\bm{k}} (\tau) \psi_{\bm{k}'}^\dagger(0) \rangle,
\end{align}

\noindent where we will also be using the notation \(G(\bm{k}, \bm{k}, \tau) = G(\bm{k}, \tau)\). After a Fourier transform, the imaginary time propagators can be expressed through the Fourier coefficients \(G(\bm{k}, i\omega_n)\), with fermionic Matsubara frequencies \(\omega_n = (2n+1)\pi / \beta\). The \(8 \times 8\) matrix can in general be spanned by the Pauli matrix outer products 

\begin{equation}
\rho_\alpha \otimes \tau_\beta \otimes \sigma_\gamma,
\end{equation}

\noindent where \(\alpha,\beta,\gamma \in \{0,1,2,3\}\) and the Pauli matrix \(\rho_\alpha\) acts on the momentum sector degree of freedom, \(\tau_\beta\) on the particle/hole degree of freedom, and \(\sigma_\gamma\) on the spin degree of freedom. 

We also introduce the magnon spinor 

\begin{equation}
B_{\bm{q}} = 
\begin{pmatrix}
M_{\bm{q}}^R
& (M_{-\bm{q}}^R)^\dagger 
& M_{\bm{q}}^U
& (M_{-\bm{q}}^U)^\dagger 
\end{pmatrix}^T\!,
\end{equation}

\noindent where each magnon operator in the spinor corresponds to the destruction of an excitation with momentum \(\bm{q}\) and spin \(-1\), or the creation of an excitation with momentum \(-\bm{q}\) and spin \(+1\). The magnon operator propagators can now be collected in the magnon propagator matrix

\begin{equation}
D_{\gamma \gamma'}(\bm{q}, \tau) = 
- \langle T_\tau B_{\bm{q}}^\gamma(\tau) B_{-\bm{q}}^{\gamma'}(0) \rangle.
\end{equation}

\noindent After a Fourier transform, the propagator matrix takes the form

\begin{align}
D(q) 
= 
\begin{pmatrix}
0 
& \mathcal{D}^{RR}(q)
& 0 
& \mathcal{D}^{RU}(q)
\\
\mathcal{D}^{RR}(-q)
& 0 
& \mathcal{D}^{UR}(-q)
& 0
\\
0 
& \mathcal{D}^{UR}(q)
& 0 
& \mathcal{D}^{UU}(q)
\\
\mathcal{D}^{RU}(-q)
& 0 
& \mathcal{D}^{UU}(-q)
& 0
\end{pmatrix},
\end{align}

\noindent in terms of the previously introduced propagators \(\mathcal{D}^{\kappa\kappa'}\). Here, $q = (\bm{q}, i\nu_m)$ is a three-vector containing both momentum and the Matsubara frequency. As the magnon propagators respect time-reversal and inversion symmetry, we have \(\mathcal{D}^{\kappa \kappa'}(-q) = \mathcal{D}^{\kappa\kappa'}(q)\). Further, the magnon propagators also satisfy \(\mathcal{D}^{RU}(q) = \mathcal{D}^{UR}(q)\).\\
\indent In spinor notation for the magnon and electron operators, the interaction Hamiltonian can be written on the form 

\begin{equation}
H_\mathrm{int} = \frac{V}{4} \sum_{\substack{\bm{k} \in \Box\\ \bm{q} \in \diamondsuit}} \sum_{\alpha \beta\gamma}
g_\gamma^{\alpha\beta} B_{\bm{q}}^\gamma \psi_{\bm{k} + \bm{q} \alpha}^\dagger \psi_{\bm{k} \beta}, 
\end{equation}

\noindent where the sum over \(\bm{k}\) runs over the full Brillouin zone, the sum over \(\bm{q}\) runs over the reduced Brillouin zone, and the index \(\gamma\) corresponds to the various operators in the magnon spinor \(B_{\bm{q}}^\gamma\). The matrices \(g_\gamma\) are given by 

\begin{subequations}
\begin{align}
g_1 = f_1 \otimes \rho_0, \qquad g_2 = f_2 \otimes \rho_0, 
\\
g_3 = f_1 \otimes \rho_1, \qquad g_4 = f_2 \otimes \rho_1,
\end{align}
\end{subequations}

\noindent where we have introduced the \(4 \times 4\) matrices 

\begin{subequations}
\begin{align}
f_1 
= \frac{1}{2} (\sigma_1 \tau_0 - i \sigma_2 \tau_3), 
\\
f_2  
= \frac{1}{2} (\sigma_1 \tau_0 + i \sigma_2 \tau_3), 
\end{align}
\end{subequations}

\noindent acting on the spin and particle/hole degrees of freedom to simplify the notation.

\begin{figure}[b]
    \centering
    \includegraphics[width=0.95\columnwidth,trim= 1.0cm 18.5cm 0.5cm 0.2cm,clip=true]{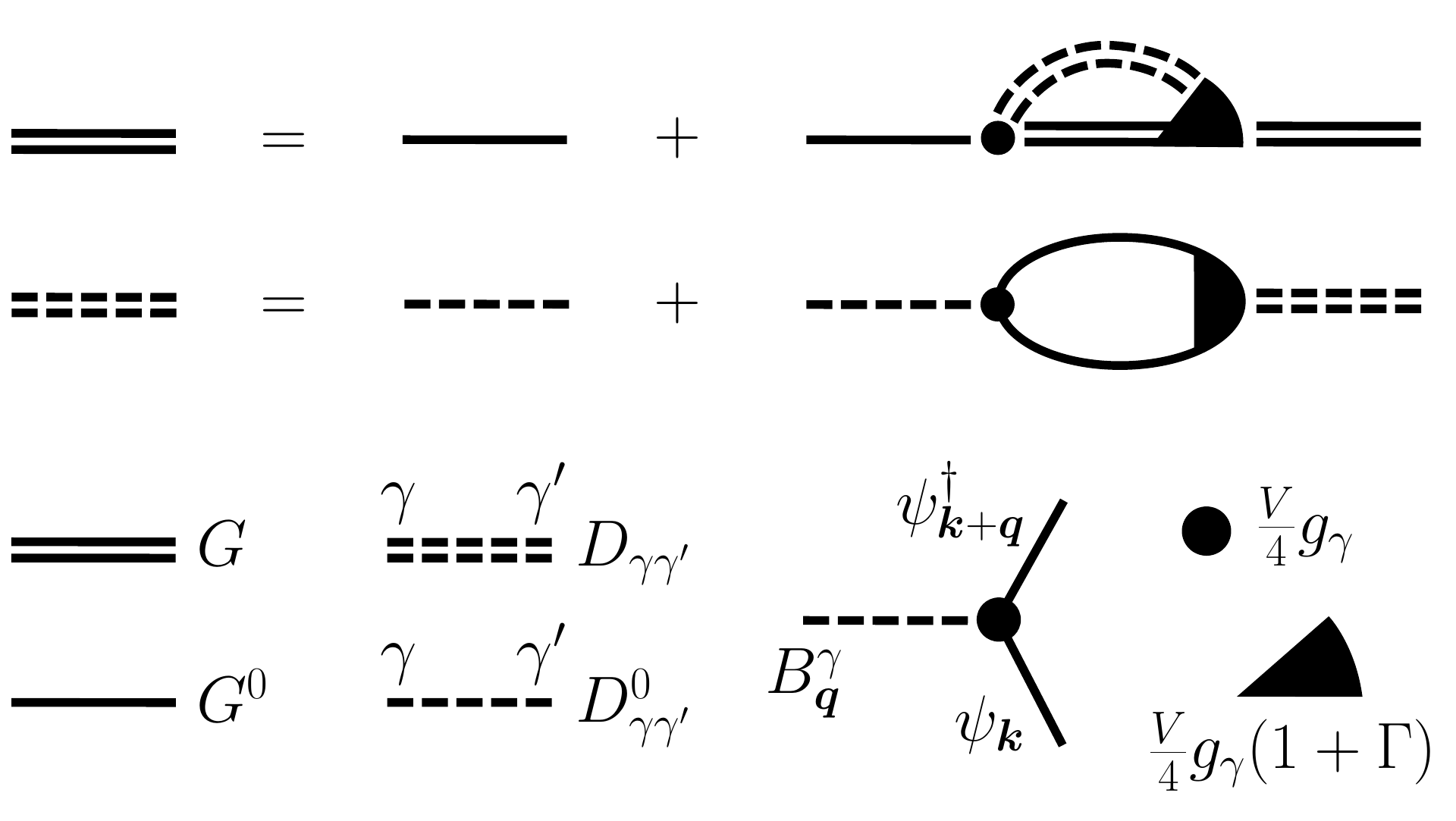}
    \caption{Feynman diagram expansion for interacting electron and magnon propagators. Each vertex is associated with a factor \(Vg_\gamma/4\), and electron and magnon propagators \(G\) and \(D\) are represented by solid and dashed lines.}
    \label{fig_diagramExpansion}
\end{figure}

\subsection{\(S\)-matrix expansion}

Starting from the non-interacting electron Hamiltonian and the spinor form of the interaction, we may now apply the \(S\)-matrix expansion and use Wicks theorem to obtain a Feynman diagram expansion for the electron Green's function \(G(\bm{k}, i\omega_n)\), as shown in Fig. \ref{fig_diagramExpansion}.

The resulting equation can be solved for the electron Green's function to obtain the Dyson equation

\begin{equation}\label{eq_dyson}
G^{-1}(k) = G_0^{-1}(k) - \Sigma(k),
\end{equation}

\noindent where \(\Sigma(k)\) is the self-energy, and \(G_0(k)\) is the non-interacting electron Green's function given by

\begin{equation}
G_0^{-1}(\bm{k}, i\omega_n) = i\omega_n \rho_0 \tau_0 \sigma_0  - \epsilon_{\bm{k}} \rho_3 \tau_3\sigma_0  + \mu \rho_0 \tau_3 \sigma_0.
\end{equation}

In the following, we neglect vertex corrections, which are discussed more in Appendix \! \ref{Section:VertexCorrections}. We may then consider only sunset type diagrams in the self-energy. Performing the \(S\)-matrix expansion, we extract the self-energy

\begin{align}\label{eq_selfConsistent}
\hspace{-0.31cm}\Sigma(k) = -\frac{V^2}{2 \beta} \sum_{k'} \sum_{\gamma \gamma'} \theta_{\bm{k} - \bm{k}'} D_{\gamma \gamma'}(k - k') g_{\gamma} G(k') g_{\gamma'},
\end{align}

\noindent as evident from the diagrammatic representation in Fig.~\ref{fig_diagramExpansion} up to signs and prefactors. Here, \(\theta_{\bm{q}}\) is defined by 

\begin{equation}
\theta_{\bm{q}} =  
\left\{
\begin{array}{lr} 
1, & \bm{q} \in \mathrm{RBZ} \\
0, & \bm{q} \in \mathrm{QBZ} 
\end{array}
\right\},
\end{equation}

\noindent and ensures that the magnon propagator momentum \(\bm{q} = \bm{k} - \bm{k}'\) is restricted to the reduced Brillouin zone (RBZ)~\footnote{In general, the electron can be scattered with any momentum, however, upon explicitly introducing regular (\(R\)) and Umklapp (\(U\)) magnon operators, all magnon scattering processes can be described within the reduced Brillouin zone.}. Here, QBZ refers to the conjugate Brillouin zone which, together with the RBZ, comprises the full electron Brillouin zone.

In the discussion so far, we have been using a Nambu spinor \(\psi_{\bm{k}}\) containing electrons at both \(\bm{k}\) and \(\bm{k} + \bm{Q}\). Thus, the \(8 \times 8\) matrix Green's function \(G(k)\) may in general have correlations between electrons at momenta \(\bm{k}\) and \(\bm{k}+ \bm{Q}\). 
In the following, we assume that the processes close to the Fermi surface dominate the self-energy. Away from half-filling, we may then neglect the correlations which are off-diagonal in the momentum sector, as they are suppressed by the large electronic energy at momentum \(\bm{k} + \bm{Q}\) when \(\bm{k}\) is close to the Fermi surface. This is discussed in more detail in Appendix~\ref{Section:OffDiagonalInMomentum}. The Green's function \(G(k)\) and the self-energy \(\Sigma(k)\) then reduce to two uncoupled blocks of size \(4 \times 4\) which are related by \(\bm{k} \rightarrow \bm{k} + \bm{Q}\). In the following, we therefore consider only one of the two blocks. 

\subsection{Eliashberg equations}
\label{Section:EliashbergTheory:EliashbergEquations}

To derive the Eliashberg equations, we decompose the self-energy matrix into contributions corresponding to the various basis matrices \(\sigma_\alpha \otimes \tau_\beta\) for Hermitian \(4 \times 4\) matrices. We set

\begin{align}
\Sigma = 
(1 - Z) i \omega_n  \sigma_0 \tau_0 
+ \chi \sigma_0 \tau_3  
+ \phi_s   \sigma_2 \tau_2
+ \phi_t   \sigma_1 \tau_1,
\end{align}

\noindent where \(Z\) is the electron renormalization, \(\chi\) is the quasiparticle energy shift, \(\phi_s\) is the spin singlet pairing amplitude, and \(\phi_t\) the amplitude for unpolarized spin triplet pairing.

Among the 16 possible terms on the form \(\sigma_\alpha \otimes \tau_\beta\), we have kept only 4.  Of the remaining 12 combinations, the 8 which do not conserve spin cannot occur because they are incompatible with the spin structure of the self energy diagram. The combinations \(\tau_3 \sigma_3\) and \(\tau_0 \sigma_3\) are disregarded because they introduce spin-dependent quasiparticle renormalization, which is not expected to be present due to the spin symmetry of the fermions in the system. Finally, we could have introduced terms \(\tilde{\phi_s} \tau_1 \sigma_2\) and \(\tilde{\phi}_t \sigma_1 \tau_2\). However, the associated fields \(\tilde{\phi}_s\) and \(\tilde{\phi}_t\) would play exactly the same roles as \(\phi_s\) and \(\phi_t\), and we therefore set them to zero.\\
\indent Due to symmetry relations between the electron correlations in the Nambu spinor Green's function matrix \(G(k)\)~\cite{Linder2019}, the normal Green's function fields satisfy

\begin{align}
Z(-k) &= Z(k), 
\hspace{0.6cm}
Z(\bm{k}, i\omega_n) = Z(\bm{k}, -i\omega_n)^*,
\\
\chi(-k) &= \chi(k),
\qquad
\chi(\bm{k}, i\omega_n) = \chi(\bm{k}, -i\omega_n)^*, 
\label{eq_fieldSymmetryRelation2}
\end{align}

\noindent and the anomalous correlations satisfy

\begin{align}
\phi_s(-k) &= +\phi_s(k), 
\hspace{0.65cm} 
\phi_s(\bm{k}, i\omega_n) 
= \phi_s(\bm{k}, -i\omega_n)^*,
\\
\phi_t(-k) &= -\phi_t(k),
\qquad
\phi_t(\bm{k}, i\omega_n) 
= \phi_t(\bm{k}, -i\omega_n)^*.
\end{align}

\indent We may now derive equations for the fields \(Z,\chi, \phi_s, \phi_t\) by inserting the form for \(\Sigma\) into the Dyson equation, inverting the inverse \(G^{-1}(k)\) and inserting \(G(k)\) into the self-energy in Eq.\! \eqref{eq_selfConsistent}. Comparing term by term, we then obtain the equations

\begin{subequations}
\begin{align}
[1 - Z(k)] i \omega_n &= 
- V^2 \frac{1}{\beta} \sum_{k'} \mathcal{D}(k-k') \frac{i\omega_{n'} Z(k')}{\Theta(k')},
\\
\chi(k) &= 
- V^2 \frac{1}{\beta} \sum_{k'} \mathcal{D}(k-k') \frac{\xi_{\bm{k}'} + \chi(k')}{\Theta(k')},
\\
\phi_s(k) &= 
- V^2 \frac{1}{\beta} \sum_{k'} \mathcal{D}(k-k') \frac{ \phi_s(k') }{\Theta(k')},
\label{eq_eliashbergSpinSinglet}
\\
\phi_t(k) &= 
+V^2 \frac{1}{\beta} \sum_{k'} \mathcal{D}(k-k') \frac{ \phi_t(k') }{\Theta(k')},
\label{eq_eliashbergSpinTriplet}
\end{align}
\end{subequations}

\noindent under the assumption that a single symmetry channel dominates, so that either \(\phi_s=0\) or \(\phi_t=0\)~\footnote{One may derive equations also when this is not the case, but they will be somewhat more involved, as there will be interference terms in the submatrix determinant \(\Theta\)}. We have also introduced the combined magnon propagator 

\begin{align}
\mathcal{D}(q) =  \theta_{\bm{q}} \mathcal{D}^{RR}(\bm{q},i\nu_m) 
+ \theta_{\bm{q} +  \bm{Q}} \mathcal{D}^{UU}(\bm{q}+\bm{Q},i\nu_m),
\end{align}

\noindent where the argument \(\bm{q}\) can now take on values in the full electron Brillouin zone. The submatrix determinant \(\Theta(k)\) is given by 

\begin{align}
\Theta(k) &= 
[ i\omega_n Z(k) ]^2 
- \tilde{\xi}_{k}^2 
- |\phi_{s,t}(k)|^2,
\end{align}

\noindent with anomalous correlation \(\phi_{s,t}\) depending on whether we consider a singlet or triplet instability, and where have introduced \(\tilde{\xi}_k = \xi_{\bm{k}} + \chi(k)\). In the following, we will assume that the quasiparticle energy shift \(\chi\) is small compared to the electron bandwidth, and that it can be neglected. Note the opposite signs on the right hand side of the equations for \(\phi_s\) and \(\phi_t\). This occurs because the spin flips in the vertices of the self-energy diagrams introduce a sign change for the spin singlet amplitude, but not for the spin triplet amplitude.

\section{Fermi surface averaged equations}\label{Section:FS_average}

When the electron energy scale is large compared to the magnon energy scale, the regions close to the Fermi surface dominate the momentum sums in the Eliashberg equations. We assume that the quasiparticle renormalization field close to the Fermi surface is weakly dependent on momentum, so that we may write \(Z(\bm{k}, i\omega_n) = Z(i\omega_n)\). Furthermore, for a single dominant pairing symmetry channel, we assume that the anomalous correlations can be written in the product form \(\phi_{s,t}(\bm{k},i\omega_n) = \psi(\bm{k})\phi_{s,t}(i\omega_n)\), where we assume some simple functional form  \(\psi(\bm{k})\) for the momentum dependence of the relevant anomalous correlation.\\
\indent Since we expect regions close to the Fermi surface to dominate the momentum sum, we may split it into a perpendicular and a parallel part, and neglect the perpendicular momentum dependence of the magnon propagator. 
Close to the critical temperature, one may furthermore linearize the Eliashberg equations in the anomalous correlations. Converting the perpendicular momentum integration into an energy integral, one then obtains

\begin{subequations}
\begin{align}
(1 - Z) i \omega_n \!=\! 
\frac{1}{\beta N_F} \sum_{\omega_{n'}}\lambda_1(i\omega_n - i\omega_{n'})
i\omega_{n'} Z'\!
\int\! d\xi \frac{N(\xi)}{\Theta(\xi,i\omega_{n'})},
\\
\phi_{s,t} = -\frac{1}{\beta N_F}\sum_{\omega_{n'}}\lambda_2^{s,t}(i\omega_n - i\omega_{n'})\,\phi_{s,t}'
\int\! d\xi \,\frac{N(\xi)}{\Theta(\xi,i\omega_{n'})}.
\end{align}
\end{subequations}

\noindent We have here introduced the dimensionless electron-magnon coupling strength \(\lambda_1(i\omega_n - i\omega_{n'})\) occurring in the quasiparticle renormalization equations and the modified coupling strength \(\lambda_2^{s,t}(i\omega_n - i\omega_{n'})\) occurring in the anomalous correlation equations. We have further denoted \(Z(k)\) by \(Z\) and \(Z(k')\) by \(Z'\), with similar notation also for the remaining fields, and denoted the electron density of states by $N(\xi)$, which takes on the value $N_F$ at the Fermi level. The dimensionless coupling strengths are given by 

\begin{align}
\lambda_1(i\omega_n - i\omega_{n'}) =  -  \frac{V^2}{N_F}\sum_{\bm{k}\bm{k}'}\delta(\xi_{\bm{k}} )\delta (\xi_{\bm{k}'})\mathcal{D}(k-k'),
\end{align}

\begin{align}
\lambda_2^{s,t}(i\omega_{n} - i\omega_{n'}) = -\zeta_{s,t} \frac{1}{\langle\psi^2(\bm{k})\rangle_{\textrm{FS}}}\frac{V^2}{N_F}\sum_{\bm{k}\bm{k}'}\delta(\xi_{\bm{k}} )\delta (\xi_{\bm{k}'})
\nonumber\\
\psi(\bm{k})\mathcal{D}(k-k')\psi(\bm{k}'),
\end{align}

\noindent where \(\zeta_s = -1\) for spin singlet and \(\zeta_t = +1\) for spin triplet is the sign associated with a spin flip in the anomalous pairing. The brackets $\langle \,\,\,\,\rangle_{\textrm{FS}}$ denote a Fermi surface average.\\  
\indent In the following, we assume that the density of states can be approximated by a constant in the dominant region close to the Fermi surface. We may then perform the energy integral analytically to obtain 

\begin{align}
(1 - Z) i \omega_n &= 
- \frac{i\pi}{\beta} \sum_{\omega_{n'}}\lambda_1(i\omega_n - i\omega_{n'})\,\operatorname{sgn} (\omega_{n'}),
\label{eq_fsAveragedEliashberg_Z}
\\
\phi_{s,t} &= +\frac{\pi}{\beta}\sum_{\omega_{n'}}\lambda_2^{s,t}(i\omega_n - i\omega_{n'}) \,\frac{ \phi'_{s,t} }{|\omega_{n'} Z'|}.
\label{eq_fsAveragedEliashberg_phi}
\end{align}
We next assume that the magnon propagator \(\mathcal{D}\) can be replaced by the non-interacting propagator \(\mathcal{D}_{0}\). Solving the Eliashberg equations is then reduced to calculating dimensionless coupling strengths \(\lambda_{1,2}\), and solving eigenvalue problems in the Matsubara frequencies. In Sec.\! \ref{Section:MagnonRenormalization}, we investigate the effect of including the magnon self-energy.

In addition to introducing the dimensionless coupling strengths \(\lambda_{1,2}\), we may follow the conventional routine and also introduce frequency dependent functions \(\alpha_{1,2}^2 F(\omega)\) defined such that 

\begin{align}
\lambda_{1,2}(i\omega_n - i\omega_{n'}) = \!\int \! \textrm{d}\omega\,\alpha_{1,2}^2 F(\omega)\frac{2 \omega}{(\omega_n - \omega_{n'})^2 + \omega^2}.
\end{align}

\noindent Comparing with the definition of \(\lambda_{1,2}\), this gives 

\begin{align}
\alpha_1^2 F(\omega) = \frac{V^2}{N_F}\sum_{\bm{k}\bm{k}'}
\delta(\xi_{\bm{k}} ) \delta (\xi_{\bm{k}'}) \delta(\omega - \omega_{\bm{\bm{k}-\bm{k}'}}) \mathcal{A}_{e}(\bm{k} - \bm{k}'),\\
\alpha_2^2 F(\omega) = \zeta_{s,t}\frac{1}{\langle\psi^2(\bm{k})\rangle_{FS}}\frac{V^2}{N_F}\sum_{\bm{k}\bm{k}'}\delta(\xi_{\bm{k}} )\delta (\xi_{\bm{k}'}) \delta(\omega - \omega_{\bm{\bm{k}-\bm{k}'}})
\nonumber \\
\psi(\bm{k})\mathcal{A}_e(\bm{k} - \bm{k}')\psi(\bm{k}'),
\end{align}

\noindent where the boosting factor 

\begin{equation}
\mathcal{A}_e(\bm{q}) = \theta_{\bm{q}} A_e^{RR}(\bm{q}) + \theta_{\bm{q} + \bm{Q}} A_e^{UU}(\bm{q} + \bm{Q}),
\end{equation}

\noindent has been defined analogously to \(\mathcal{D}(q)\).

The Eliashberg functions \(\alpha^2_{1,2} F(\omega)\) and the electron-magnon coupling strengths \(\lambda_{1,2}(i\nu_m)\) are central quantities in the Fermi surface averaged  Eliashberg equations. Through the approximate formula

\begin{equation}
T_c^\mathrm{AD} = \frac{\omega_\mathrm{log} }{1.2}
\exp \left( 
- \frac{1.04 [ 1+ \lambda_1(0)]}{\lambda_2(0) } \right),
\label{eq_AllenDynes}
\end{equation}
they can therefore be used to qualitatively understand the critical temperatures resulting from actually solving the Eliashberg equations. The above formula was suggested by Allen and Dynes~\cite{AllenDynes1975} for weak and intermediate electron-boson coupling. We have set the Coulomb pseudo-potential to zero, and use the logarithmic average

\begin{equation}
\omega_\mathrm{log} = \omega_a \exp \left[
\frac{2}{\lambda_2(0)} \int d\omega \ln \left( \frac{\omega}{\omega_a}\right) \frac{\alpha_2^2F(\omega)}{\omega}
\right]
\end{equation}

\noindent as the effective cutoff frequency, where \(\omega_a\) is an arbitrary frequency scale. 

\section{Solving the Eliashberg equations}\label{Section:solve}

We now solve the Fermi surface averaged equations using realistic material parameters, as detailed in Appendix \ref{Section:Parameters}. We set \(\Omega_A=1\), and use \(\Omega_B \equiv \Omega \in [0,1]\) to tune the sublattice coupling asymmetry.\\
\indent In order to compute the dimensionless coupling strengths $\lambda_{1,2}$, the momentum sums are transformed into integrals over momenta on the Fermi surface. The quasiparticle renormalization field \(Z(i\omega_n)\) can then be calculated using Eq. \eqref{eq_fsAveragedEliashberg_Z}. Subsequently, we use Eq. \eqref{eq_fsAveragedEliashberg_phi} to determine the critical temperature for the superconducting instability by finding the temperature for which the largest eigenvalue of the eigenvalue problem becomes \(1\) \cite{Armadillo1, Armadillo2}.  This gives the critical temperature \(T_c\) of the superconducting instability. 
\begin{figure}[t]
    \centering
    \includegraphics[width=0.95\linewidth]{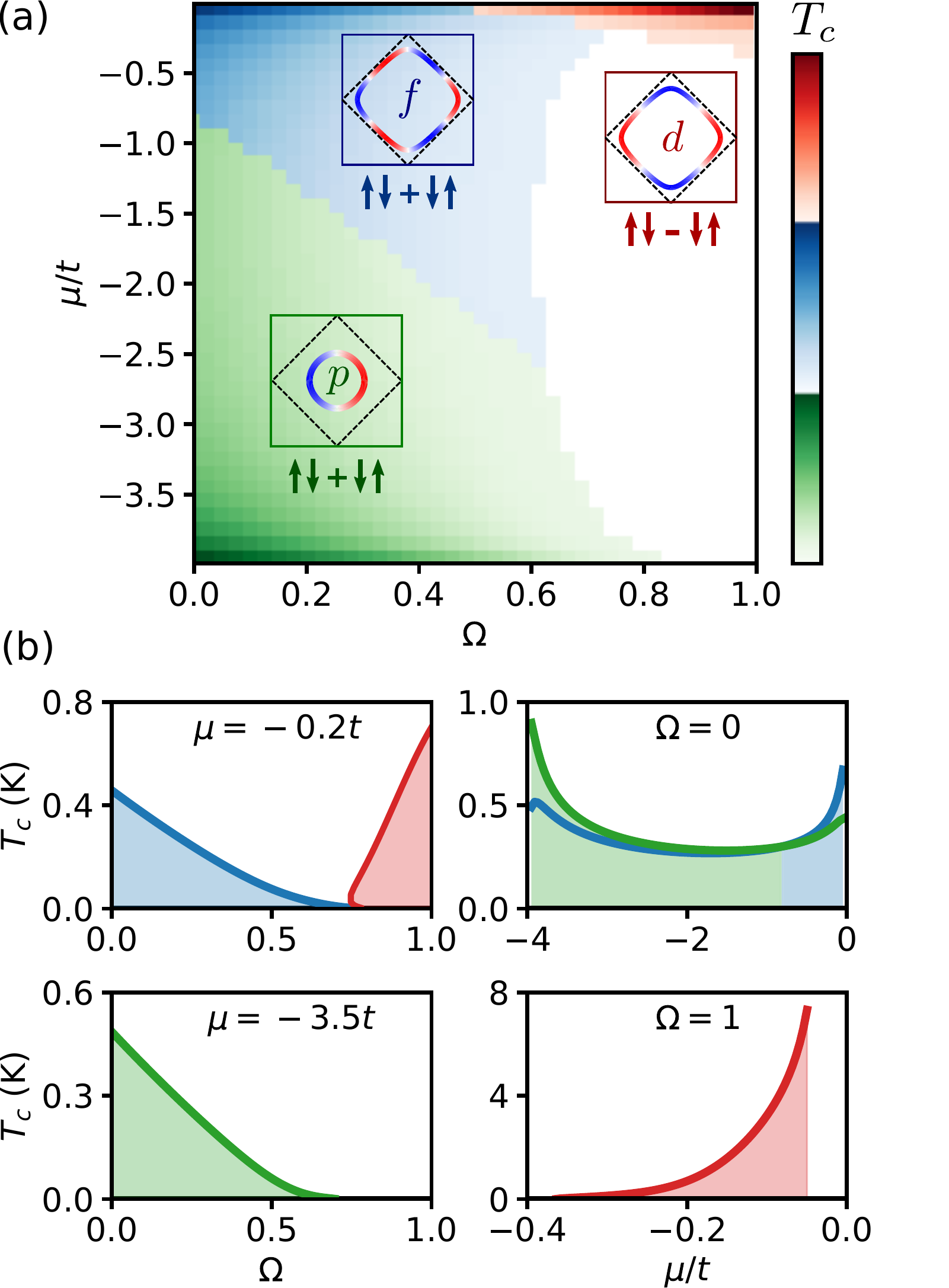}
    \caption{(a) Phase diagram in terms of sublattice coupling asymmetry \(\Omega = \Omega_B / \Omega_A\) and chemical potential \(\mu\) below half filling. We find spin triplet \(p\)-wave, spin triplet \(f\)-wave, and spin singlet \(d\)-wave phases. The phase diagram is colored according to the critical temperature normalized to the largest value in the phase diagram within the same phase. Parameter regimes supporting multiple superconducting instabilities are colored according to the phase with the largest critical temperature. The insets show the spin structure and momentum structure on the Fermi surface for the various phases. The various subfigures in (b) show the critical temperature \(T_c\) as function of \(\Omega\) (left) and \(\mu\) (right) along different lines in the phase diagram. }
    \label{fig_results}
\end{figure}
We consider three different Ansätze for the superconducting pairing, namely even frequency spin triplet \(p\)-wave pairing, even frequency spin triplet \(f\)-wave pairing and even frequency spin singlet \(d\)-wave pairing. These pairings dominate in different parts of the parameter space of our model. Other pairing symmetries like even frequency spin singlet $s$-wave and different odd frequency variants were not found to give rise to superconductivity. Fig.\! \ref{fig_results} (a) presents the phase diagram for our model in the \(\Omega\)-\(\mu\)-plane, where critical temperature normalized to the maximum value within each phase is indicated by color intensity. The type of pairing is indicated by choice of color (green/blue/red), where regimes supporting multiple solutions are colored according to the phase with the largest critical temperature. 

In the following, we discuss the different superconducting phases in the phase diagram in more detail.

\subsection{Spin triplet \(p\)-wave and \(f\)-wave pairing}

For the even frequency spin triplet \(p\)-wave and \(f\)-wave pairings, we consider anomalous pairing momentum dependence on the form 

\begin{subequations}
\begin{align}
\psi_{p}(\bm{k}) &= \cos \phi_{\bm{k}},
\\
\psi_{f}(\bm{k}) &= \cos 3 \phi_{\bm{k}},
\end{align}
\end{subequations}

\noindent where \(\phi_{\bm{k}}\) is the polar angle between the quasimomentum \(\bm{k}\) on the Fermi surface and the $x$-axis. These momentum dependencies are shown in the insets of the phase diagram.\\
\indent As expected, and in agreement with the results of Ref.~\cite{Erlandsen2019_Enhancement}, we find even frequency spin triplet \(p\)-wave superconductivity for small Fermi surfaces and large sublattice coupling asymmetry, corresponding to small \(\mu\) and \(\Omega\). For small Fermi surfaces, all processes between points on the Fermi surface are of the regular type. Since the magnon energy is smallest for small \(\bm{q}\), minimizing the denominator of the magnon propagator, the dominant contribution to the momentum sums in the Eliashberg equations originate from small \(\bm{q}\). Without sublattice coupling asymmetry (i.e. \(\Omega=1\)), coherence factor interference effects suppress the boosting factor \(A_e^{RR}(\bm{q})\), whereas \(\Omega=0\) removes these interference effects completely and makes \(p\)-wave superconductivity possible.\\
\indent Setting \(\Omega=0\), we also find an even frequency spin triplet \(f\)-wave solution in the entire chemical potential range we have considered. As shown in Fig.\! \ref{fig_results} (b), the critical temperature of the \(p\)-wave solution is larger than the critical temperature of the \(f\)-wave solution for small Fermi surfaces. For Fermi surfaces approaching half filling, however, the situation is reversed due to emergence of subleading Umklapp processes. 
The interaction providing spin triplet pairing is attractive for scattering processes between \(\bm{k}\) and \(\bm{k}'\) only when \(\phi(\bm{k}, i\omega_n)\) and \(\phi(\bm{k}', i\omega_n)\) have the same sign. Consider now the scattering processes between points on the Fermi surface where the momentum transfer is closest to $\bm{Q}$, bringing the electron from one side of the Fermi surface to the opposing side. From the $f$-wave and $p$-wave momentum structure of the anomalous correlations shown in the insets of Fig.~\ref{fig_results} (a), it is clear that these processes are always repulsive in the \(p\)-wave phase and typically attractive in the \(f\)-wave phase. This explains why the \(f\)-wave phase has a higher critical temperature than the \(p\)-wave phase upon approaching half-filling. As discussed in more detail in Sec.~\ref{Section:ExperimentalConsiderations}, the combination of $\Omega = 0$ and the presence of Umklapp processes may, however, be challenging to access experimentally.\\ 
\begin{figure}[t]
    \centering
    \includegraphics[width=0.96\linewidth]{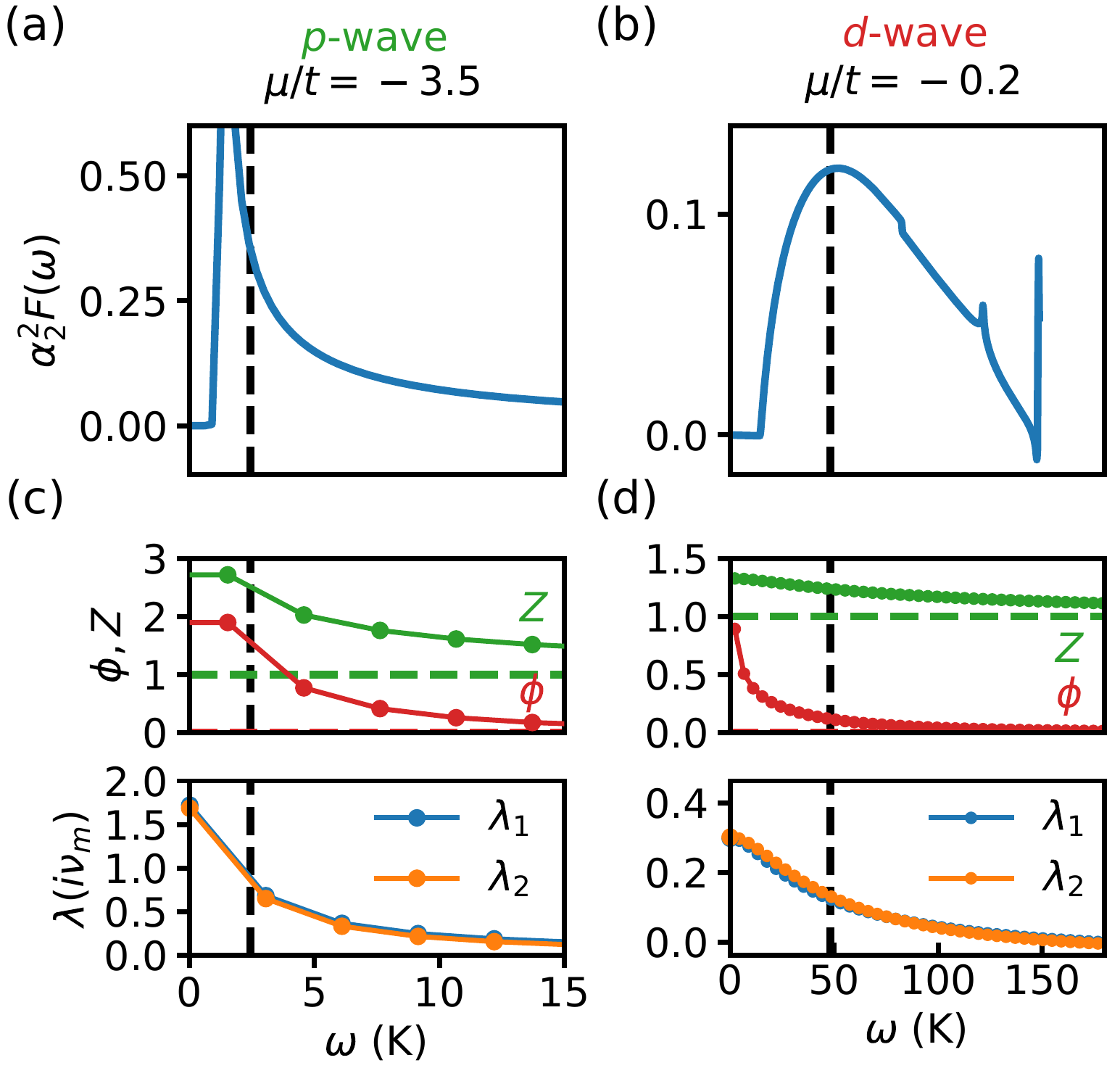}
    \caption{Eliashberg functions and Eliashberg equation solutions for the $p$-wave regime (\(\mu/t=-3.5\) and \(\Omega=0\)) to the left and $d$-wave regime (\(\mu/t=-0.2\) and \(\Omega=1\)) to the right. Subfigures (a) and (b) show the Eliashberg function \(\alpha_2^2F(\omega)\). Subfigures (c) and (d) show the dimensionless electron-magnon coupling strengths  \(\lambda_{1,2}(i\nu_m)\) as well as the Matsubara frequency dependence of the quasiparticle renormalization \(Z(i\omega_n)\) and the anomalous correlation \(\phi(i\omega_n)\) at the critical temperatures for the respective superconducting instabilities.
    The logarithmic average \(\omega_\mathrm{log}\) is shown with vertical dashed lines. }
    \label{fig_frequencyDependence}
\end{figure}
\indent Compared with the results of Ref.~\cite{Erlandsen2019_Enhancement}, we find significantly lower critical temperatures for the $p$-wave phase. We attribute this difference to the magnon energy cutoff. As long-wavelength processes dominate, the characteristic magnon frequency in the pairing interaction is much smaller than the upper cutoff on the magnon spectrum. Since the characteristic frequency serves as the energy scale for the critical temperature, the critical temperature is then significantly reduced, which is captured in the Eliashberg theory analysis.\\ 
\indent More quantitatively, this argument can be understood in terms of the Allen-Dynes formula of Eq. \eqref{eq_AllenDynes}. Since the boosting factor \(A_e^{RR}(\bm{q})\) is peaked for small momenta \(\bm{q}\), and the electron-magnon coupling strength $V$ is momentum-independent, the electron-magnon coupling function \(\alpha^2_2F(\omega)\) is peaked at small frequencies. This is shown in Fig.\! \ref{fig_frequencyDependence} (a), where the logarithmic average \(\omega_\mathrm{log}\) is indicated with a dashed line. The effective magnon frequency for the superconducting pairing is therefore significantly reduced compared to the largest magnon frequency in the system. 
Further, the lower panel of Fig.~\ref{fig_frequencyDependence} (b) shows  \(\lambda_{1,2}(i\nu_m)\), which decays quickly beyond the effective cutoff. Solving the Eliashberg equations gives the solutions for the anomalous correlation \(\phi(i\omega_n)\), which also decays quickly beyond the cutoff, and the quasiparticle renormalization \(Z(i\omega_n)\), which decays to \(1\).\\

\subsection{Spin singlet \(d\)-wave pairing}
In the Eliashberg equations, the difference between the spin triplet case in Eq.\! \eqref{eq_eliashbergSpinTriplet} and the singlet case in Eq.\! \eqref{eq_eliashbergSpinSinglet} is the sign. Thus, the small momentum process pairing potential that was attractive for spin triplet pairing becomes repulsive for spin singlet pairing. To obtain singlet pairing attraction, we therefore need to rely on dominant processes with a relative sign between the anomalous pairing \(\phi_s(k)\) on the left-hand-side and right-hand-side of the equation. Since small momentum processes cannot provide this sign change, we need to rely on Umklapp processes. As an $s$-wave Ansatz does not change sign around the Fermi surface, we instead choose the \(d\)-wave Ansatz

\begin{equation}
\psi_d(\bm{k} ) = \frac{1}{2\pi} (\cos k_x - \cos k_y ),
\end{equation}

\noindent shown in the inset of the phase diagram in Fig.\! \ref{fig_results} (a). Since the \(d\)-wave phase relies on Umklapp processes, it occurs for chemical potentials \(\mu\) approaching half-filling in the phase diagram. Furthermore, the Umklapp processes benefit from the coherence factor interference in the boosting factor \(A_e^{UU}(\bm{q})\), which is maximized for \(\Omega=1\). Crucially, these interferences also suppress the competing regular processes with small momentum \(\bm{q}\), which would otherwise prevent spin singlet superconductivity. The \(d\)-wave phase therefore occurs only for large \(\Omega\) in the phase diagram. This picture is also verified in Fig.~\ref{fig_results} (b), which shows the critical temperature for the spin singlet \(d\)-wave phase as function of coupling asymmetry, and as function of chemical potential at \(\Omega=1\).\\
\indent The electron-magnon coupling strength function \(\alpha_2^2F(\omega)\) is shown in Fig.\! \ref{fig_frequencyDependence} (b). With \(\Omega=1\), the regular small momentum processes are suppressed, and away from half-filling, the Umklapp processes between points on the Fermi surface require the magnons to carry momentum which differs from \(\bm{Q}\) by a finite amount. Therefore, \(\alpha^2_2 F(\omega)\) takes on significant values only beyond a relatively large lower frequency cutoff. This cutoff corresponds to the magnon energy associated with the smallest momentum necessary to bring an Umklapp scattered electron with incoming momentum \(\bm{k}\) from \(\bm{k} + \bm{Q}\) and back to the Fermi surface. Moreover, it should be noted that 
this smallest momentum depends on where on the Fermi surface the electron was situated to begin with. At the lowest relevant frequencies in $\alpha_2^2 F(\omega)$, only a few momenta \(\bm{k}\) bring \(\bm{k} + \bm{Q}\) to a position where the momentum transfer necessary to get back to the Fermi surface is associated with a magnon energy that is small enough to match the frequency $\omega$. The function $\alpha_2^2 F(\omega)$ then only obtains contributions from a few points $\bm{k}$ that bring \(\bm{k} + \bm{Q}\) close enough to the Fermi surface. As the frequency $\omega$ increases, $\alpha_2^2 F(\omega)$ obtains contributions from more points $\bm{k}$ as the restriction on how close $\bm{k} + \bm{Q}$ needs to be to the Fermi surface is relaxed. Therefore, \(\alpha_2^2 F(\omega)\) is not peaked at small frequencies. The situation should be contrasted with the $p$-wave case, where regular scattering on the Fermi surface with vanishing momentum is possible regardless of where on the Fermi surface the initial electron is situated. Denoting the magnon spectrum gap by \(\omega_0\), $\alpha_2^2 F (\omega \rightarrow \omega_0)$ therefore receives large contributions from $\bm{k} - \bm{k}' \approx 0$ regardless of where on the Fermi surface $\bm{k}$ is situated. \\ 
\indent The reduced reliance of the $d$-wave pairing on processes with small magnon energy gives rise to a larger effective magnon frequency \(\omega_\mathrm{log}\). This larger characteristic magnon frequency suppresses the magnon propagator occurring in \(\lambda_{1,2}(i\nu_m)\) for small Matsubara frequencies, but also increases the frequency scale over which the magnon propagator decays compared to the $p$-wave regime. Together with a large density of states close to half filling, this causes the significant critical temperatures that are observed for the \(d\)-wave regime in Fig.\! \ref{fig_results} (b).
As shown in Fig.~\ref{fig_frequencyDependence} (d), the dimensionless electron-magnon coupling strength \(\lambda_{1,2}(i\nu_m)\) decays to zero beyond the effective cutoff frequency, whereas \(\phi(i\omega_n)\) has a crossover from behaviour \(1/\omega_n\) to \(1/\omega_n^3\).\\  

\subsection{Effect of frustration}

\begin{figure}[t]
    \centering
    \includegraphics[width=0.95\linewidth]{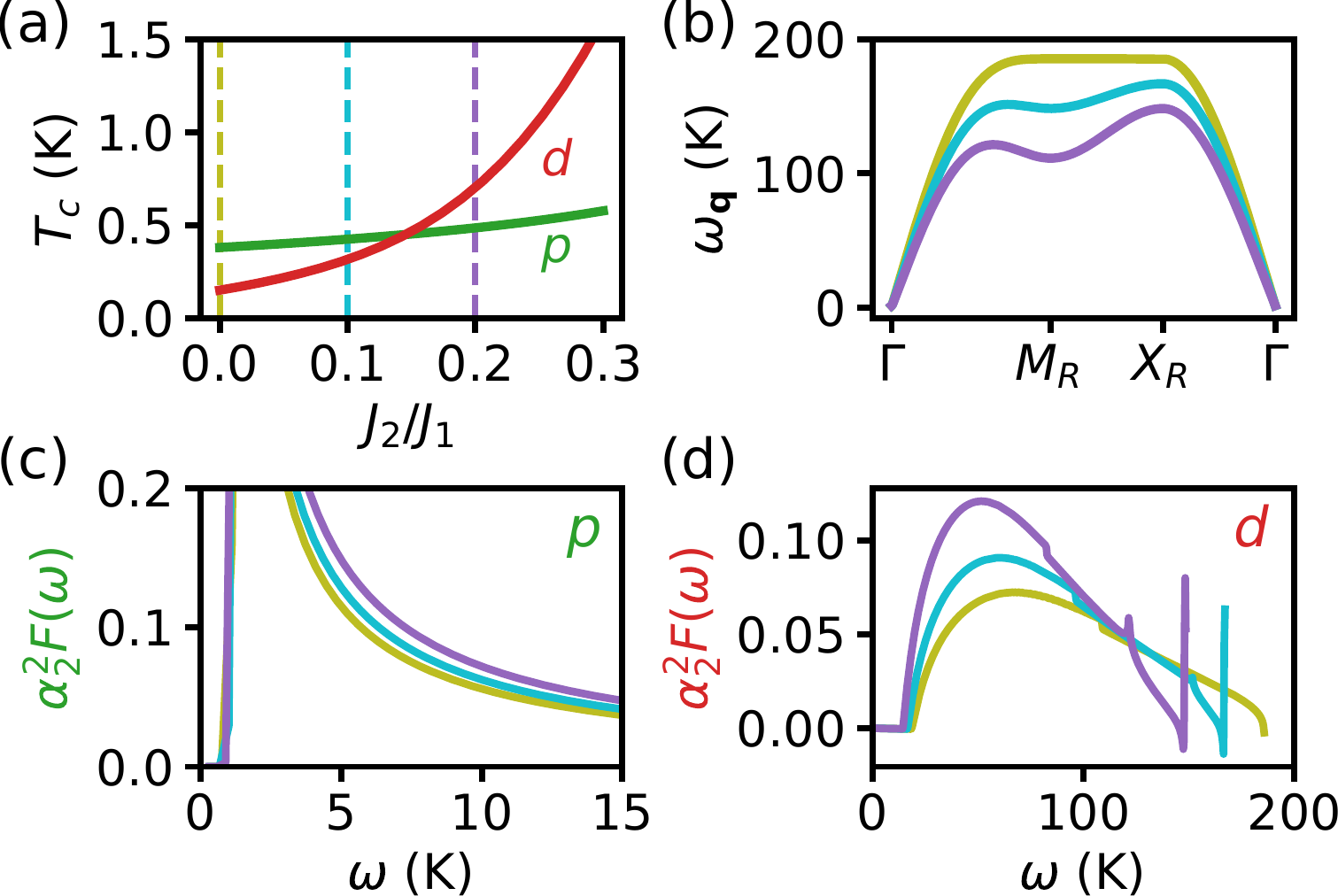}
    \caption{Effect of frustrating the antiferromagnet with an next-to-nearest neighbour exchange coupling \(J_2\). (a) Critical temperature for the \(p\)-wave  (at \(\mu=-3.5t\) and \(\Omega=0\)) and \(d\)-wave (at \(\mu=-0.2t\) and \(\Omega=1\)) instabilities as function of \(J_2\) frustrating the antiferromagnet. (b) Magnon spectrum for different values of \(J_2\), as indicated by the vertical dashed lines in (a), between Brillouin zone high symmetry points as shown in Fig. \ref{fig:BrillouinZone} (a). (c) Electron-magnon coupling function \(\alpha_2^2F(\omega)\) in the \(p\)-wave regime. (d) \(\alpha_2^2F(\omega)\) in the \(d\)-wave regime. Frustration reduces the magnon excitation energies and enhance the spin fluctuations in the system. Thus, weight is shifted from high magnon energies to low magnon energies in the electron-magnon coupling function \(\alpha_2^2F(\omega)\), and this increases the critical temperature.}
    \label{fig:frustrationTemperature}
\end{figure}

\indent Since the superconductivity in our system relies on spin fluctuations, we expect interactions in the AFMI spin model that enhance fluctuations to also enhance the critical temperature. Earlier weak-coupling studies have investigated the effect of a frustrating next-nearest neighbor exchange coupling $J_2 > 0$ in the antiferromagnet on superconductivity dominated by regular fermion-magnon scattering processes \cite{Erlandsen2020_TI, Erlandsen2020_Schwinger}. In Fig.\! \ref{fig:frustrationTemperature} (a), we show how the critical temperature increases with \(J_2\) for both the \(p\)-wave and \(d\)-wave instabilities. The effect of $J_2$ on the superconductivity can be understood in terms of the magnon excitation energies in Fig.\! \ref{fig:frustrationTemperature} (b), showing that the magnon bands are flattened as $J_2$ increases. As displayed in Fig.\! \ref{fig:frustrationTemperature} (c) and (d), this shifts weight from large to the more significant small frequencies in the electron-magnon coupling function \(\alpha_{2}^2F(\omega)\), leading to a higher critical temperature. Notably, increasing $J_2$ does not affect the gap in the magnon spectrum, meaning that the effective cutoff for the $p$-wave phase is not much affected. For the $d$-wave phase, the effective cutoff is somewhat reduced for larger $J_2$, but trading some cutoff for a larger dimensionless coupling strength $\lambda_2(0)$ is nevertheless found to be beneficial. 
As the $d$-wave phase
has a smaller dimensionless coupling strength than the $p$-wave phase, the increase of the dimensionless coupling strength arising from $J_2$ leads to a more dramatic increase in critical temperature for the $d$-wave curve in Fig.\!  \ref{fig:frustrationTemperature} (a).

\section{Magnon renormalization}
\label{Section:MagnonRenormalization}

To consider the effect of magnon renormalization, we consider the electron bubble diagram shown in Fig.\! \ref{fig_diagramExpansion}, and once again neglect vertex corrections. Performing the \(S\)-matrix expansion, one may show that magnon propagators \(D_{\gamma \gamma'}\) satisfy the Dyson equation

\begin{equation}
D^{-1}(q) = D_{0}^{-1}(q) - \Pi(q),
\end{equation}

\noindent where the polarization matrix is given by 

\begin{equation}
\Pi_{\gamma \gamma'}(q) = \frac{V^2}{4\beta} \sum_k \operatorname{Tr} \left[g_{\gamma} G(k+q) g_{\gamma'} G(k) \right].
\end{equation}

\noindent From the matrix structure of the matrices \(g_\gamma\), it follows that \(\Pi_{\gamma \gamma'}\) takes the form 

\begin{align}
\Pi(q) = 
\begin{pmatrix}
0 
& \Pi^{RR}(q)
& 0 
& \Pi^{RU}(q)
\\
\Pi^{RR}(-q)
& 0 
& \Pi^{UR}(-q)
& 0
\\
0 
& \Pi^{UR}(q)
& 0 
& \Pi^{UU}(q)
\\
\Pi^{RU}(-q)
& 0 
& \Pi^{UU}(-q)
& 0
\end{pmatrix}.
\end{align}

In principle, we should now solve the coupled equations for the electron and magnon propagators. However, to estimate the effect of magnon renormalization, we use the non-interacting electron Green's functions to calculate the polarizations. Using the previous assumption of neglecting terms in the electron Green's function which are off-diagonal in momentum sector, we may furthermore neglect the mixed process polarizations \(\Pi^{UR}\) and \(\Pi^{RU}\). This leaves the the regular and Umklapp polarizations \(\Pi^{RR}\) and \(\Pi^{UU}\), which reduce to

\begin{align}
\Pi_{0}^{RR}(q) &= \frac{V^2}{\beta} \sum_k G_0^{11}(k+q) G_0^{22}(k),
\label{eq_Pi_RR}
\\
\Pi_{0}^{UU}(q) &= \frac{V^2}{\beta} \sum_k G_0^{11}(k+q + Q) G_0^{22}(k),
\label{eq_Pi_UU}
\end{align}

\noindent where \(G_0^{11}\) and \(G_0^{22}\) are matrix elements in the non-interacting electron Green's function \(G_0\) corresponding to different spins. 

Solving the Dyson equation for the magnon propagator, one may show that the regular and Umklapp propagators become

\begin{align}
\mathcal{D}^{RR}(q) &= \left[ \left( \frac{1 - \mathcal{D}_0^{UU} \Pi_{0}^{UU}}{1 - r \mathcal{D}_0^{UU} \Pi_{0}^{UU} }  \right) - \mathcal{D}_0^{RR} \Pi_{0}^{RR} \right]^{-1}\! \mathcal{D}_0^{RR}(q),
\\
\mathcal{D}^{UU}(q) &= \left[ \left( \frac{1 - \mathcal{D}_0^{RR} \Pi_{0}^{RR}}{1 - r \mathcal{D}_0^{RR} \Pi_{0}^{RR} } \right) - \mathcal{D}_0^{UU} \Pi_{0}^{UU} \right]^{-1}\! \mathcal{D}_0^{UU}(q),
\end{align}

\noindent where we have introduced the quantity

\begin{align}
r(\bm{q}) 
= 1 - \frac{A^{UR}_e(\bm{q}) A^{RU}_e(\bm{q})}{A^{RR}_e(\bm{q}) A^{UU}_e(\bm{q})} .
\end{align}

\noindent Here, the Umklapp polarization occurs in the regular propagator and vice versa due to the presence of mixed magnon propagators. 

We note that in the special case \(\Omega= 0\) where we found spin triplet pairing, we have \(r=0\) since all the boosting factors are equal. In the opposite limit of \(\Omega=1\) where we found spin singlet $d$-wave pairing approaching half-filling, the mixed propagators vanish, so that \(r=1\) and each of the two magnon propagators \(\mathcal{D}^{\kappa\kappa}(q)\) are just renormalized by the corresponding polarization \(\Pi^{\kappa\kappa}(q)\).

We may now calculate the regular and the Umklapp polarizations. Performing the Matsubara frequency sums in Eqs. \eqref{eq_Pi_RR} and \eqref{eq_Pi_UU}, we obtain the standard result

\begin{align}
\Pi_{0}^{RR}(\bm{q}, i \nu_m) &= {V^2} \sum_{\bm{p}} \Bigg(\frac{n_{\textrm{F}}(\xi_{\bm{p}}) - n_{\textrm{F}}(\xi_{\bm{p} + \bm{q}})  }{i\nu_m + \xi_{\bm{p}} - \xi_{\bm{p} + \bm{q} }}\Bigg),
\label{eq_polarization_momentumSum_RR}
\\
\Pi_{0}^{UU}(\bm{q}, i \nu_m) &= {V^2} \sum_{\bm{p}} \Bigg(\frac{n_{\textrm{F}}(\xi_{\bm{p}}) - n_{\textrm{F}}(\xi_{\bm{p} + \bm{q} + \bm{Q}})  }{i\nu_m + \xi_{\bm{p}} - \xi_{\bm{p} + \bm{q} + \bm{Q}}}\Bigg),
\label{eq_polarization_momentumSum_UU}
\end{align}

\noindent where the momentum sums are evaluated in the thermodynamic limit through numerical integration \cite{GNU}.
Using that \(\xi_{\bm{p}} = \xi_{-\bm{p}}\), one may show that the imaginary part of the polarization vanishes, so that only the real part remains. 

\begin{figure}
    \centering
    \includegraphics[width=0.9\linewidth]{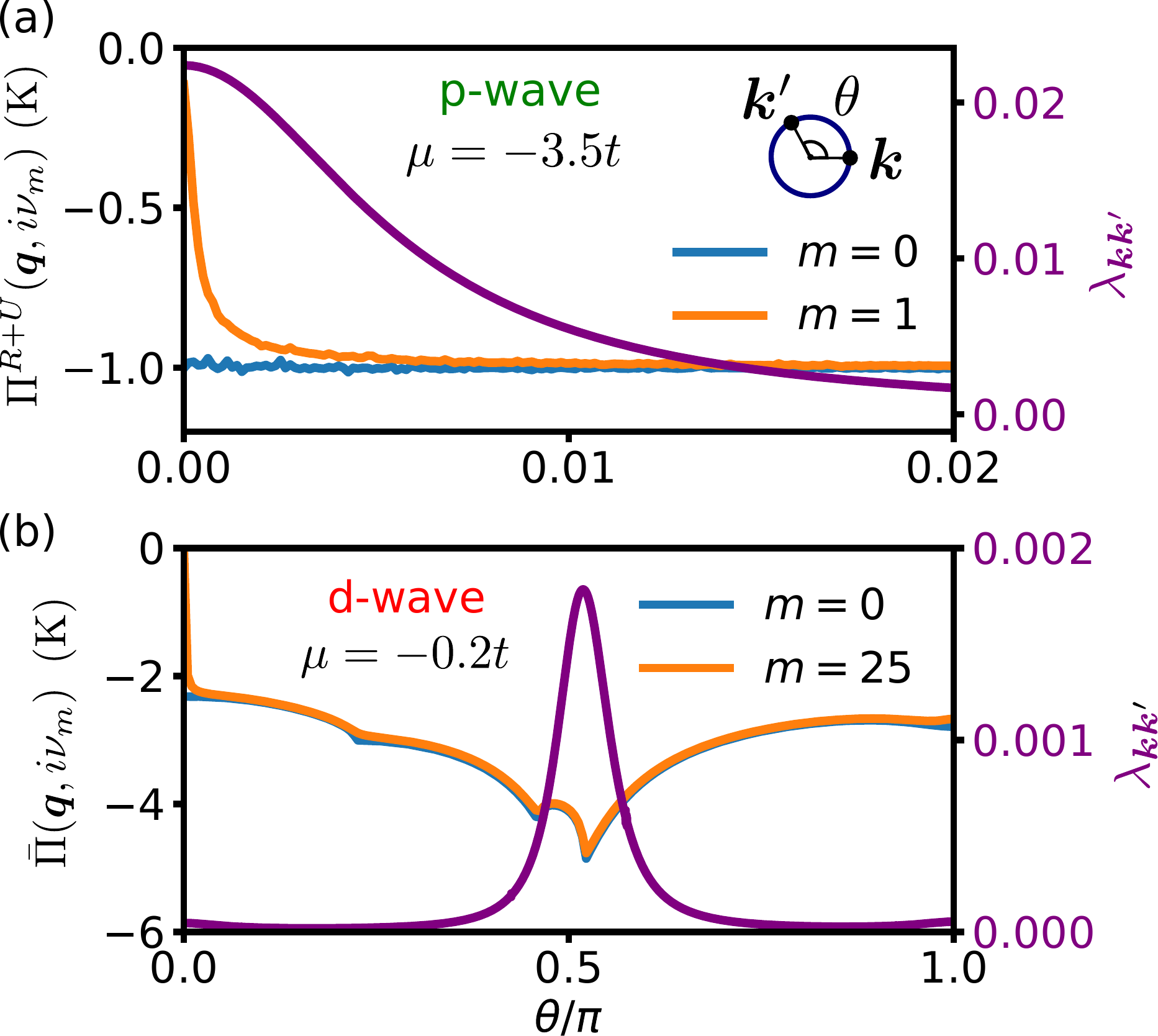}
    \caption{Polarization renormalizing the magnon propagator for magnon scattering momenta \(\bm{q}\) between points \(\bm{k}\) and \(\bm{k}'\) on the Fermi surface, corresponding to angles 0 and \(\theta\), as shown in the inset of (a). The relative contributions \(\lambda_{\bm{k}\bm{k}'}\) from the various points on the Fermi surface to \(\lambda_2(i\nu_m = 0)\) is shown in purple. (a) shows the combined polarization \(\Pi_0^{R+U}(\bm{q}, i\nu_m)\) and the contributions to \(\lambda_{2}(0)\) for \(\mu/t = -3.5\) and \(\Omega=0\), where we expect \(p\)-wave superconductivity. The dominant contributions to \(\lambda_{2}(0)\) come from small momentum processes close to \(\theta=0\). (b) shows the polarization \(\bar{\Pi}(\bm{q}, i\nu_m)\), corresponding to 
    \(\Pi^{RR}\) when \(\bm{k} - \bm{k}'=\bm{q}\) is inside, and 
    \(\Pi^{UU}\) when \(\bm{k} - \bm{k}' = \bm{q}\) is outside the reduced Brillouin zone. Dominant contributions to \(\lambda_{2}(0)\) come from Umklapp processes in vicinity to \(\bm{k} - \bm{k}' = \bm{Q}\). The temperature has been set to \(T=\SI{1}{\kelvin}\) in both subfigures.}
    \label{fig_polarizationLambda}
\end{figure}

\indent For \(\Omega=0\) and a small Fermi surface, the relevant processes are regular processes. The renormalization of the regular propagator then depends on $\Pi^{R+U}_{0}(\bm{q},i\nu_m) \equiv \Pi^{RR}_{0}(\bm{q},i\nu_m) + \Pi^{UU}_{0}(\bm{q},i\nu_m)$, where the Umklapp polarization $\Pi^{UU}_{0}$ is small. In Fig.\! \ref{fig_polarizationLambda} (a) we present
the polarization $\Pi^{R+U}_{0}(\bm{k} - \bm{k}',i\nu_m)$ together with the contributions to \(\lambda_2(i\nu_m = 0)\) from the various momenta \(\bm{k}'\) on the Fermi surface given incoming electron momentum \(\bm{k}\) as shown in the inset. The dominant contributions to $\lambda_2(i\nu_m = 0)$ arise from \(\theta \approx 0\), which corresponds to scattering processes with small momentum \(\bm{q} = \bm{k} - \bm{k}'\). In this region, the zero frequency polarization is more or less constant. Consistent with what we expect from Eq.\! \eqref{eq_polarization_momentumSum_RR}, the finite frequency polarizations approach zero as $\bm{q} \rightarrow 0$. The region where the finite frequency polarizations deviate significantly from the zero frequency polarization is, however, small compared to the region over which we expect the dominant contributions to \(\lambda_2\)~\footnote{It should be noted that although the size of the region with significant deviations from the zero frequency polarization increases with Matsubara frequency $\nu_m$, so does also the width of the region over which we expect the dominant contributions to the corresponding \(\lambda_2(i\nu_m)\). Thus, we still expect to be able to approximate the polarization by a constant for larger Matsubara frequencies \(\nu_m\).}. 
Hence we may approximate the polarization for $\Omega = 0$ and small Fermi surface by a constant value $\Pi_C \approx \Pi^{RR}_{0}(\bm{q} \rightarrow 0, i\nu_m = 0) = - N_F V^2$.\\     
\indent For \(\Omega=1\), the regular and Umklapp propagators are simply renormalized by the regular and Umklapp polarizations, respectively. Fig.\! \ref{fig_polarizationLambda} (b) therefore presents the polarization $\bar{\Pi}_0(\bm{q},i\nu_m) \equiv \theta_{\bm{q}}\Pi_{0}^{RR}(\bm{q},i\nu_m) + \theta_{\bm{q} + \bm{Q}}\Pi_{0}^{UU}(\bm{q} + \bm{Q},i\nu_m)$, which is relevant for the $d$-wave phase. Also shown are the contributions to \(\lambda_2(i\nu_m = 0)\) as in Fig.\! \ref{fig_polarizationLambda} (a). The polarization is now weakly dependent on frequency, but it varies somewhat with momentum in the relevant region. Qualitatively, it should also in this case be possible to extract the effect of magnon renormalization by setting the polarization to a constant value.\\
\indent In the following, we consider the same two special cases as above. For \(\Omega=0\) and large doping, where we found \(p\)-wave superconductivity, the relevant magnon propagator is \(\mathcal{D}^{RR}(q)\), which can be written 

\begin{align}
\mathcal{D}^{RR}(q) = -  \frac{4\omega_\mathbf{q} A_e^{RR}(\mathbf{q})}{\nu_m^2 + \omega_\mathbf{q}^2 +  4\omega_\mathbf{q} A_e^{RR}(\mathbf{q}) \Pi_{C}}.
\end{align}

\noindent Thus, the magnon frequency in the denominator has been replaced by an effective magnon frequency \(\omega_\mathbf{q}^\mathrm{eff}\), given by 

\begin{equation}
(\omega_\mathbf{q}^\mathrm{eff} )^2 = 
\omega_\mathbf{q}^2 + 4\omega_\mathbf{q} A_e^{RR}(\mathbf{q}) \Pi_{C}. 
\end{equation}

\noindent Since the polarization is negative, the effective magnon frequency may  turn imaginary, indicating that it is no longer reasonable to start out from a Néel ordered state.
At \(\bm{q}=0\), where the magnon energy  is the smallest, it happens for

\begin{equation}
    \abs{\Pi_{C}} \geq 2KS  \left( \frac{1 + K/2z_1 J_1}{1 + K/z_1 J_1} \right)\!,
\end{equation}
where $z_1$ is the number of nearest neighbors.\\
\indent For \(\Omega=1\) and large Fermi surface, where we found \(d\)-wave superconductivity, the two relevant propagators \(\mathcal{D}^{RR}\) and \(\mathcal{D}^{UU}\) are given by 

\begin{align}
\mathcal{D}^{\kappa\kappa}(q) = -  \frac{4 \omega_\mathbf{q} A_e^{\kappa\kappa}(\mathbf{q})}{\nu_m^2 + \omega_\mathbf{q}^2 + 4\omega_\mathbf{q} A_e^{\kappa \kappa}(\mathbf{q}) \Pi^{\kappa \kappa}_{0}},
\end{align}

\noindent with \(\kappa \in \{R,U\}\). Similar to the previous special case, we may now introduce an effective magnon frequency. Since the unrenormalized magnon frequency is the smallest for \(\bm{q}=0\) and the regular boosting factor \(A_e^{RR}(\bm{q} = 0)\) is suppressed due to coherence factor interference, we expect that the effective magnon frequency may primarily turn imaginary for Umklapp processes close to \(\bm{q}=0\). One may show that this happens for

\begin{equation}
\abs{\Pi^{UU}_{0}} \geq \frac{1}{2}KS.
\end{equation}

Although the coupling to the electrons may therefore in principle destroy the magnetic order in the antiferromagnet, unsurprisingly, this does not happen as long as the easy axis anisotropy is sufficiently large compared to the polarization.
\indent A picture now emerges where the easy-axis anisotropy and the magnon renormalization play opposite roles stabilizing and destabilizing the magnetic order in the antiferromagnet, respectively. Retaining magnetic order upon inclusion of magnon renormalization requires a larger easy axis anisotropy. The larger easy axis anisotropy has little effect on the numerator of the magnon propagator, but shifts the square of the magnon energies in the denominator upwards by an almost constant value with respect to momentum when $J_2/J_1$ is small. By choice of the easy-axis anisotropy, the effect of the magnon renormalization on the effective magnon frequencies can then be compensated. Superconductivity may therefore still occur at critical temperatures similar to those obtained by disregarding magnon renormalization.

\section{Discussion}
\label{Section:Discussion}

The Eliashberg equation solutions in this paper are obtained using Fermi surface averaged equations, thus neglecting the dependence of the magnon propagator and the fields appearing in the Eliashberg equations on momentum perpendicular to the Fermi surface. The justification for this is as follows: Although the magnon propagator is momentum dependent, the behaviour of the right-hand-side of the Eliashberg equations when moving $\bm{k}'$ away from the Fermi surface is still dominated by the suppression arising from the fermion energies in the denominator due to the large energy scale of the electrons. In this case there are additional variations arising from the momentum dependence of the magnon propagator. Thus, a possible avenue for further work could be to take the full momentum dependence in the Eliashberg equations into account in order to test the accuracy of our approximation.\\
\indent The results 
also rely on vertex corrections being small, so that the series of vertex diagrams can be cut off after the zeroth-order contribution. For phonon-mediated superconductivity, the smallness of the higher-order vertex diagrams is ensured by Migdal's theorem \cite{Migdal1958}, which states that higher-order diagrams are smaller by a factor $\omega_E/E_F$, where $\omega_E$ is a characteristic phonon frequency. Migdal's theorem is however known to break down for long-wavelength phonons \cite{Migdal1958, Allen1983} and in systems with strong Fermi surface nesting~\cite{Virosztek1990, Virosztek1999, Schrodi2020_2}. Moreover, for reduced spatial dimensionality, the reduction of the higher-order diagrams should be expected to be less dramatic \cite{Madhukar1977, Schrodi2020_2}. 
As the superconductivity studied in this work relies on long-wavelength magnons and/or a two-dimensional Fermi surface close to half-filling, it then seems plausible that vertex corrections could be of importance. A discussion of the effect of vertex corrections in the present system is presented in Appendix \ref{Section:VertexCorrections}. For large doping, we find that vertex corrections can become of relevant magnitude, but that the region in momentum space where the corrections are large might be small enough to limit their effect. Exactly at half-filling, the vertex corrections are expected to be quite large, but their effect can be reduced by moving away from half-filling.\\ 
\indent Upon approaching half-filling, we would also at some point expect on-set of spin density wave correlations. Exactly at half-filling, one may straightforwardly generalize the above Eliashberg theory to accommodate the expected commensurate spin density wave instability. Previously, this has been done for the phonon-induced instability~\cite{Szczesniak2005}. Below half filling, the commensurate wave-vector \(\bm{Q}\) does not connect points on the Fermi surface. We therefore expect the commensurate spin density wave to be suppressed relative to superconductivity due to the large electronic energy for processes between states which are not on the Fermi surface. However, we may still have incommensurate spin density waves, which are far more challenging to investigate theoretically. In this paper, we have been investigating the properties of the superconducting phases, and the highly non-trivial interplay between superconducting and spin density wave orders that we could potentially obtain is beyond the scope of the present study.\\
\indent Another effect that could be included is the effect of the quasiparticle energy shift $\chi$. For the present system, $\chi$ was found to be small compared to the Fermi energy for chemical potentials ranging from half-filling and down towards the vicinity of the bottom of the band. Apart from the limit where the Fermi level approaches the bottom of the band, inclusion of $\chi$ would typically amount to a small, weakly frequency-dependent, shift of the effective chemical potential in the Eliashberg equations, and it was therefore neglected in the presented calculations.\\
\indent The effect of Coulomb interactions on the electron self-energy is in general challenging to calculate \cite{Allen1983}. Their effect on the Fermi surface averaged Eliashberg equations for the anomalous pairing amplitudes is typically included by neglecting vertex corrections and including an extra repulsive and frequency-independent potential in the equations. The Coulomb repulsion will then have a limited effect on the critical temperatures as long as $\lambda_2(i\nu_m = 0)$ is somewhat larger than the Coulomb pseudopotential $\mu^*$ \cite{Allen1983}. Moreover, taking the Coulomb interaction to be momentum independent, its contributions to the gap equation will cancel for unconventional pairing symmetries like the ones considered in this work. For a momentum dependent Coulomb potential, the Coulomb contributions to the gap equation no longer cancel identically for unconventional pairing symmetries, but $\mu^*$ will still be reduced compared to the $s$-wave case.\\ 
\indent In the system setup we have considered, the antiferromagnetic order and interfacial coupling to the two antiferromagnets has been chosen such that any magnetic fields cancel. If we instead consider a single antiferromagnetic layer and \(\Omega \neq 1\), there would be a net magnetic field, as shown in Eqs. \eqref{eq_magneticField_A_L} and \eqref{eq_magneticField_B_L}. In addition, there would also be a term in the magnon propagator that is odd in frequency, as shown in Eq.\! \eqref{eq_magnonPropagator_oneLayer}. The odd part
of the propagator would renormalize and reduce the magnetic field \(h\), and produce an effective magnetic field \(\tilde{h}\). Together with the odd part of the propagator, this effective magnetic field could in principle give rise to an exotic coexistence of odd- and even frequency superconductivity~\cite{Matsumoto2012}. For the experimentally most relevant parameters, we would however expect the magnetic field to be too strong to give significant critical temperatures.\\
\indent We also note that a previously studied system consisting of a normal metal sandwiched between two ferromagnetic insulators \cite{Fjaerbu2018} gives rise to a $p$-wave phase that bears many similarities with the $p$-wave phase considered in the present study. The main difference between the two systems is the absence of the magnon coherence factors in the ferromagnetic case. The numerator of the magnon propagator (or effective potential in a weak-coupling framework) for the ferromagnet therefore scales as $\omega^{FM}_{\bm{q}} \sim K$ for long-wavelength magnons, while the numerator of the magnon propagator for the antiferromagnet scales as $A^{RR}_e(\bm{q})\, \omega_{\bm{q}} \sim J_1$. For superconductivity dominated by long-wavelength magnons, with $K/J_1 \ll 1$, the dimensionless electron-magnon coupling \(\lambda_{2}(i\nu_m=0)\) may however still be of the same magnitude in both cases, corresponding to similar dimensionless coupling constants in a weak-coupling framework. This is because the ferromagnet propagator can simply rely on having a smaller gap in the magnon spectrum, making the denominator of the propagator smaller for the long-wavelength processes. As the critical temperature in a simple weak-coupling framework only depends on the dimensionless coupling constant and the cutoff on the boson spectrum, sizeable critical temperatures can then be obtained for both ferromagnets and antiferromagnets. Within an Eliashberg framework, on the other hand, the effective cutoff frequency is determined by the characteristic magnon energies in the pairing interaction. 
Since, with ferromagnets, the large values for \(\lambda_{2}(i\nu_m=0)\) were obtained by relying on smaller magnon energies in the denominator of the propagator, the effective frequency cutoff will be smaller, and the critical temperatures obtainable with ferromagnets should be smaller than with antiferromagnets. \\
\indent In the current antiferromagnetic case, magnon renormalization was found to have little effect on the available critical temperatures. This is because the larger easy axis anisotropy $K$, required to protect magnetic order in the AFMIs, is compensated by the magnon energy renormalization in the denominator of the propagator. The larger easy-axis anisotropy has little effect on the numerator of the propagator. For the case of the ferromagnet, on the other hand, increasing $K$ so that it compensates the renormalization would also lead to a larger numerator in the propagator. Magnon renormalization could then open the way for slightly higher critical temperatures.

\section{Experimental considerations}
\label{Section:ExperimentalConsiderations}

The model employed in this study allows us to tune the interfacial coupling between the normal metal and the two different sublattices of the antiferromagnet independently. In principle, such a general asymmetric coupling could be engineered, as discussed in the introduction. However, the experimentally most promising route to realizing superconductivity in systems well described by our model appears to be through fully compensated and uncompensated interfaces, where the conduction band electrons in the normal metal are coupled to only one AFMI sublattice ($\Omega = 0$), or equally to both AFMI sublattices ($\Omega = 1$). 
There is however a significant difference between our model and the intended realization with an uncompensated interface for the case \(\Omega=0\). In the intended realization, the square lattice of the normal metal matches the exposed sublattice of the antiferromagnet, and not the square lattice of the antiferromagnet itself, as in our model. Thus, the electron Brillouin zone coincides with the Brillouin zone of the antiferromagnet. Although it is possible to imagine a compensated interface where the magnons at the interface live in a smaller Brillouin zone than the electrons, this would not be the typical case.\\
\indent Within our model, Umklapp processes are included for both $\Omega = 0$ and $\Omega = 1$. In the intended realization for \(\Omega=0\), however, Umklapp processes are absent. For a small Fermi surface, the effect of Umklapp processes in our model is small, since all processes between points on the Fermi surface are of the regular type. The $p$-wave phase we expect for uncompensated interfaces and large doping is therefore well represented by our model. For small doping, however, the $f$-wave phase of our model takes precedence over the \(p\)-wave phase precisely because of the Umklapp processes. The \(f\)-wave phase is therefore of less experimental relevance, and we expect that the $p$-wave phase would dominate regardless of doping for a normal metal sandwiched between two uncompensated interfaces.\\
\indent For the $p$-wave phase with $\Omega = 0$, having a trilayer heterostructure in order to cancel all magnetic fields seems necessary, as the critical temperature is significantly reduced compared to previous predictions \cite{Erlandsen2019_Enhancement}. For the $d$-wave phase with $\Omega = 1$, the result of coupling to a single antiferromagnet would lead to the presence of a staggered field. As a staggered field might be less detrimental to superconductivity than a uniform spin-splitting \cite{Cebula2000}, a bilayer heterostructure might be a viable option in this case.\\
\indent When it comes to choice of parameters, it is clear that a strategy of simply taking a very small gap in the magnon spectrum in order to increase the dimensionless coupling strength $\lambda_2(0)$ has its limitations, as this leads to a very small effective cutoff frequency and slow increase in critical temperature with dimensionless coupling strength. In order to realize superconductivity in this system it is then essential that the constant prefactor that appears in the gap equation is sufficiently large. A sizeable interfacial exchange coupling and electron density of states is then preferable. Moreover, as the effective induced interaction experienced by the electrons in the normal metal might be reduced with the thickness of the normal metal \cite{Fjaerbu2018}, the metallic layer should be kept quite thin.\\
\indent The easy-axis anisotropy governs the size of the gap in the magnon spectrum, and appears to play a crucial role in realizing superconductivity. A sufficiently large gap in the magnon spectrum could be important for both the $p$-wave and $d$-wave phases in order to stabilize the antiferromagnet. The $p$-wave phase does, however, rely more heavily on fine-tuning of the easy-axis anisotropy in order to produce a nonzero, but sufficiently small, effective magnon gap producing a sizeable critical temperature. This could make the $p$-wave phase more difficult to realize experimentally. The $d$-wave pairing receives contributions from a wider range of magnon energies, and the critical temperature is therefore more robust to a shift of the magnon energies. For larger Fermi surfaces, a larger easy-axis anisotropy is however needed to preserve magnetic order in the antiferromagnet, which could in itself be an experimental complication. However, using a magnetically more stable three-dimensional antiferromagnet instead of the two-dimensional magnet considered in our model, could potentially lead to a reduction in the easy-axis anisotropy required to stabilize the magnets. \\
\indent In contrast to earlier results, the present study indicates that the $d$-wave phase may be able to produce higher critical temperatures than the $p$-wave phase. However, the $d$-wave phase is, in our model, dependent on proximity to half-filling, where it e.g.\! needs to compete with spin-density wave order. This competition may push the superconducting phase down towards lower filling-fractions associated with lower critical temperatures. 
It should also be noted that since the $d$-wave phase relies on Umklapp processes, it is more sensitive to the detailed structure of the Fermi surface. In comparison with the $p$-wave phase, the $d$-wave phase may therefore place stricter requirements on the electron band structure of the normal metal in the experimental realization.
Compared with the third option of coupling to ferromagnetic insulators, however, both phases considered in the present study seem more promising.

\section{Summary}\label{Section:Summary}
We use Eliashberg theory to study interfacially induced magnon-mediated superconductivity in a normal metal-antiferromagnet heterostructure. For large doping and uncompensated antiferromagnetic interfaces, we find $p$-wave superconductivity, while for small doping and compensated antiferromagnetic interfaces, we find $d$-wave superconductivity. This can be understood in terms of sublattice interferences suppressing and enhancing scattering processes in the system.
Although the qualitative results are in accordance with earlier weak-coupling studies, the critical temperature achievable for the $p$-wave phase is found to be significantly reduced as the characteristic magnon frequency in the pairing interaction is much smaller than the cutoff on the magnon spectrum. The $d$-wave phase, on the other hand, is found to rely less on long-wavelength magnons and can therefore potentially produce larger critical temperatures when approaching half-filling. Close to half-filling the $d$-wave instability may however have to compete with a spin-density wave instability, potentially reducing the available critical temperatures. 
A sufficiently large gap in the magnon spectrum might be necessary to stabilize the magnetic order in the antiferromagnets due to feedback from the electrons, but this is found to have limited effect on the critical temperatures. 

\section{Acknowledgements}
We acknowledge useful discussions with J.W. Wells. 
We acknowledge financial support from the Research Council of Norway Grant No.\! 262633 “Center of Excellence on Quantum Spintronics” and Grant No.\! 250985, “Fundamentals of Low-dissipative Topological Matter”.\\

\appendix

\section{Antiferromagnetic magnons}
\label{Section:Antiferromagnet}

Starting from the AFMI Hamiltonian, we introduce the linearized Holstein-Primakoff transformation~\cite{Kittel1987}

\begin{subequations}
\begin{align}
S_{i \in A, H}^+ = \sqrt{2S} a_{iH}  
\hspace{1.82cm}
S_{i \in A, L}^+ = \sqrt{2S} a_{iL}^\dagger&
\\
S_{j \in B, H}^+ = \sqrt{2S} b_{jH}^\dagger
\hspace{1.74cm} 
S_{j \in B, L}^+ = \sqrt{2S}  b_{jL}&
\\
S_{i \in A, H}^- = \sqrt{2S} a_{iH}^\dagger 
\hspace{1.82cm} 
S_{i \in A, L}^- = \sqrt{2S} a_{iL}&
\\
S_{j \in B, H}^- = \sqrt{2S} b_{jH} 
\hspace{1.74cm} 
S_{j \in B, L}^- = \sqrt{2S} b_{jL}^\dagger&
\\
S_{i \in A, H}^z = S - a_{iH}^\dagger a_{iH}
\hspace{0.575cm}
S_{i \in A, L}^z = - S + a_{iL}^\dagger a_{iL}&
\\
S_{j \in B, H}^z = - S + b_{jH}^\dagger b_{jH}
\hspace{0.5cm}
S_{j \in B, L}^z = S - b_{jL}^\dagger b_{jL}&,
\end{align}
\end{subequations}

\noindent where we have assumed oppositely aligned antiferromagnetic order in the spin space $z$-direction for the two antiferromagnets. 

Inserting this into the AFMI Hamiltonian and expressing it in terms of sublattice magnon Fourier modes \(a_{\bm{q}\eta}\) and \(b_{\bm{q}\eta}\), the AFMI Hamiltonian takes the form 

\begin{align}
H_\mathrm{AFMI} = \sum_{\bm{q},\eta} 
&C_{\bm{q}} (a_{\bm{q}\eta}^\dagger a_{\bm{q}\eta}  + b_{\bm{q}\eta}^\dagger b_{\bm{q}\eta} ) 
\nonumber \\
&+D_{\bm{q}} (a_{\bm{q}\eta} b_{-\bm{q}\eta} + a_{\bm{q}\eta}^\dagger b_{-\bm{q}\eta}^\dagger ), 
\end{align}

\noindent where \(C_{\bm{q}}\) and \(D_{\bm{q}}\) are given by 

\begin{subequations}
\begin{align}
C_{\bm{q}} &= 2 z_1 J_1 S - 2 z_2 J_2 S ( 1 - \tilde{\gamma}_{\bm{q}} ) + 2KS, 
\\
D_{\bm{q}} &= 2 z_1 J_1 S \gamma_{\bm{q}},
\end{align}
\end{subequations}

\noindent and we have defined

\begin{align}
\gamma_{\bm{q}} = \frac{1}{z_1} \sum_{\pmb{\delta}_1} e^{i \bm{q} \cdot \pmb{\delta}_1 },
\qquad
\tilde{\gamma}_{\bm{q}} = \frac{1}{z_2} \sum_{\pmb{\delta}_2} e^{i \bm{q} \cdot \pmb{\delta}_2 }.
\end{align}

\noindent Here, \(z_1\) and \(z_2\) are the number of nearest and next-nearest neighbour vectors, which are summed over and denoted by \(\pmb{\delta}_1\) and \(\pmb{\delta}_2\). The Hamiltonian is diagonalized through the Bogoliubov transform

\begin{subequations}
\begin{align}
a_{\bm{q}\eta} &= u_{\bm{q}} \alpha_{\bm{q}\eta} + v_{\bm{q}} \beta_{-\bm{q}\eta}^\dagger, \\
b_{-\bm{q}\eta}^\dagger &= u_{\bm{q}} \beta_{-\bm{q}\eta}^\dagger 
+ v_{\bm{q}} \alpha_{\bm{q}\eta},
\end{align}
\end{subequations}

\noindent where the coherence factors \(u_{\bm{q}}\) and \(v_{\bm{q}}\) can be written as 

\begin{equation}
u_{\bm{q}} = \cosh \theta_{\bm{q}}, \qquad \qquad 
v_{\bm{q}} = \sinh \theta_{\bm{q}},
\end{equation}

\noindent in terms of the hyperbolic angle  

\begin{equation}
\theta_{\bm{q}} = -\frac{1}{2} \tanh^{-1} \left( \frac{D_{\bm{q}}}{C_{\bm{q}}} \right).
\end{equation}

The resulting magnon spectrum is 

\begin{equation}
\omega_{\bm{q}} = \sqrt{C_{\bm{q}}^2 - D_{\bm{q}}^2}.
\end{equation}

\noindent By expressing the inverse hyperbolic tangent in terms of a logarithm, one may show the relations 

\begin{subequations}
\begin{align}
u_{\bm{q}}^2 + v_{\bm{q}}^2 
&= +C_{\bm{q}}/\omega_{\bm{q}}, 
\label{eq_qafm_uvRelation_a}
\\
2 u_{\bm{q}}v_{\bm{q}} 
&= - D_{\bm{q}} / \omega_{\bm{q}},
\label{eq_qafm_uvRelation_b}
\end{align}
\end{subequations}

\noindent for the coherence factor combinations which appear in the magnon propagator. 

Whereas \(u_{\bm{q}}\) is positive, \(v_{\bm{q}}\) is  typically negative. Furthermore, we notice that when \(K\) and \(J_2\) are small compared to \(J_1\), \(|\theta_{\bm{q}}|\) becomes large when \(\bm{q} \rightarrow 0\), as \(D_{\bm{q}}\) approaches \(C_{\bm{q}}\). This causes \(u_{\bm{q}}\) to grow large and positive and \(v_{\bm{q}}\) to grow large and negative in this limit.

\section{Interfacial coupling Hamiltonian}
\label{sec:Coupling}

In the main text, we presented expressions for the interfacial coupling and the magnon propagators under the assumption that the two antiferromagnets couple to the normal metal with equal strength. In this appendix, we generalize the results beyond this assumption.

The interfacial coupling Hamiltonian describing the coupling to a single antiferromagnetic insulator labelled by \(\eta\) can be written \(H_\mathrm{int}^{\eta} = H_\mathrm{int}^{h, \eta} + H_\mathrm{int}^{V, \eta}\), where the magnetic exchange field contributions \(H_\mathrm{int}^{h, \eta} = H_\mathrm{int}^{h,A,\eta} + H_\mathrm{int}^{h,B, \eta}\) from the two sublattices are 

\begin{subequations}
\begin{align}
&\hspace{-0.1cm}H_\mathrm{int}^{h,A, H} = 
- \bar{J}\, \Omega_A^H S\! \sum_{\bm{k} \in \square, \sigma}\! \sigma \!
\left( c_{\bm{k}\sigma}^\dagger c_{\bm{k} \sigma} +c_{\bm{k}+\bm{Q},\sigma}^\dagger c_{\bm{k} \sigma} 
\right)\!,
\label{eq_magneticField_A_H}
\\
&\hspace{-0.1cm}H_\mathrm{int}^{h,B,H} = + \bar{J}\, \Omega_B^H S\! \sum_{\bm{k} \in \square, \sigma}\! \sigma \!
\left( 
c_{\bm{k}\sigma}^\dagger c_{\bm{k} \sigma} -c_{\bm{k}+\bm{Q},\sigma}^\dagger c_{\bm{k} \sigma} 
\right)\!,
\label{eq_magneticField_B_H}
\\
&\hspace{-0.1cm}H_\mathrm{int}^{h,A, L} = 
+ \bar{J}\, \Omega_A^L S\! \sum_{\bm{k} \in \square, \sigma}\! \sigma \!
\left( c_{\bm{k}\sigma}^\dagger c_{\bm{k} \sigma} +c_{\bm{k}+\bm{Q},\sigma}^\dagger c_{\bm{k} \sigma} 
\right)\!,
\label{eq_magneticField_A_L}
\\
&\hspace{-0.1cm}H_\mathrm{int}^{h,B,L} = - \bar{J}\, \Omega_B^L S\! \sum_{\bm{k} \in \square, \sigma}\! \sigma \!
\left( 
c_{\bm{k}\sigma}^\dagger c_{\bm{k} \sigma} - c_{\bm{k}+\bm{Q},\sigma}^\dagger c_{\bm{k} \sigma} 
\right)\!,
\label{eq_magneticField_B_L}
\end{align}
\end{subequations}

\noindent and where the exchange coupling strengths \(\bar{J} \Omega_{\Upsilon}^{\eta}\) are in general different for the two antiferromagnets. Coupling to only one antiferromagnet can be realized by e.g. setting \(\Omega_{\Upsilon}^L = 0\).  Assuming \(\Omega^H_{\Upsilon} = \Omega^L_{\Upsilon}\), however, all magnetic fields cancel. 

The electron-magnon interaction is given by 

\begin{align}
H_\mathrm{int}^{V,\eta} = 
V \sum_{\substack{\bm{k} \in \Box\\ \bm{q} \in \diamondsuit}} 
\Big[
M_{\bm{q}\eta}^R c_{\bm{k} + \bm{q}, \downarrow}^\dagger c_{\bm{k},\uparrow} 
+ M_{\bm{q}\eta}^U c_{\bm{k} + \bm{q}+\bm{Q}, \downarrow}^\dagger c_{\bm{k},\uparrow} \nonumber \\
+ (M_{-\bm{q}\eta}^R)^\dagger c_{\bm{k} + \bm{q}, \uparrow}^\dagger c_{\bm{k},\downarrow} 
+(M_{-\bm{q}\eta}^U)^\dagger c_{\bm{k} + \bm{q}+\bm{Q}, \uparrow}^\dagger c_{\bm{k},\downarrow} 
\Big],
\end{align}

\noindent where we have defined magnon operators \(M^{\kappa}_{\bm{q}\eta}\) associated with the antiferromagnet \(\eta\) as 

\begin{subequations}
\begin{align}
M^{\kappa}_{\bm{q}H} &=  \Omega^{H}_A a_{\bm{q} H}  + \kappa\, \Omega^{H}_B  b_{-\bm{q} H}^\dagger, 
\\
M^{\kappa}_{\bm{q}L} &= \Omega^{L}_A a_{-\bm{q} L}^\dagger + \kappa\, \Omega^{L}_B b_{\bm{q}L}, 
\end{align}
\end{subequations}

\noindent so that the operator \(M_{\bm{q}}^{\kappa}\) introduced in the main text is given by \(M_{\bm{q}}^{\kappa} = M_{\bm{q}H}^{\kappa} + M_{\bm{q}L}^{\kappa}\). Expressing the magnon operators in terms of the eigenmagnon operators resulting from the Bogoliubov transformation, the corresponding magnon propagators are

\begin{align}
\begin{aligned}
\mathcal{D}_{0,\eta}^{\kappa \kappa'}(\bm{q}, i\omega_n) = 
& - A_{e,\eta}^{\kappa \kappa'}(\bm{q}) \frac{2\omega_{\bm{q}} }{\omega_n^2 + \omega_{\bm{q}}^2}\\
&- A_{o,\eta}^{\kappa \kappa'}\frac{ 2 i \omega_n  }{\omega_n^2 + \omega_{\bm{q}}^2}.
\label{eq_magnonPropagator_oneLayer}
\end{aligned}
\end{align}

\noindent Here, the first term is even under the three-vector transformation \(q \rightarrow - q\), and the second term is odd. The expressions for $A_{e,\eta}^{\kappa \kappa'}(\bm{q})$ can be obtained from Eq.\! \eqref{eq_evenBoostingFactors} in the main text by the simple generalization $A_{e}^{\kappa\kappa'}(\bm{q}) \rightarrow A_{e,\eta}^{\kappa\kappa'}(\bm{q})$ and $\Omega_{\Upsilon} \rightarrow \Omega^{\eta}_{\Upsilon}$.  The odd part prefactor \(A_{o,\eta}^{\kappa\kappa'}\) is \(\bm{q}\)-independent, and given by

\begin{align}
A^{\kappa\kappa'}_{o, \eta} &=
\frac{1}{2} \eta
\left[ (\Omega_A^\eta)^2 - \kappa \kappa' (\Omega_B^\eta)^2 \right],
\end{align}
\noindent  where we associate the index $\eta$ with the values $H \rightarrow 1$ and $L \rightarrow -1$. We notice that the odd part of the propagator has different signs for the two antiferromagnets, so that their contributions cancel out when we couple equally to the two antiferromagnets.

\section{Suppression of electron correlations which are off-diagonal in momentum}
\label{Section:OffDiagonalInMomentum}

In this appendix, we argue that terms in the electron Green's function which are off-diagonal in momentum are suppressed as long as the electron propagator renormalization is small compared to the difference in electron energies at the momenta \(\bm{k}\) and \(\bm{k} + \bm{Q}\). 

The self-energy is in general an \(8 \times 8\) matrix in the momentum, particle/hole, and spin degrees of freedom. The self-energy can then be written 

\begin{equation}
\Sigma = \begin{pmatrix}
\Sigma_{11} & \Sigma_{12} \\
\Sigma_{21} & \Sigma_{22} 
\end{pmatrix},
\end{equation}

\noindent where \(\Sigma_{ij}\) is now a \(4\times 4\) submatrix in the particle/hole and spin degrees of freedom corresponding to momentum sector \((i,j)\). The self-energy is related to the Green's function through the Dyson equation, so that 

\begin{equation}
G^{-1}(k) = 
\begin{pmatrix}
G_0^{-1}(k) - \Sigma_{11}(k) & -\Sigma_{12}(k) \\
- \Sigma_{21}(k) & G_0^{-1}(k+Q) - \Sigma_{22}(k)\\
\end{pmatrix}.
\end{equation}

\noindent Away from half-filling, both \(\bm{k}\) and \(\bm{k}+\bm{Q}\) cannot both be on the Fermi surface, so one of the submatrices on the diagonal will have a term proportional to the identity matrix and prefactor of the same order as the electron energy scale. We now assume that \(\bm{k}\) is close to the Fermi surface, so that this applies to $G_0^{-1}(k+Q)$. 
To obtain the Green's function \(G(k)\), we then make use of the following matrix inversion identity~\cite{Horn2012}: Given a matrix \(G^{-1}\) which can be partitioned into submatrices and written on the form

\begin{equation}
G^{-1} = \begin{pmatrix} N_{11} & N_{12} \\ N_{21} & N_{22}  \end{pmatrix}, 
\end{equation}

\noindent where \(N_{11}\) and \(N_{22}\) are invertible matrices~\footnote{The identity can also be formulated more generally, but this will be sufficient for our purposes.}, the inverse can similarly be expressed 

\begin{equation}
G = 
\begin{pmatrix}
M_{11} & M_{12} \\
M_{21} & M_{22}
\end{pmatrix},
\end{equation}

\noindent with submatrices

\begin{subequations}
\begin{align}
M_{11} &= (N_{11} - N_{12} N_{22}^{-1} N_{21} )^{-1}, \\
M_{12} &= -(N_{11} - N_{12} N_{22}^{-1} N_{21} )^{-1} N_{12} N_{22}^{-1},
\\
M_{21} &= -N_{22}^{-1} N_{21} (N_{11} - N_{12} N_{22}^{-1} N_{21} )^{-1},
\\
M_{22} &= (N_{22} - N_{21} N_{11}^{-1} N_{12})^{-1}.
\end{align}
\end{subequations}

In our case, \(N_{22}\) can now be thought of as an electronic energy that is much larger than the other submatrices, which have contributions from the self-energy and the non-interacting Green's function close to the Fermi surface. As long as the renormalization is small compared to the electron energy scale in the problem, \(N_{22}^{-1} N_{21}\) is then small, and \(M_{21}\) and \(M_{12}\) are suppressed relative to \(M_{11}\). By similar reasoning, \(M_{22}\) is also small, and \(M_{11}\) can be approximated by \(N_{11}^{-1}\).\\ 
\indent When \(\bm{k}+\bm{Q}\) is close to the Fermi surface, we may similarly neglect \(M_{11}\) and the off-diagonal terms, but not \(M_{22} \approx N^{-1}_{22}\). For a general \(\bm{k}\), we may therefore neglect the off-diagonal terms, which is exactly what we use in the main text. 

\section{Vertex corrections}
\label{Section:VertexCorrections}
\begin{figure}[t] 
    \begin{center}
        \includegraphics[width=0.95\columnwidth,trim= 0.0cm 19.7cm 0.5cm 1.1cm,clip=true]{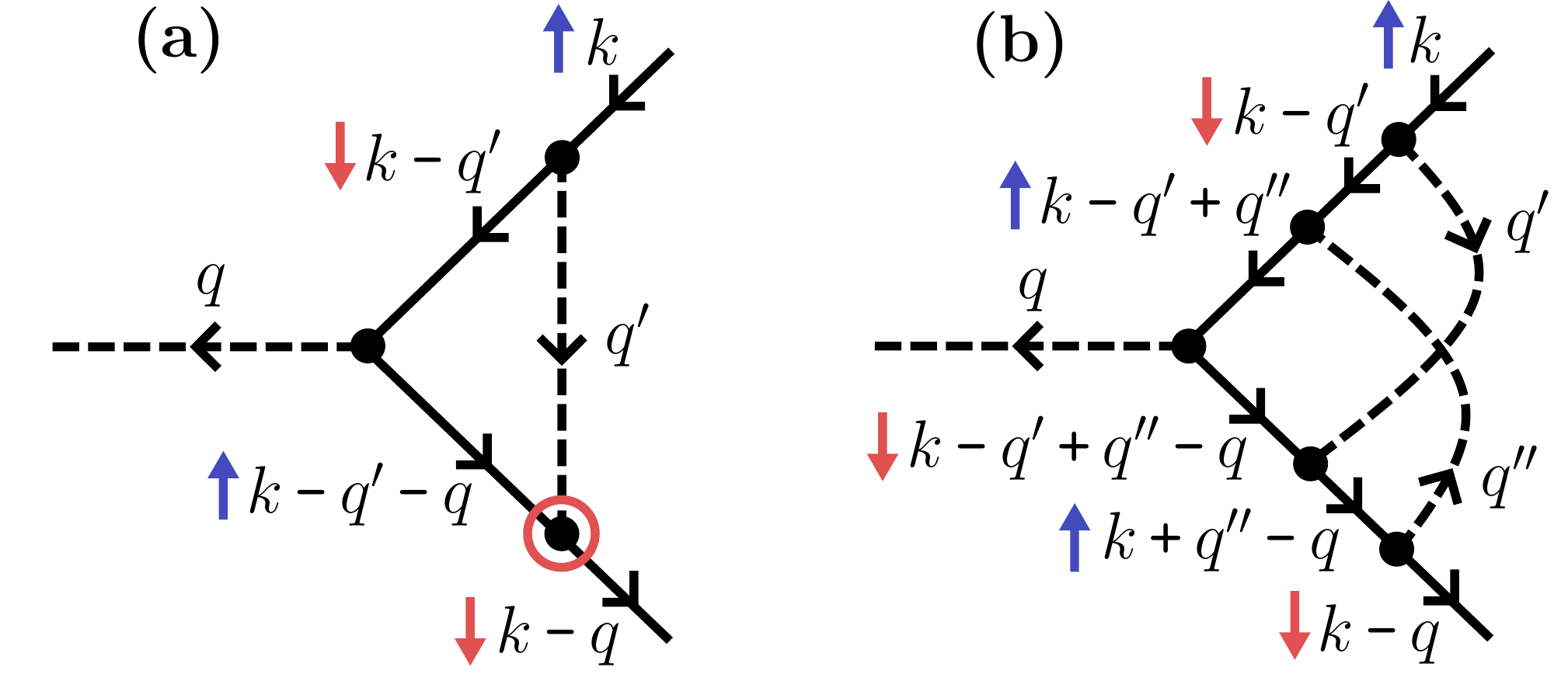}
    \end{center}
    \caption{(a) Typical lowest-order vertex correction, which vanishes in this case due to conservation of spin. (b) Lowest-order vertex correction for our model.}
    \label{fig:ho_vertex}
\end{figure}

\indent In order to obtain some insight into the importance of vertex corrections, we will attempt to estimate the magnitude of the lowest-order vertex corrections. Focusing on regular processes for the time being, a magnon-equivalent of the lowest-order vertex correction for phonon-mediated superconductivity is presented in Fig.\! \ref{fig:ho_vertex} (a). Due to conservation of spin, this diagram vanishes for our system. Starting with the upper vertex of the vertical magnon line, we see that the electron spin is flipped from $\uparrow$ to $\downarrow$, meaning that the outgoing magnon carries a spin $+1$. In the lower vertex of the vertical magnon line, this spin needs to be returned to the electrons, but the incoming electron already has spin $\uparrow$ instead of spin $\downarrow$, and spin can therefore not be conserved in this vertex. Including Umklapp processes, the momentum structure of Fig.\! \ref{fig:ho_vertex} (a) will differ, but the spin structure stays the same. The lowest-order vertex corrections are therefore of the type represented by the diagram in Fig.\! \ref{fig:ho_vertex} (b). As the diagram in (b) is of higher order, the effect of vertex corrections should then be expected to be smaller than what would have been the case if the diagram in (a) had not vanished.\\
\indent The diagram in Fig.\! \ref{fig:ho_vertex} (b) represents a correction to the electron-magnon vertex $\frac{V}{4}g_{\gamma} \rightarrow \frac{V}{4}g_{\gamma}(1 + \Gamma)$, where 

\begin{align}
\begin{aligned}
    &\Gamma(k, q) \sim \frac{V^4}{\beta^2}\!\sum_{q', q''}\!\mathcal{D}^{RR}_{0}(q')\,\mathcal{D}^{RR}_{0}(q'')\,G^{11}_{0}(k + q'' - q)\\
    &\times G^{22}_{0}(k - q' + q'' - q)\,G^{11}_{0}(k - q' + q'')\, G^{22}_{0}(k - q').
\end{aligned}
\label{eq:Gamma}
\end{align}
A quick estimate for $\Gamma$ can be obtained in the following way \cite{Coleman2015}. We approximate the magnon propagators as

\begin{align}
    \mathcal{D}^{RR}_{0}(q) \sim -\frac{A^{RR}_e(0)}{\omega_{c}},
\end{align}
for Matsubara frequency $q_m$ less than some cutoff frequency $\omega_c \!\sim\! \omega_0$, where $\omega_0$ is the magnon gap. For $q_m > \omega_c$, we take the magnon propagator to be zero. The number of terms that should be included in each of the Matsubara sums is then roughly $\beta \omega_c$. When performing the sums over momentum, the fermions will typically be away from the Fermi surface. We then approximate the momentum sums with the number of lattice sites $N$, and the electron Green's functions by $G^{aa}_{0} \sim 1/E_F$, where $E_F$ is the Fermi energy, which is taken as a measure of the electron energy scale $\sim 1\,\textrm{eV}$. We then obtain  

\begin{align}
\begin{aligned}
    \Gamma \sim \Bigg(\frac{V^2 N A^{RR}_e(0)}{E^2_F}\Bigg)^2 \sim \Bigg(\frac{1}{100}\Bigg)^2,
\end{aligned}
\label{eq_vertexCorrection_simplestEstimate}
\end{align}
where we have inserted typical values for the relevant energy scales, and taken $\Omega = 0$ which is suitable for the case of a relatively small Fermi surface where regular processes dominate. This estimate would indicate that vertex corrections are typically small. It does however not take into account that there can be large contributions arising from fermions being close to the Fermi surface when $\bm{q} \rightarrow 0$. In order to pick up such contributions, we need to perform a more detailed estimate.\\
\indent Starting from Eq.\! \eqref{eq:Gamma}, we can perform the Matsubara sums. Following Ref.\! \cite{Roy2014}, we focus on the term that arises from the poles of the boson propagators, limiting the number of factors with fermion energies in the denominator. At zero temperature, this term becomes

\begin{align}
\begin{aligned}
    &\Gamma_{1}(k, q) = -4V^4\!\sum_{\bm{q}', \bm{q}''}\!A^{RR}_{e}(\bm{q}')A^{RR}_{e} (\bm{q}'')\\
    &\times\!\Bigg(\frac{1}{\omega_{\bm{q}'} + ik_n - \xi_{\bm{k} - \bm{q}'}}\Bigg)\!\Bigg(\frac{1}{\omega_{\bm{q}''} + \xi_{\bm{k} + \bm{q}'' - \bm{q}} - i(k_n - q_m)}\Bigg)\\
    &\times\!\Bigg(\frac{1}{\omega_{\bm{q}'} + i(k_n - q_m) - \omega_{\bm{q}''} - \xi_{\bm{k} - \bm{q}' + \bm{q}'' - \bm{q}}}\Bigg)\\
    &\times\!\Bigg(\frac{1}{\omega_{\bm{q}'} + ik_n -\omega_{\bm{q}''} - \xi_{\bm{k} - \bm{q}' + \bm{q}''}}\Bigg).
\end{aligned}
\label{eq_vertexCorrectionExpression}
\end{align}
\begin{figure}[b]
    \centering
    \includegraphics[width=0.8\columnwidth,trim= 2.5cm 20.0cm 4.0cm 0.5cm,clip=true]{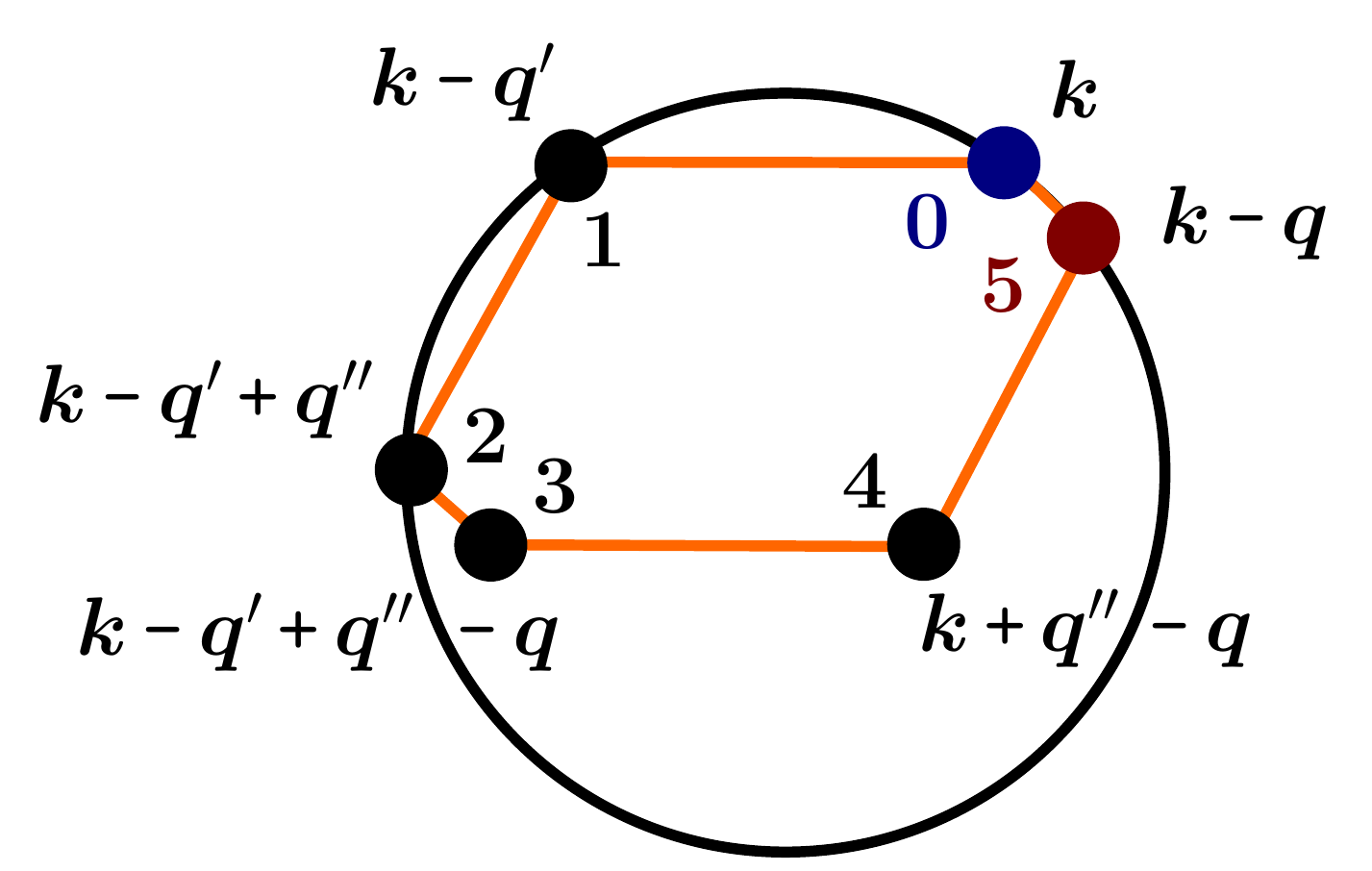}
    \caption{The momentum scattering processes in the simplest non-vanishing vertex correction of Fig.\! \ref{fig:ho_vertex} (b) can be represented as a hexagon. Opposing sides in the hexagon are parallel and equally long due to conservation of momentum. The Fermi surface (assuming large doping) is shown as a circle.}
    \label{fig_vertexCorrectionScattering}
\end{figure}
To estimate this term in the limit of small \(\bm{q}\), we need to analyze which regions of the Brillouin zone that dominate the momentum sums. The momentum scattering processes in the vertex correction diagram can be represented by a hexagon where opposing sides are parallel and equally long due to conservation of momentum, as shown in Fig. \ref{fig_vertexCorrectionScattering}. Each vertex in the hexagon represents the momentum of an electron propagator in the Feynman diagram. We consider processes where \(\bm{k}\) and \(\bm{k} - \bm{q}\) are close to the Fermi surface. Consider the variables \(\bm{q}'\) and \( \bm{q}' - \bm{q}'' \equiv \bm{\pi}\) to be the integration variables of our momentum sums. For small $\bm{q}$, the vertices 2 and 3 in Fig.~\ref{fig_vertexCorrectionScattering} are reasonably close to each other. The dominant contributions to the diagram should therefore arise when \(\bm{k} - \bm{q}' + \bm{q}''\) is close to the Fermi surface. With \(\bm{q}' - \bm{q}''\) fixed, the position of the remaining two vertices 1 and 4 is fixed by choosing \(\bm{q}'\). Taking vertex 1 to be close to the Fermi surface, vertex 4 will now typically end up away from the Fermi surface. The number of terms in the sum over $\bm{q}'$ where vertex 1 is close to the Fermi surface is of order \(N_F \omega_c\). The Green's functions corresponding to points 1 and 4 in the hexagon can then be approximated by \(1/\omega_c\) and \(1/E_F\), respectively. By also approximating the boosting factors by their maximum values \(A_e^{RR}(\bm{0})\) and the remaining magnon energies by $\omega_c$, we may then approximate the vertex correction by

\begin{align}
\begin{aligned}
\Gamma_{1}(k, q) = - \frac{4N_F [V^2 A^{RR}_{e}(\bm{0})]^2}{E_F} \!\sum_{\pmb{\pi}} 
\!\Bigg(\frac{1}{ ik_n - \xi_{\bm{k} - \pmb{\pi}}}\Bigg)
\\
\times\!\Bigg(\frac{1}{ i(k_n - q_m)  - \xi_{\bm{k} - \pmb{\pi} - \bm{q}}}\Bigg).
\end{aligned}
\label{eq:Gamma_step1}
\end{align}
Alternatively, one can attempt to further restrict the sum over $\bm{q}'$ in order to keep all the fermions close to the Fermi surface, producing a similar result as in Eq. \eqref{eq:Gamma_step1}.\\
\indent Introducing \(\bm{p} = \bm{k} - \pmb{\pi}\), the diagram can now be calculated by Taylor expanding in small \(\bm{q}\), using

\begin{align}
\xi_{\bm{p} - \bm{q}} \approx \xi_{\bm{p}} - (\nabla \xi_{\bm{p}})\cdot \bm{q} = \xi_{\bm{p}} - v_F q \cos\big(\theta_{\bm{q}}(\bm{p})\big),
\end{align}

\noindent where \(v_F\) is the Fermi velocity and \(\theta_{\bm{q}}(\bm{p})\) is the angle between \(\nabla \xi_{\bm{p}}\) and \(\bm{q}\). Writing the momentum integral in terms of polar coordinates and integrating out the radial momentum, we then obtain 

\begin{align}
\begin{aligned}
&\Gamma_1(k,q) \approx - i \frac{4 [N_F V^2 A^{RR}_{e}(\bm{0})]^2}{E_F} 
\\
&\!\times \Bigg[\Theta(k_n) - \Theta(k_n - q_m) \Bigg]
\int^{2\pi}_{0} \!\textrm{d}\theta\, \Bigg(\frac{1}{iq_m - v_F q\cos(\theta)}\Bigg)\,.
\end{aligned}
\end{align}

\noindent Performing also the angular integration, one may show that the vertex correction contribution for nonzero bosonic Matsubara frequency is of order

\begin{align}
\begin{aligned}
    &\Gamma_{1}(k, q) \sim \Bigg(\frac{ N_F V^2 A^{RR}_e(0)}{E_F}\Bigg)^2 \!\Bigg(\frac{E_F}{\sqrt{v^2_F q^2 + q^2_m}}\Bigg).
\end{aligned}
\label{eq:gamma_est}
\end{align}

In short, this result can be interpreted as follows: Dominant contributions to Eq.\! \eqref{eq_vertexCorrectionExpression} arise from \(N_F \omega_c\) terms in each of the momentum sums where two of the electron propagators then are of order \(1/\omega_c\) as these electrons are close to the Fermi surface. One of the electron propagators is replaced by a factor \(1/E_F\), as the electron in this case is not close to the Fermi surface.  The last propagator momentum is reasonably close to the Fermi surface due to the small momentum scattering \(\bm{q}\). This propagator is found to be of order \(1/\sqrt{(v_F q)^2 + q_m^2}\), where the square root can be interpreted as an interpolation between the frequency and the momentum energy scales for the scattering process with three-momentum \((\bm{q}, q_m)\).\\
\indent For $\bm{q} \rightarrow 0$, $q_m \sim 1\,\textrm{K}$, and typical values for the remaining energy scales, the expression in Eq.\! \eqref{eq:gamma_est} is found to be of order $1$, indicating that vertex corrections could become important for long-wavelength magnons. As $v_F q$ in the denominator grows quickly with $q$, the momentum region where our estimate for the vertex corrections is of importance is quite limited. Whereas the above expression is quickly reduced when $q$ surpasses $q_m/v_F $, the corresponding momentum cutoff for the magnon propagator depends on the magnon group velocity close to the bottom of the band, meaning that the momentum region where the estimated vertex corrections are of importance is typically significantly smaller than the momentum region where we obtain large contributions to the Eliashberg equations. A more rigorous treatment of the vertex corrections would treat both momentum sums in detail and could potentially give rise to contributions that are larger and/or less quickly reduced with increasing $\bm{q}$.\\
\indent As the diagram in Fig.\! \ref{fig:ho_vertex} (a) vanishes for our model, one might imagine that it could be possible to obtain significant vertex corrections by going to higher-order in magnon operators in the electron-magnon interaction, giving rise to electron-magnon scattering processes without spin flips. Including higher-order terms in the interaction Hamiltonian arising from the $z$-component of the antiferromagnetic spins, one may construct a diagram like the one in Fig.\! \ref{fig:ho_vertex} (a) where the vertical magnon line has been replaced with a magnon loop and the vertices of the magnon loop does not involve an electron spin flip, conserving the electron spin in the diagram. Performing estimates like those presented above, such diagrams are found to be of similar magnitude and displaying a similar suppression with increasing momentum $\bm{q}$ as the diagram in Fig.\! \ref{fig:ho_vertex} (b).\\
\indent For larger Fermi surfaces and $\Omega = 1$, the regular processes are of little importance and the physics is dominated by Umklapp processes. Modifying the diagram of Fig.\! \ref{fig:ho_vertex} (b) to only include Umklapp processes, all spin-$\downarrow$ electron propagators therefore attain an additional momentum shift $\bm{Q}$. Below half-filling, for $\bm{k}$ on the Fermi surface, placing $\bm{k} - \bm{q} + \bm{Q}$ on the Fermi surface now requires a finite momentum $\bm{q}$.
Contrary to the case with regular processes and \(\bm{q} \rightarrow 0\), choosing the hexagon vertex $2$ reasonably close to the Fermi surface does therefore not necessarily mean that the hexagon vertex $3$ is also reasonably close. \\
\indent Exactly at half-filling, the Fermi surface is perfectly nested, and the electron momenta can all be chosen reasonably close to the Fermi surface for a wide range of integration momenta and relevant values of $\bm{q}$. Thus, we would get large vertex corrections~\cite{Virosztek1990, Virosztek1999, Djajaputra1999}. Moving away from half-filling, the nesting of the Fermi surface is no longer perfect. Our simplest estimate in Eq.~\eqref{eq_vertexCorrection_simplestEstimate} will then eventually be restored, where \(A_e^{RR}(\bm{0})\) needs to be replaced with the maximum value of \(A_e^{UU}(\bm{q})\) for scattering processes on the Fermi surface. Thus, we would expect vertex corrections to become unimportant sufficiently far away from half-filling.

\vspace{0.8cm}

\section{Material parameters}
\label{Section:Parameters}

Typical parameter values are shown in Table~\ref{tab_parameterValues}. The electron density of states is calculated by numerical evaluation of the elliptical integral in Ref.\! \cite{Piasecki2008}.

\newpage

\begin{table}[htb]
\begin{center}
\caption{Parameter values used in the numerical results. We refer to the main text for an explanation of their meanings.}
\label{tab_parameterValues}
\begin{tabular}{cl}\\
\hline
Quantity & Value \\
\hline
\(J_1\) & \SI{2}{\milli\electronvolt} \\
\(J_2\) & \(0.2 \, J_1\) \\
\(K\) & \num{1e-4} \(J_1\) \\
\(\bar{J}\) & \SI{15}{\milli\electronvolt} \\
\(S\) & \(1\) \\
\( t\) & \SI{1}{\electronvolt} \\
\hline
\end{tabular}
\end{center}
\end{table}

\FloatBarrier

\bibliographystyle{apsrev4-1}  
                               
\addcontentsline{toc}{chapter}{\bibname}
\bibliography{Refs}  

\begin{thebibliography}{61}%
\makeatletter
\providecommand \@ifxundefined [1]{%
 \@ifx{#1\undefined}
}%
\providecommand \@ifnum [1]{%
 \ifnum #1\expandafter \@firstoftwo
 \else \expandafter \@secondoftwo
 \fi
}%
\providecommand \@ifx [1]{%
 \ifx #1\expandafter \@firstoftwo
 \else \expandafter \@secondoftwo
 \fi
}%
\providecommand \natexlab [1]{#1}%
\providecommand \enquote  [1]{``#1''}%
\providecommand \bibnamefont  [1]{#1}%
\providecommand \bibfnamefont [1]{#1}%
\providecommand \citenamefont [1]{#1}%
\providecommand \href@noop [0]{\@secondoftwo}%
\providecommand \href [0]{\begingroup \@sanitize@url \@href}%
\providecommand \@href[1]{\@@startlink{#1}\@@href}%
\providecommand \@@href[1]{\endgroup#1\@@endlink}%
\providecommand \@sanitize@url [0]{\catcode `\\12\catcode `\$12\catcode
  `\&12\catcode `\#12\catcode `\^12\catcode `\_12\catcode `\%12\relax}%
\providecommand \@@startlink[1]{}%
\providecommand \@@endlink[0]{}%
\providecommand \url  [0]{\begingroup\@sanitize@url \@url }%
\providecommand \@url [1]{\endgroup\@href {#1}{\urlprefix }}%
\providecommand \urlprefix  [0]{URL }%
\providecommand \Eprint [0]{\href }%
\providecommand \doibase [0]{http://dx.doi.org/}%
\providecommand \selectlanguage [0]{\@gobble}%
\providecommand \bibinfo  [0]{\@secondoftwo}%
\providecommand \bibfield  [0]{\@secondoftwo}%
\providecommand \translation [1]{[#1]}%
\providecommand \BibitemOpen [0]{}%
\providecommand \bibitemStop [0]{}%
\providecommand \bibitemNoStop [0]{.\EOS\space}%
\providecommand \EOS [0]{\spacefactor3000\relax}%
\providecommand \BibitemShut  [1]{\csname bibitem#1\endcsname}%
\let\auto@bib@innerbib\@empty
\bibitem [{\citenamefont {Bardeen}\ \emph {et~al.}(1957)\citenamefont
  {Bardeen}, \citenamefont {Cooper},\ and\ \citenamefont
  {Schrieffer}}]{Bardeen1957}%
  \BibitemOpen
  \bibfield  {author} {\bibinfo {author} {\bibfnamefont {J.}~\bibnamefont
  {Bardeen}}, \bibinfo {author} {\bibfnamefont {L.~N.}\ \bibnamefont {Cooper}},
  \ and\ \bibinfo {author} {\bibfnamefont {J.~R.}\ \bibnamefont {Schrieffer}},\
  }\href {\doibase 10.1103/PhysRev.108.1175} {\bibfield  {journal} {\bibinfo
  {journal} {Phys. Rev.}\ }\textbf {\bibinfo {volume} {108}},\ \bibinfo {pages}
  {1175} (\bibinfo {year} {1957})}\BibitemShut {NoStop}%
\bibitem [{\citenamefont {Schlawin}\ \emph {et~al.}(2019)\citenamefont
  {Schlawin}, \citenamefont {Cavalleri},\ and\ \citenamefont
  {Jaksch}}]{Schlawin2019}%
  \BibitemOpen
  \bibfield  {author} {\bibinfo {author} {\bibfnamefont {F.}~\bibnamefont
  {Schlawin}}, \bibinfo {author} {\bibfnamefont {A.}~\bibnamefont {Cavalleri}},
  \ and\ \bibinfo {author} {\bibfnamefont {D.}~\bibnamefont {Jaksch}},\ }\href
  {\doibase 10.1103/PhysRevLett.122.133602} {\bibfield  {journal} {\bibinfo
  {journal} {Phys. Rev. Lett.}\ }\textbf {\bibinfo {volume} {122}},\ \bibinfo
  {pages} {133602} (\bibinfo {year} {2019})}\BibitemShut {NoStop}%
\bibitem [{\citenamefont {Kavokin}\ and\ \citenamefont
  {Lagoudakis}(2016)}]{Kavokin2016}%
  \BibitemOpen
  \bibfield  {author} {\bibinfo {author} {\bibfnamefont {A.}~\bibnamefont
  {Kavokin}}\ and\ \bibinfo {author} {\bibfnamefont {P.}~\bibnamefont
  {Lagoudakis}},\ }\href {\doibase 10.1038/nmat4646} {\bibfield  {journal}
  {\bibinfo  {journal} {Nature Materials}\ }\textbf {\bibinfo {volume} {15}},\
  \bibinfo {pages} {599} (\bibinfo {year} {2016})}\BibitemShut {NoStop}%
\bibitem [{\citenamefont {Laussy}\ \emph {et~al.}(2010)\citenamefont {Laussy},
  \citenamefont {Kavokin},\ and\ \citenamefont {Shelykh}}]{Laussy2010}%
  \BibitemOpen
  \bibfield  {author} {\bibinfo {author} {\bibfnamefont {F.~P.}\ \bibnamefont
  {Laussy}}, \bibinfo {author} {\bibfnamefont {A.~V.}\ \bibnamefont {Kavokin}},
  \ and\ \bibinfo {author} {\bibfnamefont {I.~A.}\ \bibnamefont {Shelykh}},\
  }\href {\doibase 10.1103/PhysRevLett.104.106402} {\bibfield  {journal}
  {\bibinfo  {journal} {Phys. Rev. Lett.}\ }\textbf {\bibinfo {volume} {104}},\
  \bibinfo {pages} {106402} (\bibinfo {year} {2010})}\BibitemShut {NoStop}%
\bibitem [{\citenamefont {Takada}(1978)}]{Takada1978}%
  \BibitemOpen
  \bibfield  {author} {\bibinfo {author} {\bibfnamefont {Y.}~\bibnamefont
  {Takada}},\ }\href {\doibase 10.1143/JPSJ.45.786} {\bibfield  {journal}
  {\bibinfo  {journal} {Journal of the Physical Society of Japan}\ }\textbf
  {\bibinfo {volume} {45}},\ \bibinfo {pages} {786} (\bibinfo {year}
  {1978})}\BibitemShut {NoStop}%
\bibitem [{\citenamefont {Scalapino}(1999)}]{Scalapino1999}%
  \BibitemOpen
  \bibfield  {author} {\bibinfo {author} {\bibfnamefont {D.~J.}\ \bibnamefont
  {Scalapino}},\ }\href {\doibase 10.1023/A:1022559920049} {\bibfield
  {journal} {\bibinfo  {journal} {Journal of Low Temperature Physics}\ }\textbf
  {\bibinfo {volume} {117}},\ \bibinfo {pages} {179} (\bibinfo {year}
  {1999})}\BibitemShut {NoStop}%
\bibitem [{\citenamefont {Moriya}\ and\ \citenamefont
  {Ueda}(2003)}]{Moriya2003}%
  \BibitemOpen
  \bibfield  {author} {\bibinfo {author} {\bibfnamefont {T.}~\bibnamefont
  {Moriya}}\ and\ \bibinfo {author} {\bibfnamefont {K.}~\bibnamefont {Ueda}},\
  }\href {\doibase 10.1088/0034-4885/66/8/202} {\bibfield  {journal} {\bibinfo
  {journal} {Reports on Progress in Physics}\ }\textbf {\bibinfo {volume}
  {66}},\ \bibinfo {pages} {1299} (\bibinfo {year} {2003})}\BibitemShut
  {NoStop}%
\bibitem [{\citenamefont {Berk}\ and\ \citenamefont
  {Schrieffer}(1966)}]{Berk1966}%
  \BibitemOpen
  \bibfield  {author} {\bibinfo {author} {\bibfnamefont {N.~F.}\ \bibnamefont
  {Berk}}\ and\ \bibinfo {author} {\bibfnamefont {J.~R.}\ \bibnamefont
  {Schrieffer}},\ }\href {\doibase 10.1103/PhysRevLett.17.433} {\bibfield
  {journal} {\bibinfo  {journal} {Phys. Rev. Lett.}\ }\textbf {\bibinfo
  {volume} {17}},\ \bibinfo {pages} {433} (\bibinfo {year} {1966})}\BibitemShut
  {NoStop}%
\bibitem [{\citenamefont {Doniach}\ and\ \citenamefont
  {Engelsberg}(1966)}]{Doniach1966}%
  \BibitemOpen
  \bibfield  {author} {\bibinfo {author} {\bibfnamefont {S.}~\bibnamefont
  {Doniach}}\ and\ \bibinfo {author} {\bibfnamefont {S.}~\bibnamefont
  {Engelsberg}},\ }\href {\doibase 10.1103/PhysRevLett.17.750} {\bibfield
  {journal} {\bibinfo  {journal} {Phys. Rev. Lett.}\ }\textbf {\bibinfo
  {volume} {17}},\ \bibinfo {pages} {750} (\bibinfo {year} {1966})}\BibitemShut
  {NoStop}%
\bibitem [{\citenamefont {Cyrot}(1986)}]{cyrot1986}%
  \BibitemOpen
  \bibfield  {author} {\bibinfo {author} {\bibfnamefont {M.}~\bibnamefont
  {Cyrot}},\ }\href {\doibase https://doi.org/10.1016/0038-1098(86)90458-8}
  {\bibfield  {journal} {\bibinfo  {journal} {Solid State Communications}\
  }\textbf {\bibinfo {volume} {60}},\ \bibinfo {pages} {253} (\bibinfo {year}
  {1986})}\BibitemShut {NoStop}%
\bibitem [{\citenamefont {Scalapino}\ \emph {et~al.}(1986)\citenamefont
  {Scalapino}, \citenamefont {Loh},\ and\ \citenamefont
  {Hirsch}}]{Scalapino1986}%
  \BibitemOpen
  \bibfield  {author} {\bibinfo {author} {\bibfnamefont {D.~J.}\ \bibnamefont
  {Scalapino}}, \bibinfo {author} {\bibfnamefont {E.}~\bibnamefont {Loh}}, \
  and\ \bibinfo {author} {\bibfnamefont {J.~E.}\ \bibnamefont {Hirsch}},\
  }\href {\doibase 10.1103/PhysRevB.34.8190} {\bibfield  {journal} {\bibinfo
  {journal} {Phys. Rev. B}\ }\textbf {\bibinfo {volume} {34}},\ \bibinfo
  {pages} {8190} (\bibinfo {year} {1986})}\BibitemShut {NoStop}%
\bibitem [{\citenamefont {Miyake}\ \emph {et~al.}(1986)\citenamefont {Miyake},
  \citenamefont {Schmitt-Rink},\ and\ \citenamefont {Varma}}]{Miyake1986}%
  \BibitemOpen
  \bibfield  {author} {\bibinfo {author} {\bibfnamefont {K.}~\bibnamefont
  {Miyake}}, \bibinfo {author} {\bibfnamefont {S.}~\bibnamefont
  {Schmitt-Rink}}, \ and\ \bibinfo {author} {\bibfnamefont {C.~M.}\
  \bibnamefont {Varma}},\ }\href {\doibase 10.1103/PhysRevB.34.6554} {\bibfield
   {journal} {\bibinfo  {journal} {Phys. Rev. B}\ }\textbf {\bibinfo {volume}
  {34}},\ \bibinfo {pages} {6554} (\bibinfo {year} {1986})}\BibitemShut
  {NoStop}%
\bibitem [{\citenamefont {Monthoux}\ \emph {et~al.}(1991)\citenamefont
  {Monthoux}, \citenamefont {Balatsky},\ and\ \citenamefont
  {Pines}}]{Monthoux1991}%
  \BibitemOpen
  \bibfield  {author} {\bibinfo {author} {\bibfnamefont {P.}~\bibnamefont
  {Monthoux}}, \bibinfo {author} {\bibfnamefont {A.~V.}\ \bibnamefont
  {Balatsky}}, \ and\ \bibinfo {author} {\bibfnamefont {D.}~\bibnamefont
  {Pines}},\ }\href {\doibase 10.1103/PhysRevLett.67.3448} {\bibfield
  {journal} {\bibinfo  {journal} {Phys. Rev. Lett.}\ }\textbf {\bibinfo
  {volume} {67}},\ \bibinfo {pages} {3448} (\bibinfo {year}
  {1991})}\BibitemShut {NoStop}%
\bibitem [{\citenamefont {Monthoux}\ and\ \citenamefont
  {Pines}(1992)}]{Monthoux1992}%
  \BibitemOpen
  \bibfield  {author} {\bibinfo {author} {\bibfnamefont {P.}~\bibnamefont
  {Monthoux}}\ and\ \bibinfo {author} {\bibfnamefont {D.}~\bibnamefont
  {Pines}},\ }\href {\doibase 10.1103/PhysRevLett.69.961} {\bibfield  {journal}
  {\bibinfo  {journal} {Phys. Rev. Lett.}\ }\textbf {\bibinfo {volume} {69}},\
  \bibinfo {pages} {961} (\bibinfo {year} {1992})}\BibitemShut {NoStop}%
\bibitem [{\citenamefont {Abanov}\ \emph {et~al.}(2003)\citenamefont {Abanov},
  \citenamefont {Chubukov},\ and\ \citenamefont {Schmalian}}]{Abanov2003}%
  \BibitemOpen
  \bibfield  {author} {\bibinfo {author} {\bibfnamefont {A.}~\bibnamefont
  {Abanov}}, \bibinfo {author} {\bibfnamefont {A.~V.}\ \bibnamefont
  {Chubukov}}, \ and\ \bibinfo {author} {\bibfnamefont {J.}~\bibnamefont
  {Schmalian}},\ }\href {\doibase 10.1080/0001873021000057123} {\bibfield
  {journal} {\bibinfo  {journal} {Advances in Physics}\ }\textbf {\bibinfo
  {volume} {52}},\ \bibinfo {pages} {119} (\bibinfo {year} {2003})}\BibitemShut
  {NoStop}%
\bibitem [{\citenamefont {Kargarian}\ \emph {et~al.}(2016)\citenamefont
  {Kargarian}, \citenamefont {Efimkin},\ and\ \citenamefont
  {Galitski}}]{Kargarian2016}%
  \BibitemOpen
  \bibfield  {author} {\bibinfo {author} {\bibfnamefont {M.}~\bibnamefont
  {Kargarian}}, \bibinfo {author} {\bibfnamefont {D.~K.}\ \bibnamefont
  {Efimkin}}, \ and\ \bibinfo {author} {\bibfnamefont {V.}~\bibnamefont
  {Galitski}},\ }\href {\doibase 10.1103/PhysRevLett.117.076806} {\bibfield
  {journal} {\bibinfo  {journal} {Phys. Rev. Lett.}\ }\textbf {\bibinfo
  {volume} {117}},\ \bibinfo {pages} {076806} (\bibinfo {year}
  {2016})}\BibitemShut {NoStop}%
\bibitem [{\citenamefont {Gong}\ \emph {et~al.}(2017)\citenamefont {Gong},
  \citenamefont {Kargarian}, \citenamefont {Stern}, \citenamefont {Yue},
  \citenamefont {Zhou}, \citenamefont {Jin}, \citenamefont {Galitski},
  \citenamefont {Yakovenko},\ and\ \citenamefont {Xia}}]{Gong2017}%
  \BibitemOpen
  \bibfield  {author} {\bibinfo {author} {\bibfnamefont {X.}~\bibnamefont
  {Gong}}, \bibinfo {author} {\bibfnamefont {M.}~\bibnamefont {Kargarian}},
  \bibinfo {author} {\bibfnamefont {A.}~\bibnamefont {Stern}}, \bibinfo
  {author} {\bibfnamefont {D.}~\bibnamefont {Yue}}, \bibinfo {author}
  {\bibfnamefont {H.}~\bibnamefont {Zhou}}, \bibinfo {author} {\bibfnamefont
  {X.}~\bibnamefont {Jin}}, \bibinfo {author} {\bibfnamefont {V.~M.}\
  \bibnamefont {Galitski}}, \bibinfo {author} {\bibfnamefont {V.~M.}\
  \bibnamefont {Yakovenko}}, \ and\ \bibinfo {author} {\bibfnamefont
  {J.}~\bibnamefont {Xia}},\ }\href
  {https://advances.sciencemag.org/content/3/3/e1602579} {\bibfield  {journal}
  {\bibinfo  {journal} {Science Advances}\ }\textbf {\bibinfo {volume} {3}}
  (\bibinfo {year} {2017})}\BibitemShut {NoStop}%
\bibitem [{\citenamefont {Rohling}\ \emph {et~al.}(2018)\citenamefont
  {Rohling}, \citenamefont {Fj\ae{}rbu},\ and\ \citenamefont
  {Brataas}}]{Fjaerbu2018}%
  \BibitemOpen
  \bibfield  {author} {\bibinfo {author} {\bibfnamefont {N.}~\bibnamefont
  {Rohling}}, \bibinfo {author} {\bibfnamefont {E.~L.}\ \bibnamefont
  {Fj\ae{}rbu}}, \ and\ \bibinfo {author} {\bibfnamefont {A.}~\bibnamefont
  {Brataas}},\ }\href {\doibase 10.1103/PhysRevB.97.115401} {\bibfield
  {journal} {\bibinfo  {journal} {Phys. Rev. B}\ }\textbf {\bibinfo {volume}
  {97}},\ \bibinfo {pages} {115401} (\bibinfo {year} {2018})}\BibitemShut
  {NoStop}%
\bibitem [{\citenamefont {Hugdal}\ \emph {et~al.}(2018)\citenamefont {Hugdal},
  \citenamefont {Rex}, \citenamefont {Nogueira},\ and\ \citenamefont
  {Sudb\o{}}}]{Hugdal2018}%
  \BibitemOpen
  \bibfield  {author} {\bibinfo {author} {\bibfnamefont {H.~G.}\ \bibnamefont
  {Hugdal}}, \bibinfo {author} {\bibfnamefont {S.}~\bibnamefont {Rex}},
  \bibinfo {author} {\bibfnamefont {F.~S.}\ \bibnamefont {Nogueira}}, \ and\
  \bibinfo {author} {\bibfnamefont {A.}~\bibnamefont {Sudb\o{}}},\ }\href
  {\doibase 10.1103/PhysRevB.97.195438} {\bibfield  {journal} {\bibinfo
  {journal} {Phys. Rev. B}\ }\textbf {\bibinfo {volume} {97}},\ \bibinfo
  {pages} {195438} (\bibinfo {year} {2018})}\BibitemShut {NoStop}%
\bibitem [{\citenamefont {Fj\ae{}rbu}\ \emph {et~al.}(2019)\citenamefont
  {Fj\ae{}rbu}, \citenamefont {Rohling},\ and\ \citenamefont
  {Brataas}}]{Fjaerbu2019}%
  \BibitemOpen
  \bibfield  {author} {\bibinfo {author} {\bibfnamefont {E.~L.}\ \bibnamefont
  {Fj\ae{}rbu}}, \bibinfo {author} {\bibfnamefont {N.}~\bibnamefont {Rohling}},
  \ and\ \bibinfo {author} {\bibfnamefont {A.}~\bibnamefont {Brataas}},\ }\href
  {\doibase 10.1103/PhysRevB.100.125432} {\bibfield  {journal} {\bibinfo
  {journal} {Phys. Rev. B}\ }\textbf {\bibinfo {volume} {100}},\ \bibinfo
  {pages} {125432} (\bibinfo {year} {2019})}\BibitemShut {NoStop}%
\bibitem [{\citenamefont {Erlandsen}\ \emph {et~al.}(2019)\citenamefont
  {Erlandsen}, \citenamefont {Kamra}, \citenamefont {Brataas},\ and\
  \citenamefont {Sudb\o{}}}]{Erlandsen2019_Enhancement}%
  \BibitemOpen
  \bibfield  {author} {\bibinfo {author} {\bibfnamefont {E.}~\bibnamefont
  {Erlandsen}}, \bibinfo {author} {\bibfnamefont {A.}~\bibnamefont {Kamra}},
  \bibinfo {author} {\bibfnamefont {A.}~\bibnamefont {Brataas}}, \ and\
  \bibinfo {author} {\bibfnamefont {A.}~\bibnamefont {Sudb\o{}}},\ }\href
  {\doibase 10.1103/PhysRevB.100.100503} {\bibfield  {journal} {\bibinfo
  {journal} {Phys. Rev. B}\ }\textbf {\bibinfo {volume} {100}},\ \bibinfo
  {pages} {100503} (\bibinfo {year} {2019})}\BibitemShut {NoStop}%
\bibitem [{\citenamefont {Erlandsen}\ and\ \citenamefont
  {Sudb\o{}}(2020)}]{Erlandsen2020_Schwinger}%
  \BibitemOpen
  \bibfield  {author} {\bibinfo {author} {\bibfnamefont {E.}~\bibnamefont
  {Erlandsen}}\ and\ \bibinfo {author} {\bibfnamefont {A.}~\bibnamefont
  {Sudb\o{}}},\ }\href {\doibase 10.1103/PhysRevB.102.214502} {\bibfield
  {journal} {\bibinfo  {journal} {Phys. Rev. B}\ }\textbf {\bibinfo {volume}
  {102}},\ \bibinfo {pages} {214502} (\bibinfo {year} {2020})}\BibitemShut
  {NoStop}%
\bibitem [{\citenamefont {Erlandsen}\ \emph {et~al.}(2020)\citenamefont
  {Erlandsen}, \citenamefont {Brataas},\ and\ \citenamefont
  {Sudb\o{}}}]{Erlandsen2020_TI}%
  \BibitemOpen
  \bibfield  {author} {\bibinfo {author} {\bibfnamefont {E.}~\bibnamefont
  {Erlandsen}}, \bibinfo {author} {\bibfnamefont {A.}~\bibnamefont {Brataas}},
  \ and\ \bibinfo {author} {\bibfnamefont {A.}~\bibnamefont {Sudb\o{}}},\
  }\href {\doibase 10.1103/PhysRevB.101.094503} {\bibfield  {journal} {\bibinfo
   {journal} {Phys. Rev. B}\ }\textbf {\bibinfo {volume} {101}},\ \bibinfo
  {pages} {094503} (\bibinfo {year} {2020})}\BibitemShut {NoStop}%
\bibitem [{\citenamefont {Hugdal}\ and\ \citenamefont
  {Sudb\o{}}(2020)}]{Hugdal2020}%
  \BibitemOpen
  \bibfield  {author} {\bibinfo {author} {\bibfnamefont {H.~G.}\ \bibnamefont
  {Hugdal}}\ and\ \bibinfo {author} {\bibfnamefont {A.}~\bibnamefont
  {Sudb\o{}}},\ }\href {\doibase 10.1103/PhysRevB.102.125429} {\bibfield
  {journal} {\bibinfo  {journal} {Phys. Rev. B}\ }\textbf {\bibinfo {volume}
  {102}},\ \bibinfo {pages} {125429} (\bibinfo {year} {2020})}\BibitemShut
  {NoStop}%
\bibitem [{\citenamefont {Kamra}\ \emph {et~al.}(2019)\citenamefont {Kamra},
  \citenamefont {Thingstad}, \citenamefont {Rastelli}, \citenamefont {Duine},
  \citenamefont {Brataas}, \citenamefont {Belzig},\ and\ \citenamefont
  {Sudb\o{}}}]{Kamra2019_Antiferromagnetic}%
  \BibitemOpen
  \bibfield  {author} {\bibinfo {author} {\bibfnamefont {A.}~\bibnamefont
  {Kamra}}, \bibinfo {author} {\bibfnamefont {E.}~\bibnamefont {Thingstad}},
  \bibinfo {author} {\bibfnamefont {G.}~\bibnamefont {Rastelli}}, \bibinfo
  {author} {\bibfnamefont {R.~A.}\ \bibnamefont {Duine}}, \bibinfo {author}
  {\bibfnamefont {A.}~\bibnamefont {Brataas}}, \bibinfo {author} {\bibfnamefont
  {W.}~\bibnamefont {Belzig}}, \ and\ \bibinfo {author} {\bibfnamefont
  {A.}~\bibnamefont {Sudb\o{}}},\ }\href {\doibase 10.1103/PhysRevB.100.174407}
  {\bibfield  {journal} {\bibinfo  {journal} {Phys. Rev. B}\ }\textbf {\bibinfo
  {volume} {100}},\ \bibinfo {pages} {174407} (\bibinfo {year}
  {2019})}\BibitemShut {NoStop}%
\bibitem [{\citenamefont {Nogués}\ and\ \citenamefont
  {Schuller}(1999)}]{Nogues1999}%
  \BibitemOpen
  \bibfield  {author} {\bibinfo {author} {\bibfnamefont {J.}~\bibnamefont
  {Nogués}}\ and\ \bibinfo {author} {\bibfnamefont {I.~K.}\ \bibnamefont
  {Schuller}},\ }\href {\doibase 10.1016/S0304-8853(98)00266-2} {\bibfield
  {journal} {\bibinfo  {journal} {Journal of Magnetism and Magnetic Materials}\
  }\textbf {\bibinfo {volume} {192}},\ \bibinfo {pages} {203 } (\bibinfo {year}
  {1999})}\BibitemShut {NoStop}%
\bibitem [{\citenamefont {Nogués}\ \emph {et~al.}(2005)\citenamefont
  {Nogués}, \citenamefont {Sort}, \citenamefont {Langlais}, \citenamefont
  {Skumryev}, \citenamefont {Suriñach}, \citenamefont {Muñoz},\ and\
  \citenamefont {Baró}}]{Nogues2005}%
  \BibitemOpen
  \bibfield  {author} {\bibinfo {author} {\bibfnamefont {J.}~\bibnamefont
  {Nogués}}, \bibinfo {author} {\bibfnamefont {J.}~\bibnamefont {Sort}},
  \bibinfo {author} {\bibfnamefont {V.}~\bibnamefont {Langlais}}, \bibinfo
  {author} {\bibfnamefont {V.}~\bibnamefont {Skumryev}}, \bibinfo {author}
  {\bibfnamefont {S.}~\bibnamefont {Suriñach}}, \bibinfo {author}
  {\bibfnamefont {J.}~\bibnamefont {Muñoz}}, \ and\ \bibinfo {author}
  {\bibfnamefont {M.}~\bibnamefont {Baró}},\ }\href {\doibase
  10.1016/j.physrep.2005.08.004} {\bibfield  {journal} {\bibinfo  {journal}
  {Physics Reports}\ }\textbf {\bibinfo {volume} {422}},\ \bibinfo {pages} {65
  } (\bibinfo {year} {2005})}\BibitemShut {NoStop}%
\bibitem [{\citenamefont {Stamps}(2000)}]{Stamps2000}%
  \BibitemOpen
  \bibfield  {author} {\bibinfo {author} {\bibfnamefont {R.~L.}\ \bibnamefont
  {Stamps}},\ }\href {http://stacks.iop.org/0022-3727/33/i=23/a=201} {\bibfield
   {journal} {\bibinfo  {journal} {Journal of Physics D: Applied Physics}\
  }\textbf {\bibinfo {volume} {33}},\ \bibinfo {pages} {R247} (\bibinfo {year}
  {2000})}\BibitemShut {NoStop}%
\bibitem [{\citenamefont {Takei}\ \emph {et~al.}(2014)\citenamefont {Takei},
  \citenamefont {Halperin}, \citenamefont {Yacoby},\ and\ \citenamefont
  {Tserkovnyak}}]{Takei2014}%
  \BibitemOpen
  \bibfield  {author} {\bibinfo {author} {\bibfnamefont {S.}~\bibnamefont
  {Takei}}, \bibinfo {author} {\bibfnamefont {B.~I.}\ \bibnamefont {Halperin}},
  \bibinfo {author} {\bibfnamefont {A.}~\bibnamefont {Yacoby}}, \ and\ \bibinfo
  {author} {\bibfnamefont {Y.}~\bibnamefont {Tserkovnyak}},\ }\href {\doibase
  10.1103/PhysRevB.90.094408} {\bibfield  {journal} {\bibinfo  {journal} {Phys.
  Rev. B}\ }\textbf {\bibinfo {volume} {90}},\ \bibinfo {pages} {094408}
  (\bibinfo {year} {2014})}\BibitemShut {NoStop}%
\bibitem [{\citenamefont {Fj\ae{}rbu}\ \emph {et~al.}(2017)\citenamefont
  {Fj\ae{}rbu}, \citenamefont {Rohling},\ and\ \citenamefont
  {Brataas}}]{Fjaerbu2017}%
  \BibitemOpen
  \bibfield  {author} {\bibinfo {author} {\bibfnamefont {E.~L.}\ \bibnamefont
  {Fj\ae{}rbu}}, \bibinfo {author} {\bibfnamefont {N.}~\bibnamefont {Rohling}},
  \ and\ \bibinfo {author} {\bibfnamefont {A.}~\bibnamefont {Brataas}},\ }\href
  {\doibase 10.1103/PhysRevB.95.144408} {\bibfield  {journal} {\bibinfo
  {journal} {Phys. Rev. B}\ }\textbf {\bibinfo {volume} {95}},\ \bibinfo
  {pages} {144408} (\bibinfo {year} {2017})}\BibitemShut {NoStop}%
\bibitem [{\citenamefont {Vonsovski{\u{\i}}}\ and\ \citenamefont
  {Izyumov}(1963)}]{Vonsovskii1963}%
  \BibitemOpen
  \bibfield  {author} {\bibinfo {author} {\bibfnamefont {S.~V.}\ \bibnamefont
  {Vonsovski{\u{\i}}}}\ and\ \bibinfo {author} {\bibfnamefont {Y.~A.}\
  \bibnamefont {Izyumov}},\ }\href {\doibase 10.1070/pu1963v005n05abeh003452}
  {\bibfield  {journal} {\bibinfo  {journal} {Soviet Physics Uspekhi}\ }\textbf
  {\bibinfo {volume} {5}},\ \bibinfo {pages} {723} (\bibinfo {year}
  {1963})}\BibitemShut {NoStop}%
\bibitem [{\citenamefont {Chen}\ and\ \citenamefont
  {Goddard}(1988)}]{Chen1988}%
  \BibitemOpen
  \bibfield  {author} {\bibinfo {author} {\bibfnamefont {G.}~\bibnamefont
  {Chen}}\ and\ \bibinfo {author} {\bibfnamefont {W.~A.}\ \bibnamefont
  {Goddard}},\ }\href {http://www.jstor.org/stable/1700317} {\bibfield
  {journal} {\bibinfo  {journal} {Science}\ }\textbf {\bibinfo {volume}
  {239}},\ \bibinfo {pages} {899} (\bibinfo {year} {1988})}\BibitemShut
  {NoStop}%
\bibitem [{\citenamefont {Shimahara}(1994)}]{Shimahara1994}%
  \BibitemOpen
  \bibfield  {author} {\bibinfo {author} {\bibfnamefont {H.}~\bibnamefont
  {Shimahara}},\ }\href {\doibase 10.1143/JPSJ.63.1861} {\bibfield  {journal}
  {\bibinfo  {journal} {Journal of the Physical Society of Japan}\ }\textbf
  {\bibinfo {volume} {63}},\ \bibinfo {pages} {1861} (\bibinfo {year}
  {1994})}\BibitemShut {NoStop}%
\bibitem [{\citenamefont {Kajiwara}\ \emph {et~al.}(2010)\citenamefont
  {Kajiwara}, \citenamefont {Harii}, \citenamefont {Takahashi}, \citenamefont
  {Ohe}, \citenamefont {Uchida}, \citenamefont {Mizuguchi}, \citenamefont
  {Umezawa}, \citenamefont {Kawai}, \citenamefont {Ando}, \citenamefont
  {Takanashi}, \citenamefont {Maekawa},\ and\ \citenamefont
  {Saitoh}}]{Kajiwara2010}%
  \BibitemOpen
  \bibfield  {author} {\bibinfo {author} {\bibfnamefont {Y.}~\bibnamefont
  {Kajiwara}}, \bibinfo {author} {\bibfnamefont {K.}~\bibnamefont {Harii}},
  \bibinfo {author} {\bibfnamefont {S.}~\bibnamefont {Takahashi}}, \bibinfo
  {author} {\bibfnamefont {J.}~\bibnamefont {Ohe}}, \bibinfo {author}
  {\bibfnamefont {K.}~\bibnamefont {Uchida}}, \bibinfo {author} {\bibfnamefont
  {M.}~\bibnamefont {Mizuguchi}}, \bibinfo {author} {\bibfnamefont
  {H.}~\bibnamefont {Umezawa}}, \bibinfo {author} {\bibfnamefont
  {H.}~\bibnamefont {Kawai}}, \bibinfo {author} {\bibfnamefont
  {K.}~\bibnamefont {Ando}}, \bibinfo {author} {\bibfnamefont {K.}~\bibnamefont
  {Takanashi}}, \bibinfo {author} {\bibfnamefont {S.}~\bibnamefont {Maekawa}},
  \ and\ \bibinfo {author} {\bibfnamefont {E.}~\bibnamefont {Saitoh}},\ }\href
  {\doibase 10.1038/nature08876} {\bibfield  {journal} {\bibinfo  {journal}
  {Nature}\ }\textbf {\bibinfo {volume} {464}},\ \bibinfo {pages} {262}
  (\bibinfo {year} {2010})}\BibitemShut {NoStop}%
\bibitem [{\citenamefont {Mazzola}\ \emph {et~al.}(2017)\citenamefont
  {Mazzola}, \citenamefont {Frederiksen}, \citenamefont {Balasubramanian},
  \citenamefont {Hofmann}, \citenamefont {Hellsing},\ and\ \citenamefont
  {Wells}}]{Mazzola2017}%
  \BibitemOpen
  \bibfield  {author} {\bibinfo {author} {\bibfnamefont {F.}~\bibnamefont
  {Mazzola}}, \bibinfo {author} {\bibfnamefont {T.}~\bibnamefont
  {Frederiksen}}, \bibinfo {author} {\bibfnamefont {T.}~\bibnamefont
  {Balasubramanian}}, \bibinfo {author} {\bibfnamefont {P.}~\bibnamefont
  {Hofmann}}, \bibinfo {author} {\bibfnamefont {B.}~\bibnamefont {Hellsing}}, \
  and\ \bibinfo {author} {\bibfnamefont {J.~W.}\ \bibnamefont {Wells}},\ }\href
  {\doibase 10.1103/PhysRevB.95.075430} {\bibfield  {journal} {\bibinfo
  {journal} {Phys. Rev. B}\ }\textbf {\bibinfo {volume} {95}},\ \bibinfo
  {pages} {075430} (\bibinfo {year} {2017})}\BibitemShut {NoStop}%
\bibitem [{\citenamefont {Coleman}(2015)}]{Coleman2015}%
  \BibitemOpen
  \bibfield  {author} {\bibinfo {author} {\bibfnamefont {P.}~\bibnamefont
  {Coleman}},\ }\href {\doibase 10.1017/CBO9781139020916} {\emph {\bibinfo
  {title} {Introduction to Many-Body Physics}}}\ (\bibinfo  {publisher}
  {Cambridge University Press},\ \bibinfo {year} {2015})\BibitemShut {NoStop}%
\bibitem [{\citenamefont {Takahashi}\ \emph {et~al.}(2010)\citenamefont
  {Takahashi}, \citenamefont {Saitoh},\ and\ \citenamefont
  {Maekawa}}]{Takahashi2010}%
  \BibitemOpen
  \bibfield  {author} {\bibinfo {author} {\bibfnamefont {S.}~\bibnamefont
  {Takahashi}}, \bibinfo {author} {\bibfnamefont {E.}~\bibnamefont {Saitoh}}, \
  and\ \bibinfo {author} {\bibfnamefont {S.}~\bibnamefont {Maekawa}},\ }\href
  {http://stacks.iop.org/1742-6596/200/i=6/a=062030} {\bibfield  {journal}
  {\bibinfo  {journal} {Journal of Physics: Conference Series}\ }\textbf
  {\bibinfo {volume} {200}},\ \bibinfo {pages} {062030} (\bibinfo {year}
  {2010})}\BibitemShut {NoStop}%
\bibitem [{\citenamefont {Bender}\ and\ \citenamefont
  {Tserkovnyak}(2015)}]{Bender2015}%
  \BibitemOpen
  \bibfield  {author} {\bibinfo {author} {\bibfnamefont {S.~A.}\ \bibnamefont
  {Bender}}\ and\ \bibinfo {author} {\bibfnamefont {Y.}~\bibnamefont
  {Tserkovnyak}},\ }\href {\doibase 10.1103/PhysRevB.91.140402} {\bibfield
  {journal} {\bibinfo  {journal} {Phys. Rev. B}\ }\textbf {\bibinfo {volume}
  {91}},\ \bibinfo {pages} {140402} (\bibinfo {year} {2015})}\BibitemShut
  {NoStop}%
\bibitem [{Note1()}]{Note1}%
  \BibitemOpen
  \bibinfo {note} {In general, the electron can be scattered with any momentum,
  however, upon explicitly introducing regular (\(R\)) and Umklapp (\(U\))
  magnon operators, all magnon scattering processes can be described within the
  reduced Brillouin zone.}\BibitemShut {Stop}%
\bibitem [{\citenamefont {Linder}\ and\ \citenamefont
  {Balatsky}(2019)}]{Linder2019}%
  \BibitemOpen
  \bibfield  {author} {\bibinfo {author} {\bibfnamefont {J.}~\bibnamefont
  {Linder}}\ and\ \bibinfo {author} {\bibfnamefont {A.~V.}\ \bibnamefont
  {Balatsky}},\ }\href {\doibase 10.1103/RevModPhys.91.045005} {\bibfield
  {journal} {\bibinfo  {journal} {Rev. Mod. Phys.}\ }\textbf {\bibinfo {volume}
  {91}},\ \bibinfo {pages} {045005} (\bibinfo {year} {2019})}\BibitemShut
  {NoStop}%
\bibitem [{Note2()}]{Note2}%
  \BibitemOpen
  \bibinfo {note} {One may derive equations also when this is not the case, but
  they will be somewhat more involved, as there will be interference terms in
  the submatrix determinant \(\Theta \)}\BibitemShut {NoStop}%
\bibitem [{\citenamefont {Allen}\ and\ \citenamefont
  {Dynes}(1975)}]{AllenDynes1975}%
  \BibitemOpen
  \bibfield  {author} {\bibinfo {author} {\bibfnamefont {P.~B.}\ \bibnamefont
  {Allen}}\ and\ \bibinfo {author} {\bibfnamefont {R.~C.}\ \bibnamefont
  {Dynes}},\ }\href {\doibase 10.1103/PhysRevB.12.905} {\bibfield  {journal}
  {\bibinfo  {journal} {Phys. Rev. B}\ }\textbf {\bibinfo {volume} {12}},\
  \bibinfo {pages} {905} (\bibinfo {year} {1975})}\BibitemShut {NoStop}%
\bibitem [{\citenamefont {Sanderson}\ and\ \citenamefont
  {Curtin}(2016)}]{Armadillo1}%
  \BibitemOpen
  \bibfield  {author} {\bibinfo {author} {\bibfnamefont {C.}~\bibnamefont
  {Sanderson}}\ and\ \bibinfo {author} {\bibfnamefont {R.}~\bibnamefont
  {Curtin}},\ }\href {http://arma.sourceforge.net/armadillo_joss_2016.pdf}
  {\bibfield  {journal} {\bibinfo  {journal} {Journal of Open Source Software}\
  }\textbf {\bibinfo {volume} {1}},\ \bibinfo {pages} {26} (\bibinfo {year}
  {2016})}\BibitemShut {NoStop}%
\bibitem [{\citenamefont {Sanderson}\ and\ \citenamefont
  {Curtin}(2018)}]{Armadillo2}%
  \BibitemOpen
  \bibfield  {author} {\bibinfo {author} {\bibfnamefont {C.}~\bibnamefont
  {Sanderson}}\ and\ \bibinfo {author} {\bibfnamefont {R.}~\bibnamefont
  {Curtin}},\ }in\ \href
  {https://link.springer.com/chapter/10.1007%2F978-3-319-96418-8_50} {\emph
  {\bibinfo {booktitle} {Mathematical Software -- ICMS 2018}}}\ (\bibinfo
  {publisher} {Springer International Publishing},\ \bibinfo {address} {Cham},\
  \bibinfo {year} {2018})\ pp.\ \bibinfo {pages} {422--430}\BibitemShut
  {NoStop}%
\bibitem [{\citenamefont {$\text{GNU Scientiﬁc Library}$}()}]{GNU}%
  \BibitemOpen
  \bibfield  {author} {\bibinfo {author} {\bibnamefont {$\text{GNU Scientiﬁc
  Library}$}},\ }\href {https://www.gnu.org/software/gsl/} {\bibinfo  {journal}
  {https://www.gnu.org/software/gsl/}\ }\BibitemShut {NoStop}%
\bibitem [{Note3()}]{Note3}%
  \BibitemOpen
\bibfield  {journal} {  }\bibinfo {note} {It should be noted that although the
  size of the region with significant deviations from the zero frequency
  polarization increases with Matsubara frequency $\nu _m$, so does also the
  width of the region over which we expect the dominant contributions to the
  corresponding \(\lambda _2(i\nu _m)\). Thus, we still expect to be able to
  approximate the polarization by a constant for larger Matsubara frequencies
  \(\nu _m\).}\BibitemShut {Stop}%
\bibitem [{\citenamefont {Migdal}(1958)}]{Migdal1958}%
  \BibitemOpen
  \bibfield  {author} {\bibinfo {author} {\bibfnamefont {A.~B.}\ \bibnamefont
  {Migdal}},\ }\href {http://www.jetp.ac.ru/cgi-bin/dn/e_007_06_0996.pdf}
  {\bibfield  {journal} {\bibinfo  {journal} {JETP}\ }\textbf {\bibinfo
  {volume} {34}},\ \bibinfo {pages} {1438 } (\bibinfo {year}
  {1958})}\BibitemShut {NoStop}%
\bibitem [{\citenamefont {Allen}\ and\ \citenamefont
  {Mitrović}(1983)}]{Allen1983}%
  \BibitemOpen
  \bibfield  {author} {\bibinfo {author} {\bibfnamefont {P.~B.}\ \bibnamefont
  {Allen}}\ and\ \bibinfo {author} {\bibfnamefont {B.}~\bibnamefont
  {Mitrović}},\ }\href {\doibase
  https://doi.org/10.1016/S0081-1947(08)60665-7} {\emph {\bibinfo {title}
  {Theory of Superconducting Tc}}},\ edited by\ \bibinfo {editor}
  {\bibfnamefont {H.}~\bibnamefont {Ehrenreich}}, \bibinfo {editor}
  {\bibfnamefont {F.}~\bibnamefont {Seitz}}, \ and\ \bibinfo {editor}
  {\bibfnamefont {D.}~\bibnamefont {Turnbull}},\ \bibinfo {series} {Solid State
  Physics}, Vol.~\bibinfo {volume} {37}\ (\bibinfo  {publisher} {Academic
  Press},\ \bibinfo {year} {1983})\ pp.\ \bibinfo {pages} {1--92}\BibitemShut
  {NoStop}%
\bibitem [{\citenamefont {Virosztek}\ and\ \citenamefont
  {Ruvalds}(1990)}]{Virosztek1990}%
  \BibitemOpen
  \bibfield  {author} {\bibinfo {author} {\bibfnamefont {A.}~\bibnamefont
  {Virosztek}}\ and\ \bibinfo {author} {\bibfnamefont {J.}~\bibnamefont
  {Ruvalds}},\ }\href {\doibase 10.1103/PhysRevB.42.4064} {\bibfield  {journal}
  {\bibinfo  {journal} {Phys. Rev. B}\ }\textbf {\bibinfo {volume} {42}},\
  \bibinfo {pages} {4064} (\bibinfo {year} {1990})}\BibitemShut {NoStop}%
\bibitem [{\citenamefont {Virosztek}\ and\ \citenamefont
  {Ruvalds}(1999)}]{Virosztek1999}%
  \BibitemOpen
  \bibfield  {author} {\bibinfo {author} {\bibfnamefont {A.}~\bibnamefont
  {Virosztek}}\ and\ \bibinfo {author} {\bibfnamefont {J.}~\bibnamefont
  {Ruvalds}},\ }\href {\doibase 10.1103/PhysRevB.59.1324} {\bibfield  {journal}
  {\bibinfo  {journal} {Phys. Rev. B}\ }\textbf {\bibinfo {volume} {59}},\
  \bibinfo {pages} {1324} (\bibinfo {year} {1999})}\BibitemShut {NoStop}%
\bibitem [{\citenamefont {Schrodi}\ \emph {et~al.}(2020)\citenamefont
  {Schrodi}, \citenamefont {Oppeneer},\ and\ \citenamefont
  {Aperis}}]{Schrodi2020_2}%
  \BibitemOpen
  \bibfield  {author} {\bibinfo {author} {\bibfnamefont {F.}~\bibnamefont
  {Schrodi}}, \bibinfo {author} {\bibfnamefont {P.~M.}\ \bibnamefont
  {Oppeneer}}, \ and\ \bibinfo {author} {\bibfnamefont {A.}~\bibnamefont
  {Aperis}},\ }\href {\doibase 10.1103/PhysRevB.102.024503} {\bibfield
  {journal} {\bibinfo  {journal} {Phys. Rev. B}\ }\textbf {\bibinfo {volume}
  {102}},\ \bibinfo {pages} {024503} (\bibinfo {year} {2020})}\BibitemShut
  {NoStop}%
\bibitem [{\citenamefont {Madhukar}(1977)}]{Madhukar1977}%
  \BibitemOpen
  \bibfield  {author} {\bibinfo {author} {\bibfnamefont {A.}~\bibnamefont
  {Madhukar}},\ }\href {\doibase https://doi.org/10.1016/0038-1098(77)90554-3}
  {\bibfield  {journal} {\bibinfo  {journal} {Solid State Communications}\
  }\textbf {\bibinfo {volume} {24}},\ \bibinfo {pages} {11 } (\bibinfo {year}
  {1977})}\BibitemShut {NoStop}%
\bibitem [{\citenamefont {Szcz\c{e}śniak}(2005)}]{Szczesniak2005}%
  \BibitemOpen
  \bibfield  {author} {\bibinfo {author} {\bibfnamefont {R.}~\bibnamefont
  {Szcz\c{e}śniak}},\ }\href {\doibase
  https://doi.org/10.1016/j.physleta.2004.12.070} {\bibfield  {journal}
  {\bibinfo  {journal} {Physics Letters A}\ }\textbf {\bibinfo {volume}
  {336}},\ \bibinfo {pages} {402 } (\bibinfo {year} {2005})}\BibitemShut
  {NoStop}%
\bibitem [{\citenamefont {Matsumoto}\ \emph {et~al.}(2012)\citenamefont
  {Matsumoto}, \citenamefont {Koga},\ and\ \citenamefont
  {Kusunose}}]{Matsumoto2012}%
  \BibitemOpen
  \bibfield  {author} {\bibinfo {author} {\bibfnamefont {M.}~\bibnamefont
  {Matsumoto}}, \bibinfo {author} {\bibfnamefont {M.}~\bibnamefont {Koga}}, \
  and\ \bibinfo {author} {\bibfnamefont {H.}~\bibnamefont {Kusunose}},\ }\href
  {\doibase 10.1143/JPSJ.81.033702} {\bibfield  {journal} {\bibinfo  {journal}
  {Journal of the Physical Society of Japan}\ }\textbf {\bibinfo {volume}
  {81}},\ \bibinfo {pages} {033702} (\bibinfo {year} {2012})}\BibitemShut
  {NoStop}%
\bibitem [{\citenamefont {Cebula}\ \emph {et~al.}(2000)\citenamefont {Cebula},
  \citenamefont {Zieli{\'{n}}ski},\ and\ \citenamefont
  {Biesiada}}]{Cebula2000}%
  \BibitemOpen
  \bibfield  {author} {\bibinfo {author} {\bibfnamefont {A.}~\bibnamefont
  {Cebula}}, \bibinfo {author} {\bibfnamefont {J.}~\bibnamefont
  {Zieli{\'{n}}ski}}, \ and\ \bibinfo {author} {\bibfnamefont {J.}~\bibnamefont
  {Biesiada}},\ }\href {\doibase 10.1023/A:1007723613544} {\bibfield  {journal}
  {\bibinfo  {journal} {Journal of Superconductivity}\ }\textbf {\bibinfo
  {volume} {13}},\ \bibinfo {pages} {479} (\bibinfo {year} {2000})}\BibitemShut
  {NoStop}%
\bibitem [{\citenamefont {Kittel}(1987)}]{Kittel1987}%
  \BibitemOpen
  \bibfield  {author} {\bibinfo {author} {\bibfnamefont {C.}~\bibnamefont
  {Kittel}},\ }\href@noop {} {\emph {\bibinfo {title} {Quantum Theory of
  Solids}}}\ (\bibinfo  {publisher} {John Wiley \& Sons},\ \bibinfo {year}
  {1987})\BibitemShut {NoStop}%
\bibitem [{\citenamefont {Horn}\ and\ \citenamefont
  {Johnson}(2012)}]{Horn2012}%
  \BibitemOpen
  \bibfield  {author} {\bibinfo {author} {\bibfnamefont {R.~A.}\ \bibnamefont
  {Horn}}\ and\ \bibinfo {author} {\bibfnamefont {C.~R.}\ \bibnamefont
  {Johnson}},\ }\href@noop {} {\emph {\bibinfo {title} {Matrix Analysis}}}\
  (\bibinfo  {publisher} {Cambridge University Press},\ \bibinfo {address}
  {Cambridge},\ \bibinfo {year} {2012})\BibitemShut {NoStop}%
\bibitem [{Note4()}]{Note4}%
  \BibitemOpen
  \bibinfo {note} {The identity can also be formulated more generally, but this
  will be sufficient for our purposes.}\BibitemShut {Stop}%
\bibitem [{\citenamefont {Roy}\ \emph {et~al.}(2014)\citenamefont {Roy},
  \citenamefont {Sau},\ and\ \citenamefont {Das~Sarma}}]{Roy2014}%
  \BibitemOpen
  \bibfield  {author} {\bibinfo {author} {\bibfnamefont {B.}~\bibnamefont
  {Roy}}, \bibinfo {author} {\bibfnamefont {J.~D.}\ \bibnamefont {Sau}}, \ and\
  \bibinfo {author} {\bibfnamefont {S.}~\bibnamefont {Das~Sarma}},\ }\href
  {\doibase 10.1103/PhysRevB.89.165119} {\bibfield  {journal} {\bibinfo
  {journal} {Phys. Rev. B}\ }\textbf {\bibinfo {volume} {89}},\ \bibinfo
  {pages} {165119} (\bibinfo {year} {2014})}\BibitemShut {NoStop}%
\bibitem [{\citenamefont {Djajaputra}\ and\ \citenamefont
  {Ruvalds}(1999)}]{Djajaputra1999}%
  \BibitemOpen
  \bibfield  {author} {\bibinfo {author} {\bibfnamefont {D.}~\bibnamefont
  {Djajaputra}}\ and\ \bibinfo {author} {\bibfnamefont {J.}~\bibnamefont
  {Ruvalds}},\ }\href {\doibase 10.1142/S0217979299000035} {\bibfield
  {journal} {\bibinfo  {journal} {International Journal of Modern Physics B}\
  }\textbf {\bibinfo {volume} {13}},\ \bibinfo {pages} {25} (\bibinfo {year}
  {1999})}\BibitemShut {NoStop}%
\bibitem [{\citenamefont {Piasecki}(2008)}]{Piasecki2008}%
  \BibitemOpen
  \bibfield  {author} {\bibinfo {author} {\bibfnamefont {R.}~\bibnamefont
  {Piasecki}},\ }\href@noop {} {\  (\bibinfo {year} {2008})},\ \Eprint
  {http://arxiv.org/abs/0804.1037} {arXiv:0804.1037 [cond-mat.str-el]}
  \BibitemShut {NoStop}%
\end{thebibliography}%

\end{document}